\documentclass[a4paper,11pt]{article}
\pdfoutput=1 

\usepackage{jinstpub} 

\usepackage{color}
\usepackage{enumitem}
\usepackage{threeparttable}
\usepackage[normalem]{ulem}
\usepackage{lineno}
\usepackage{graphicx}
\usepackage{subcaption}
\usepackage{mwe}
\usepackage{amsmath}

\usepackage{amssymb}
\usepackage[title]{appendix}
\title{Appendix}

\usepackage{siunitx}
\usepackage{newtxmath,newtxtext}

\usepackage{overpic}
\usepackage{longtable}
\usepackage{multirow}

\title{\boldmath \center \LARGE Improved muon energy estimation using a detailed model of multiple Coulomb scattering in the MicroBooNE LArTPC}

\collaboration{MicroBooNE Collaboration}

\author[mm]{P.~Abratenko}
\author[n]{D.~Andrade~Aldana}
\author[ll]{J.~Asaadi}
\author[kk]{A.~Ashkenazi}
\author[l]{S.~Balasubramanian}
\author[l]{B.~Baller}

\author[dd]{A.~Barnard}
\author[dd]{G.~Barr}
\author[dd]{D.~Barrow}

\author[z]{J.~Barrow}
% ...

% ...
\affiliation[l]{Fermi National Accelerator Laboratory (FNAL), Batavia, IL 60510, USA}

% ...
\affiliation[n]{Illinois Institute of Technology (IIT), Chicago, IL 60616, USA}
% ...
\affiliation[z]{University of Minnesota, Minneapolis, MN, 55455, USA}
% ...
\affiliation[dd]{University of Oxford, Oxford OX1 3RH, United Kingdom}
% ...
\affiliation[kk]{Tel Aviv University, Tel Aviv, Israel, 69978}
\affiliation[ll]{University of Texas, Arlington, TX, 76019, USA}
\affiliation[mm]{Tufts University, Medford, MA, 02155, USA}
% ...

  \emailAdd{microboone\_info@fnal.gov}
\date{}

\abstract{We present an improved technique for estimating a muon's energy by measuring the deflections along its path inside the MicroBooNE detector from multiple Coulomb scattering (MCS). This approach implements several innovations that better capture detector non-idealizations compared to previous MCS-based muon energy estimators. As a result, it achieves improved resolution, reduced bias, and better data-model agreement. Using model simulation, for fully contained events the estimated bias is within 1\% and the estimated resolution narrows from 10\% to 4.3\% as muon energy increases from 0.1\,GeV to 2\,GeV. For events with particles exiting the detector volume, at least a meter of reconstructed muon track, and a muon energy below 2\,GeV, the estimated bias is less than 2\% and the estimated resolution varies from 7\% to 17\% over muon energy. These demonstrate significant improvements over the performance of previous work using an MCS-based energy estimator at MicroBooNE~\cite{mcs_2017}, which exhibited approximately twice worse resolution and a bias of 20\% over the same energy region. Data-model goodness-of-fit studies are used to validate the estimator's performance on data, showing good agreement within model uncertainties. }


\begin{document}
\maketitle
\flushbottom

\section{Introduction}\label{sec:intro}

Exiting muons present a difficult reconstruction challenge to liquid argon time projection chamber (LArTPC) detectors, as they leave the detector volume before stopping. Traditional muon energy ($E_{\mu}$) estimators, such as those derived from the muon's reconstructed calorimetric energy or residual range, require the full muon track to accurately estimate its energy. In the MicroBooNE LArTPC detector~\cite{uboone_design}, over half of muons from $\nu_{\mu}$ charged current (CC) neutrino interactions from the Booster Neutrino Beam (BNB) exit the detector. This effect is most pronounced at high muon energies, making the muon and neutrino energy in certain phase space regions of neutrino interactions difficult to accurately reconstruct. These facts motivate the reconstruction of $E_{\mu}$ through measurements of a muon's multiple Coulomb scattering (MCS) from electromagnetic interactions with argon nuclei, as the full muon track is not required to be reconstructed under this method. MicroBooNE has previously developed a MCS-based $E_{\mu}$ estimator~\cite{mcs_2017}, although its simplified model implementation limits its accuracy and leaves room for improvement.

The MicroBooNE detector is a 10.36\,m by 2.56\,m by 2.32\,m LArTPC with an 85 tonne active mass. It achieves sub-cm-level position resolution and MeV-level detection thresholds~\cite{uboone_design} from the reconstruction of ionization charge. An applied 273\,V/cm electric field causes ionized electrons to drift towards an anode consisting of 3 wire planes offset by 3\,mm from each other. Each wire plane contains parallel wires with a 3\,mm separation, and the wires are aligned at orientations of $+60^{\circ}$, $0^{\circ}$, and $-60^{\circ}$ with respect to the vertical, giving maximal angular coverage between the three planes~\cite{uboone_design}. Thirty-two photomultiplier tubes are used to detect scintillation light and provide a prompt timing signal to isolate neutrino interactions from a background of cosmic rays. Neutrinos are generated in the BNB 470\,m upstream of the MicroBooNE detector from 8\,GeV protons hitting a beryllium target. The neutrino flux is estimated to be 93.6\% $\nu_{\mu}$, 5.9\% $\overline{\nu}_{\mu}$, 0.5\% $\nu_{e}+\overline{\nu}_{e}$, and have a mean energy of 0.8\,GeV~\cite{uboone_flux}. For the remainder of the paper, we use the term $\nu_{\mu}$ to refer to $\nu_{\mu} + \overline{\nu}_{\mu}$, and $\mu$ to refer to $\mu + \overline{\mu}$. Neutrino-argon interactions are modeled using \texttt{GENIE} v3.0.6 G18\_10a\_02\_11a tuned to T2K data~\cite{genie-tune-paper,T2K:2016jor} and the final state particle interactions are simulated in the liquid argon medium using \texttt{GEANT4}~\cite{geant_4}. Detector response modeling takes into account the impact of variations in TPC waveform, light yield and propagation, the space charge effect, and ionization recombination~\cite{recombination,sce1,sce2}, with more details available in Ref.~\cite{wiremod}.

This work studies muons created in BNB $\nu_{\mu}$ CC interactions. It uses data taken from MicroBooNE's first three runs as well as corresponding simulation. An inclusive $\nu_{\mu}$ CC selection previously developed for MicroBooNE's low energy excess search~\cite{numuCC_selection} is applied to both data and simulated samples, achieving an estimated selection efficiency of 68\% and purity of 92\% on the combined selection of fully contained (FC) and partially contained (PC) neutrino interactions. The reconstruction of these events utilizes tomographic imaging~\cite{imaging} to create a three-dimensional (3D) point cloud of ionized charge, from which precise particle trajectories consisting of $\sim$cm spaced points are determined. This work takes the reconstructed trajectory of the muon as input and separates it into 14\,cm segments, each fit with a line, to measure MCS through the angles between adjacent fitted line segments. The segment length of 14\,cm is used as it is the Bremsstrahlung radiation length of a muon in liquid argon and simplifies the relation between the muon's energy and its deflection after traveling one segment length. An additional 21\,cm length requirement is placed on the reconstructed muon so that the reconstructed track is long enough for at least two segments, allowing the final segment to be as short as 7\,cm or half a normal segment length. The reconstruction of multiple segments makes it possible to measure the effect of MCS, and results in a final selection of 103\,k events in data.

In this paper, we present a new MCS-based $E_{\mu}$ estimator for use in LArTPC experiments that introduces four key improvements over previous work~\cite{mcs_2017}. These are: a double-Gaussian scattering angle PDF that accurately models both the core and tails of the angular distribution, a separation of planar scattering angles into drift-sensitive and wire-plane-sensitive components, enabling independent treatment of their different detector resolutions, a track-orientation-dependent tuning that accounts for variations in reconstruction quality as a function of the track's alignment with the drift direction, and a new approach to determining the detector angular resolution directly from the scattering angle distribution of high-energy muons. Together, these improvements yield reduced bias, improved resolution, and better data-model agreement. In section~\ref{sec:mcs_intro} we introduce a theoretical characterization of MCS as well as non-idealizations in the measurement of MCS originating from the LArTPC detector. In section~\ref{sec:tuning} we describe the tuning of MCS angle distribution probability distribution functions (PDFs) and their use in determining $E_{\mu}$ through the process of maximum likelihood estimation (MLE). In section~\ref{sec:resolution} we report the estimated resolution and bias of the $E_{\mu}$ estimator using simulation, and in section~\ref{sec:validation} we validate the accuracy of this simulation through data-model comparisons. Finally, in section~\ref{sec:conclusion} we summarize the novel features and performance of the $E_{\mu}$ estimator. More details on the algorithm used and its tuning are provided in the appendix, and the body of the code is available on Github~\cite{mcs_github} along with a small example of it in use.
\section{Description of MCS}\label{sec:mcs_intro}

\subsection{Overview of multiple Coulomb scattering}

In the few-GeV region, muons scatter off target nuclei following a roughly inverse relation with the muon's energy as a result of the Coulomb force. MCS-based algorithms leverage this feature to estimate the muon's momentum from multiple measurements of its scattering angles.  More formally, multiple scattering theory has been developed by Moli\`ere and others~\cite{moliere} and is parametrized by the Highland formula, which estimates the distribution and associated variance of scattering angles, with revised fit parameters $S_{2}=13.6$\,MeV and $\epsilon = 0.038$ from Lynch and Dahl~\cite{highland}. From the Highland prediction, after traveling a distance $X$ through a medium with Bremsstrahlung radiation length $X_{0}$, the quantity $\sigma_{H}$ describes the width of a Gaussian probability distribution function (PDF) describing the scattering angle, $\theta$, of a particle within a plane constructed from the its outgoing velocity and a vector perpendicular to its incident velocity
\begin{equation}
    \sigma_{H} = \frac{S_{2}}{pc\beta}z\sqrt{\frac{X}{X_{0}}} \left(1+\epsilon \times \mathrm{ln}\left(\frac{Xz^{2}}{X_{0}\beta^{2}}\right) \right),
    \label{eqn:highland}
\end{equation}
where the particle's momentum is $p$, its velocity is $c\beta$, and its charge magnitude is $z$. In the case of a highly relativistic muon traveling one radiation length $X_{0}$, estimated to be 14\,cm in liquid argon, this relation simplifies to
\begin{equation}
    \sigma_{H} \approx \frac{S_{2}}{pc\beta} = \frac{S_{2}}{E_{\mu}(1-\frac{m^{2}}{E_{\mu}^{2}})},
    \label{eqn:highland_simple}
\end{equation}
where $m \approx 105.7$\,MeV is the muon's mass.

Typically the deflection is measured over a segment length rather than instantaneously, and is expected  to be Gaussian distributed with width $\frac{1}{\sqrt3} \sigma_{H}$. Since the scattering angle is computed as the difference between two reconstructed segment directions, incoming and outgoing, the variances in each segment direction are added. Under the approximation that the particle energy is the same between segments, this yields a measured scattering angle distribution with a width of $\sqrt\frac{2}{3} \sigma_{H}$. Since the fit parameter $S_{2} = 13.6$\,MeV was determined through a fit to a range of atomic nuclei, a more accurate depiction in the case of liquid argon requires a dedicated parameter tune. This can be achieved by adding a tuning function $\kappa (E_{\mu})$, which can also correct for approximations in the Highland formula as a function of muon energy and will be given a precise functional form in section~\ref{sec:tuning}. Additionally, a constant term $\sigma_{\mathrm{res}}$, assumed to cover the detector angular resolution, can be added in quadrature to give a more realistic depiction of the measured angular distribution:
\begin{equation}
    \sigma_{\mathrm{pred}}^{2}(E_{\mu}) = \frac{2}{3} \kappa^{2}(E_{\mu}) \sigma_{H}^{2}(E_{\mu}) + \sigma_{\mathrm{res}}^{2}.
    \label{eqn:highland_full}
\end{equation}

The previous work~\cite{mcs_2017} implemented a MCS-based muon energy estimator following the theoretical treatment above. In this approach, muon tracks are split into 14\,cm segments, and the planar scattering angles between segments is computed, with an arbitrary choice of orthogonal coordinates within the plane perpendicular to each segment. These angle measurements are used to estimate the starting muon kinetic energy using a maximum likelihood approach. The estimated muon starting kinetic energy is propagated to each angle location based on the distance traveled using the mean energy loss predicted by the Bethe-Bloch formula~\cite{bethe-bloch}, generating predicted kinetic energies referred to as ``local $KE_{\mu}$''. A predicted PDF is constructed by using the local $KE_{\mu}$ Highland prediction, equation~\ref{eqn:highland}, for the width of a Gaussian distribution. This PDF is modified by a scaling function $\kappa(E_{\mu})$ used to tune the Highland prediction to the detector reconstruction performance, as well as a constant term added in quadrature to describe the expected detector resolution. The PDF for each angle measurement is evaluated at the corresponding measurement value to generate a series of individual measurement likelihoods, which are multiplied together to get the total likelihood. The starting muon energy estimate is varied to find the value that maximizes the total likelihood, which is taken as the maximum likelihood estimator (MLE).

\subsection{Improvements in the modeling of muon scattering angle distributions}

We build upon the above MCS model by refining the assumed scattering PDF and accounting for LArTPC detector effects and reconstruction errors. Beginning with the theoretical description of MCS, the Highland formula assumes a Gaussian distribution of planar scattering angles and neglects the non-Gaussian tails. Although these tails only contribute a few percent of the total PDF area, when they are observed the Gaussian PDF model will significantly under-predict their likelihood relative to their true likelihood. Since the process of maximum likelihood estimation relies on the product of likelihoods from measured angles $\theta$, a single highly erroneous likelihood can heavily penalize the overall likelihood prediction and force the likelihood maximization to prefer lower $E_{\mu}$ predictions that reduce this erroneous penalty and bias the $E_{\mu}$ estimator. Given that muon tracks often contain a dozen or more angle measurements, this represents a sizable fraction of events. This has the largest impact at high energies where muon tracks are longer. Furthermore, the planar scattering angle distribution of high energy muons is modeled by narrow-width Gaussian distributions. This model approach does not have the flexibility to allow for the existence of reconstruction flaws, such as the inclusion of delta rays in the muon track, which serve to increase the observed long tail of the scattering angle distribution beyond what MCS-focused theory predicts.

We introduce an expanded scattering angle PDF model description. Rather than use a Gaussian distribution, we construct a double-Gaussian function $f(\theta;E_{\mu})$:
\begin{equation}
    f(\theta;E_{\mu}) = \frac{A(E_{\mu})}{\sigma_{1}(E_{\mu}) \sqrt{2\pi}} e^{\theta^{2}/2\sigma_{1}^{2}(E_{\mu})} + \frac{1-A(E_{\mu})}{\sigma_{2}(E_{\mu}) \sqrt{2\pi}} e^{\theta^{2}/2\sigma_{2}^{2}(E_{\mu})}.
    \label{eqn:double_gaussian}
\end{equation}
The single parameter $\sigma_{\mathrm{pred}}$ is replaced with two parameters $\sigma_{1}$ and $\sigma_{2}$ that determine the widths of each Gaussian, as well as an area fraction parameter $A$ that determines how much each Gaussian contributes to the total PDF area, with $A=1$ representing a single-Gaussian PDF. Each of these free parameters may be tuned as functions of $E_{\mu}$ similar to how $\kappa (E_{\mu})$ is tuneable in the original model description of equation~\ref{eqn:highland_full}. More details on this tuning process are given in section~\ref{sec:tuning}. The inclusion of two Gaussian terms allows the model to describe the dominant Gaussian behavior that contains $\sim$90\% of the PDF area that is well described by the Highland formula while also preserving accurate likelihood modeling in the tails. We refer to the Gaussian term described by the Highland formula that has a larger area contribution as the primary Gaussian, and we refer to the other Gaussian term with a smaller area contribution and wider width as the secondary Gaussian. In this manner, the double-Gaussian PDF enables an accurate likelihood estimation for measured $\theta$ over a wide range of $\theta$ values.

\begin{figure}
    \centering
    \includegraphics[width=0.75\linewidth]{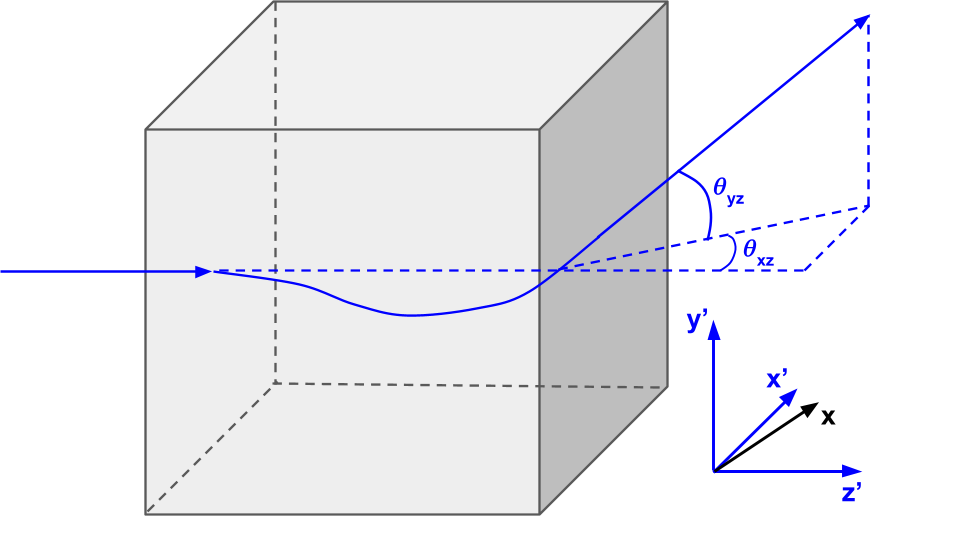}
    \caption{Illustration of a muon path in blue as it passes through a segment length of material. In this example, the incoming muon direction $z'$ is nearly perpendicular to the detector drift direction $x$, therefore $y' \approx y$ and $x' \approx x$, and the drift coordinate sensitivity is almost entirely captured in $\theta_{xz}$ and removed from $\theta_{yz}$.}
    \label{fig:coordinates}
\end{figure}

The second modification we introduce addresses an asymmetry in the reconstruction of charge within the LArTPC detector. Although MCS inherently is symmetric with respect to the angle in the transverse plane of the incident muon, the MicroBooNE detector is not. There are different reconstruction resolutions along the drift direction ($x$) and the wire-plane directions ($y,z$). Additionally, ionization charge diffuses differently between these directions due to the presence of an applied electric field along the drift direction. For these reasons, we would ideally choose axes in the transverse muon plane that separate these effects. When the muon is traveling in the $yz$ plane, complete separation of drift and wire-plane directions in the construction of local coordinate axes is possible. However, in realistic cases, the muon has a nonzero velocity in the drift direction. Therefore, the local axes $x'$ and $y'$ are defined to achieve maximal possible separation of drift and wire-plane directions within the transverse muon plane of the incident muon direction $z'$, as detailed in figure~\ref{fig:coordinates}.
\begin{align}
    y' & =   z' \times x, \\
    x' & = - z' \times y'
    \label{eqn:local_coords}.
\end{align}
The muon's two planar scattering angles are then distinguished as $\theta_{xz}$ and $\theta_{yz}$, where the prime notation on the $x', y', z'$ subscripts is dropped for simplicity. Letting an outgoing segment's direction be given as $u$, these are written as
\begin{align}
    \theta_{xz} & = \arctan(\frac{u \cdot x'}{u \cdot z'}), \\
    \theta_{yz} & = \arctan(\frac{u \cdot y'}{u \cdot z'}).
\end{align}
This definition leaves $\theta_{yz}$ minimally sensitive and $\theta_{xz}$ maximally sensitive to the detector positional resolution along the drift direction. This separation is very significant, as we find extreme differences in detector resolutions between these angles, as is discussed in more detail below. Previous studies have either relied on the 3D angle between track segments, $\theta_{3\mathrm{D}}^{2} \approx \theta_{xz}^{2} + \theta_{yz}^{2}$, or have used an arbitrary choice of $x'$, $y'$ within the transverse plane with no separate treatment of $\theta_{xz}$ and $\theta_{yz}$. Both of these approaches lose information and fail to fully utilize the higher resolution potential that is present in the majority of tracks that allow for significant separation of the drift and wire-plane directions in the local coordinate system.

This work also introduces a new approach to estimating the detector resolutions for $\theta_{xz}$ and $\theta_{yz}$. Rather than determining the resolution by finding the value that minimizes the overall bias in the $E_{\mu}$ estimator, we implement a more direct comparison to simulation through the examination of fits to measured scattering angle distributions. It is worth noting that the $E_{\mu}$ MLE is not guaranteed to be unbiased for finite sample sizes, so that even a perfectly modeled PDF may generate a non-trivial bias. Therefore, minimizing this bias may not yield accurate PDFs, harming the $E_{\mu}$ estimation in subtle ways. Instead, the approach used in this work attempts to directly capture the reconstructed distributions in simulation.

\begin{figure}[hbtp!]
     \centering
     \includegraphics[clip,trim={2.0cm 0.0cm 4.0cm 5.0cm},width=0.75\textwidth]{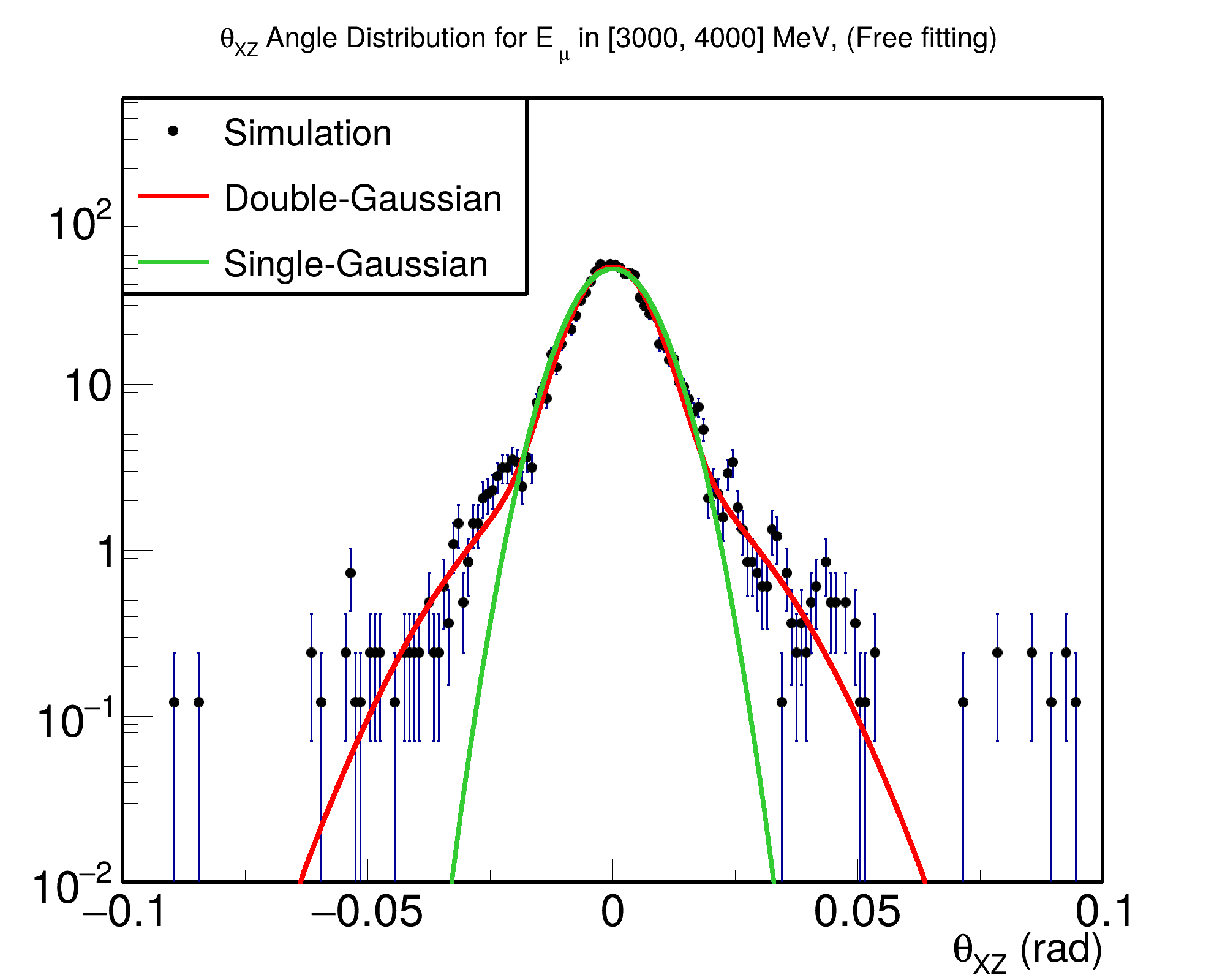}
     \put(-100,240){\Large MicroBooNE}
     \put(-90,227){\Large Simulation}
     \put(-225,260){\Large Local $KE_{\mu} \in [3,4]$\,GeV}
     \put(-335,95){\rotatebox{90}{\large Probability Density}}
     \caption{Comparison of single- and double-Gaussian PDFs to the reconstructed $\theta_{xz}$ distribution in simulation. Probability density is plotted on a log scale so differences at large angles can be seen. The average muon kinetic energy of the reconstructed segments adjacent to each angle measurement is in $[3,4]$\,GeV. Both PDFs are fit to the reconstructed distribution to show the best possible performance, and in the case of the double-Gaussian PDF to determine the detector resolutions.}
    \label{fig:theta_xz_highE}
\end{figure}
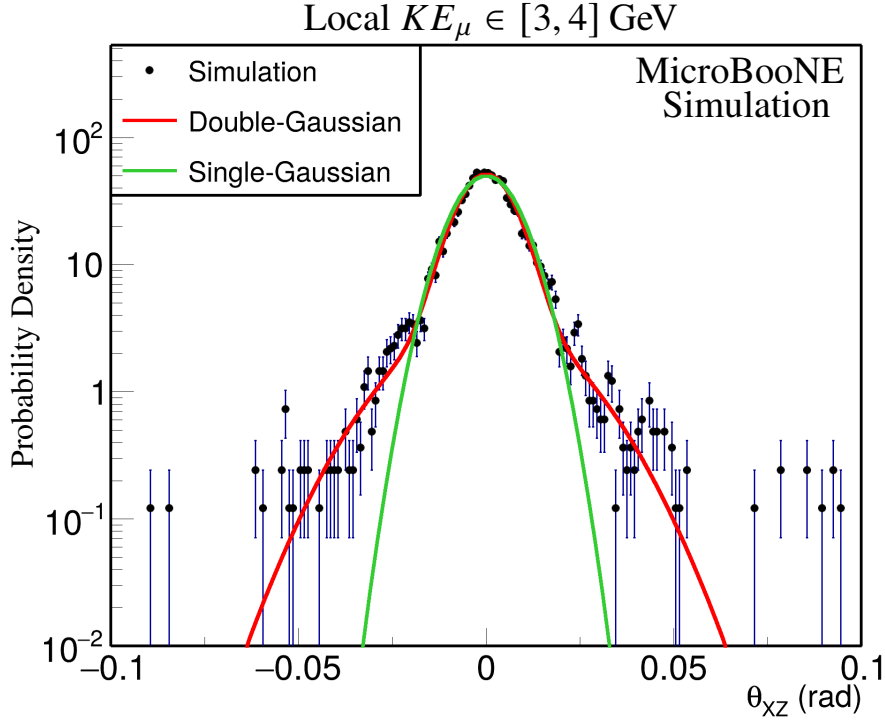

The detector resolution is computed by looking at segments of tracks where the true muon kinetic energy $KE_{\mu}$ is very high, at least 3\,GeV, and the MCS contribution to equation~\ref{eqn:highland_full} is minimized. We are then able to extract the detector resolution from fitting the PDF to the observed angle distribution with minimal dependence on the Highland formula. This is demonstrated in figure~\ref{fig:theta_xz_highE}, where the double-Gaussian fit to the measured $\theta_{xz}$ distribution yields $\sigma_{1} = 6.9$\,mrad, leading to a $\theta_{xz}$ detector resolution of $\sigma_{\text{res 1}} = 5.8$\,mrad. In a similar manner, we determine a resolution parameter $\sigma_{\text{res 2}}$ for the secondary Gaussian term in the double-Gaussian PDF, and separate resolution parameters for $\theta_{yz}$.

The final model expansion that we introduce attempts to account for differences in reconstruction performance based on the component of the track's direction along the drift direction, $v_{x}$. As this component increases in magnitude from $|v_{x}|=0$ to $|v_{x}|=1$, tracks shift from being isochronous, defined by a similar arrival time for all the ionized charge, to intermediate, to prolonged, defined by a track spanning a relatively long readout time with minimal $yz$ positional variation. There are significant differences in reconstruction performance as $|v_{x}|$ varies, associated with the different challenges in deconvolution and imaging that each case presents. However, minimal variation is observed in $\theta_{xz}$ over $|v_{x}|$, so the following treatment is only performed for $\theta_{yz}$.

\begin{figure}[htbp!]
    \centering
    \includegraphics[trim={0.0cm 3.0cm 0.0cm 1.5cm},clip,width=0.9\linewidth]{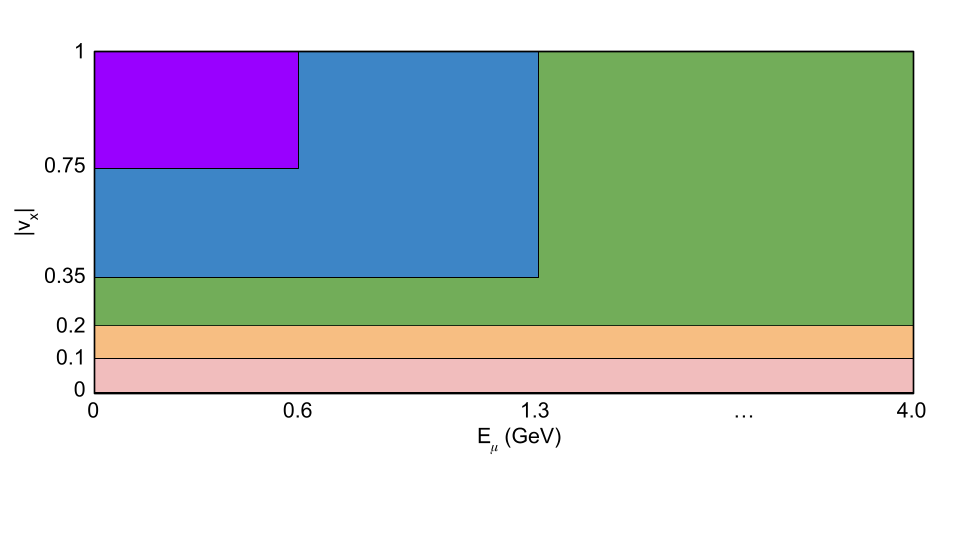}
    \caption{Illustration of $|v_{x}|$ binning for $\theta_{yz}$ tunes. Bin edges are $|v_{x}| \in [0, 0.1, 0.2, 0.35, 0.75, 1]$, and each color represents a different bin. The two largest $|v_{x}|$ bins do not extend to 4\,GeV in $E_{\mu}$ because of limited statistics in that part of the \{$E_{\mu}$, $|v_{x}|$\} phase space, limiting tuning capabilities. Instead, measured angles at high $E_{\mu}$ and high $|v_{x}|$ are placed at the next largest $|v_{x}|$ tune.}
    \label{fig:vx_binning}
\end{figure}

To accurately model the variation in the $\theta_{yz}$ distribution over $|v_{x}|$, we separate track segments into five bins in $|v_{x}|$, as illustrated in figure~\ref{fig:vx_binning}. For each $|v_{x}|$ bin we determine detector resolutions from simulation of high energy muons, as was done for $\theta_{xz}$, shown in Table~\ref{table:resolutions}, and perform five separate tunes of the fit parameters $\sigma_{1}$, $\sigma_{2}$, and $A$. When applying a likelihood function to estimate $E_{\mu}$ through MLE, for each measured angle $\theta_{yz}$ we use the tune corresponding to the bin of the average $|v_{x}|$ of the adjoining segments. There are a wide range of detector resolutions across $\theta_{xz}$, $\theta_{yz}$, and the five $|v_{x}|$ slices, demonstrating the importance of treating each case separately. Tracks with very low and very high $|v_{x}|$ correspond to isochronous and prolonged topologies respectively, where the reconstruction suffers from ambiguity in the deconvolution and imaging steps, as seen in the larger detector resolutions in these cases.

\begin{table}[htbp!]
    \caption{Detector resolutions $\sigma_{\text{res 1}}$ and $\sigma_{\text{res 2}}$ for each tune, as determined from the reconstructed angle scattering distribution at high $E_{\mu}$ in simulation.}
    \centering
    \begin{tabular}{|l|c|c|}
        \hline
        Tune & $\sigma_{\text{res 1}}$ (mrad) & $\sigma_{\text{res 2}}$ (mrad) \\
        \hline
        $\theta_{xz}$                            & $5.8 \pm 0.1$  & $18.2 \pm 0.2$ \\
        $\theta_{yz}$, $|v_{x}| \in [0,0.1]$     & $44.9 \pm 1.3$ & $150.0 \pm 3.9$ \\
        $\theta_{yz}$, $|v_{x}| \in [0.1,0.2]$   & $20.6 \pm 0.5$ & $51.5 \pm 1.6$ \\
        $\theta_{yz}$, $|v_{x}| \in [0.2,0.35]$  & $14.0 \pm 0.3$ & $39.7 \pm 2.4$ \\
        $\theta_{yz}$, $|v_{x}| \in [0.35,0.75]$ & $13.1 \pm 0.3$ & $41.8 \pm 4.3$ \\
        $\theta_{yz}$, $|v_{x}| \in [0.75,1]$    & $11.4 \pm 0.3$ & $73.5 \pm 4.9$ \\
        \hline
    \end{tabular}
    \label{table:resolutions}
\end{table}

In total, we introduce four new features to the construction and tuning of scattering angle PDFs. These allow the model to better describe, including variations in reconstruction quality, the reconstructed planar scattering angle distributions resulting from muon MCS and smearing by the detector's imperfect reconstruction of charge. The modifications to the scattering angle PDFs allow them to better describe the reconstructed distributions seen in the MicroBooNE detector, and will translate to a more effective model tuning and $E_{\mu}$ estimation described in section~\ref{sec:tuning}, reducing the bias and resolution in the estimator.  Table~\ref{table:resolutions} reports the measured resolutions for each tune, which finds detector angular resolutions for 14\,cm muon track segments. Compared to the 3\,mrad resolution used in the previous work~\cite{mcs_2017}, we find resolutions that are twice as large in the case of $\theta_{xz}$ and between 4--15 times as large in $\theta_{yz}$ depending on the track $|v_{x}|$ direction. This shows that the previous study used a resolution estimate that failed to capture the full variation in scattering angle measurements present in MicroBooNE simulation.
\section{Model tuning and energy estimation}\label{sec:tuning}

As previously discussed, the Highland formula has been fit to a range of different target nuclei, leading to the overall normalization parameter value of $S_{2}=13.6$\,MeV. To better match this prediction to liquid argon, and to capture other model approximations and reconstruction imperfections in the simulation that may vary with $E_{\mu}$, the $S_{2}$ parameter should be tuned to MicroBooNE simulation through the use of the arbitrary tuning function $\kappa(E_{\mu})$ in equation~\ref{eqn:highland_full}. Furthermore, the extension of the PDF from a theoretically-grounded single-Gaussian model to the mixed theoretical and effective description of the double-Gaussian creates new function parameters that lack a theoretical underpinning. As a result, rather than use a single tuning function $\kappa(E_{\mu})$, each parameter in equation~\ref{eqn:double_gaussian} is scaled by its own tuning function. These factors motivate a dedicated tuning effort so that the double-Gaussian PDF accurately models the reconstructed scattering angle distributions in simulation.

\subsection{Iterative tuning procedure}

The most direct way to tune the parameters $\sigma_{1}$, $\sigma_{2}$, and $A$ would be to construct pseudo-arbitrary functional forms, such as polynomials, for $\sigma_{1}(E_{\mu})$, $\sigma_{2}(E_{\mu})$, and $A(E_{\mu})$, and compare the predicted PDF distribution in bins of $E_{\mu}$ against the corresponding reconstructed scattering angle distributions in simulation. This would allow for a simultaneous tune of each polynomial parameter to the full range of muon energies observed in the detector. However, such a tune is complicated by a number of factors. There is partial degeneracy in the parameter space, where different values of \{$A$, $\sigma_{2}$\} can generate very similar PDFs. It is also possible to generate a likelihood curve with multiple local maxima in $E_{\mu}$ at fixed $\theta$ even when $\sigma_{1}$, $\sigma_{2}$, and $A$ are all strictly decreasing functions of $E_{\mu}$. Such a feature would both be theoretically undesirable---physics tells us higher energy muons generate tighter scattering angle distributions---and mathematically inconvenient. Maximum likelihood estimation works best with a single maximum, otherwise local maxima can increase ambiguity in the energy estimation and greatly increase the variance in the estimate.

For these reasons, a more directly controlled iterative tuning procedure is adopted rather than a simpler simultaneous tune to all three parameters. Throughout the iterative tuning procedure, the previously-determined detector resolutions shown in table~\ref{table:resolutions} are held constant. Instead of tuning through direct comparisons to the scattering angle distributions, double-Gaussian function fits are separately performed to each distribution to generate sets of fit parameters and then the tunes for $\sigma_{1}(E_{\mu})$, $A(E_{\mu})$, and $\sigma_{2}(E_{\mu})$ are performed in that order one at a time on the fit parameters. Here the term ``fit'' is used to refer to the $\chi^{2}$ minimization of the double-Gaussian functions in comparison to reconstructed scattering angle distributions, while ``tune'' is reserved for the $\chi^{2}$ minimization of $\sigma_{1}(E_{\mu})$, $\sigma_{2}(E_{\mu})$, and $A(E_{\mu})$. Between each tune, the double-Gaussian parameters are re-fit, holding any previously-tuned parameters at those fixed values. The tune of $\sigma_{1}(E_{\mu})$ is performed first as it closely follows the Highland prediction that provides a good first-order description of the scattering angle distribution. The tune of $A(E_{\mu})$ is performed second as it is highly constrained by $\sigma_{1}$ and the maximum amplitude of the angle distribution, leaving $\sigma_{2}(E_{\mu})$ to be tuned last. A more detailed description of the tuning procedure is provided in the appendix.

Reconstructed scattering angles from both FC and PC muon tracks are assigned a bin according to the average projected $KE_{\mu}$ at the midpoints of the segments that form the angle. This $E_{\mu}$ projection is determined by taking the muon's true starting energy and propagating an energy estimate along the reconstructed track, decreasing the energy in accordance with the mean energy loss predicted by the Bethe-Bloch formula~\cite{bethe-bloch}. Ideally, each $E_{\mu}$ bin would have a roughly constant Highland formula prediction across the bin range, but limited simulation statistics force some high energy bins to be wider than is desired.

\subsection{Comparison of tuned model prediction to simulation}

\begin{figure}[hbtp!]
     \centering
     \includegraphics[clip,trim={2.0cm 0.0cm 5.3cm 4.9cm},width=0.32\textwidth]{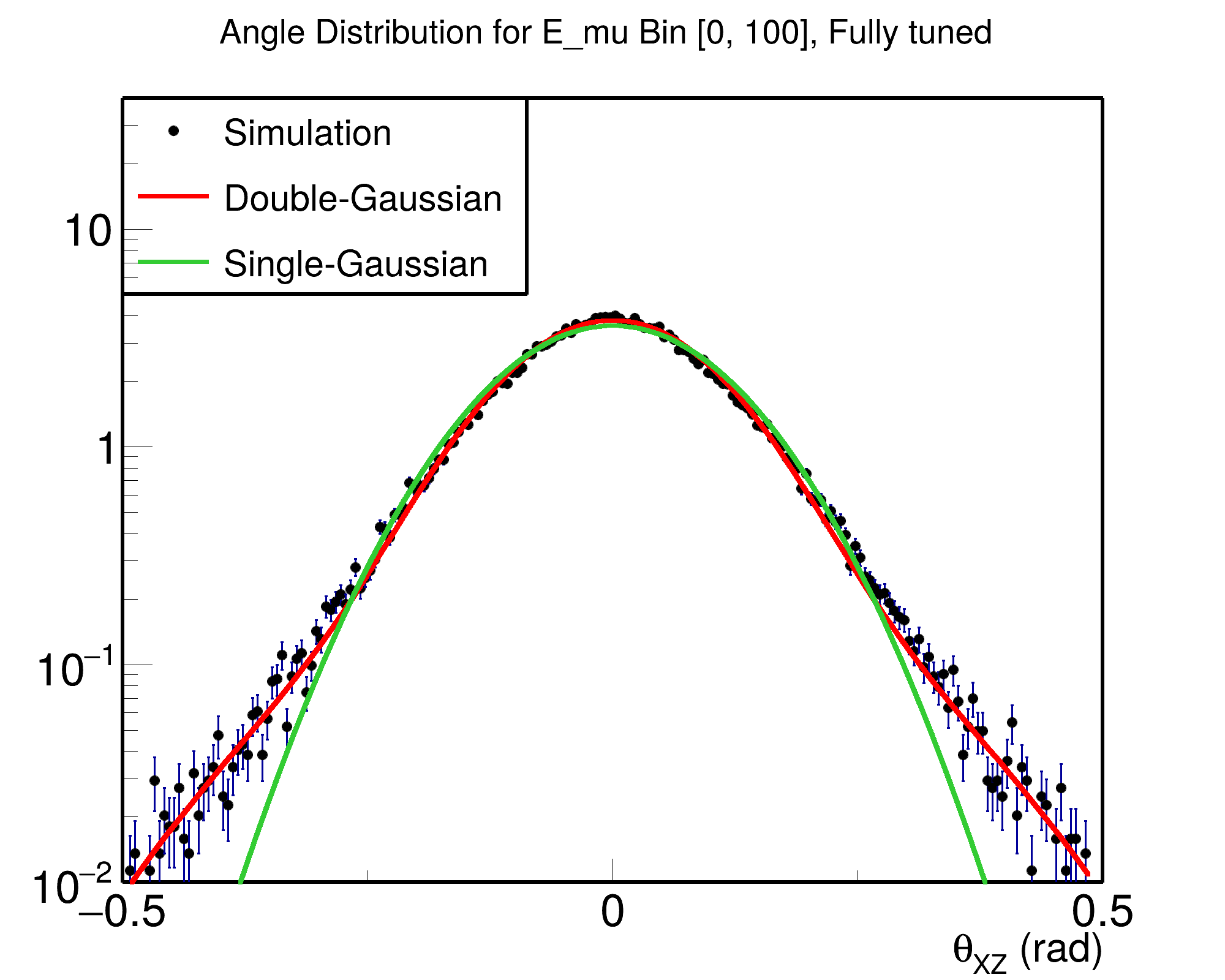}
     \includegraphics[clip,trim={2.0cm 0.0cm 5.3cm 4.9cm},width=0.32\textwidth]{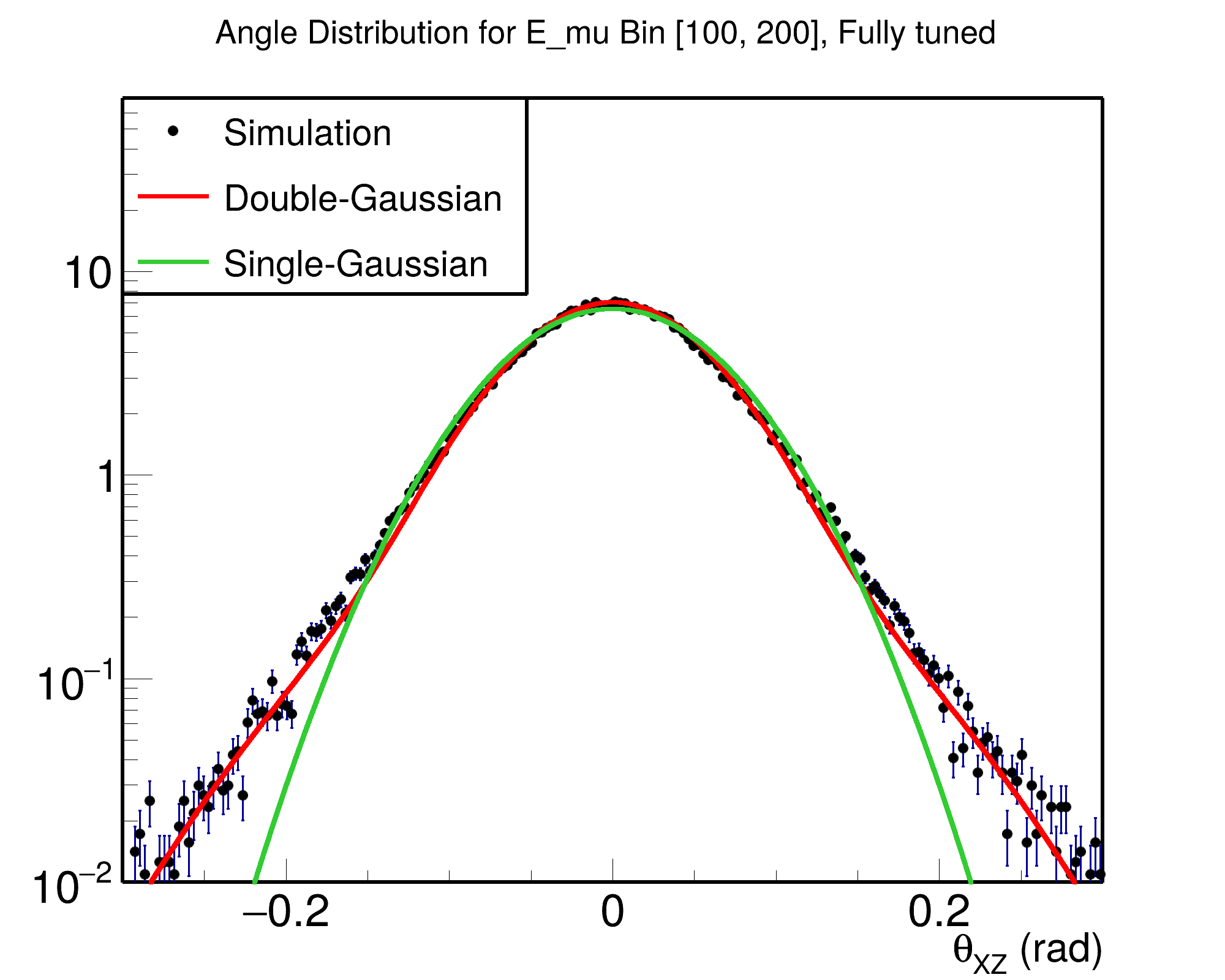}
     \vspace{0.3cm}
     \put(-255,116){\footnotesize Local $KE_{\mu} \in [0,0.1]$\,GeV}
     \put(-120,116){\footnotesize Local $KE_{\mu} \in [0.1,0.2]$\,GeV}
     \put(  20,116){\footnotesize Local $KE_{\mu} \in [0.2,0.3]$\,GeV}
     \put(-60,102){\footnotesize MicroBooNE}
     \put(-53, 91){\footnotesize Simulation}
     \includegraphics[clip,trim={2.0cm 0.0cm 5.3cm 4.9cm},width=0.32\textwidth]{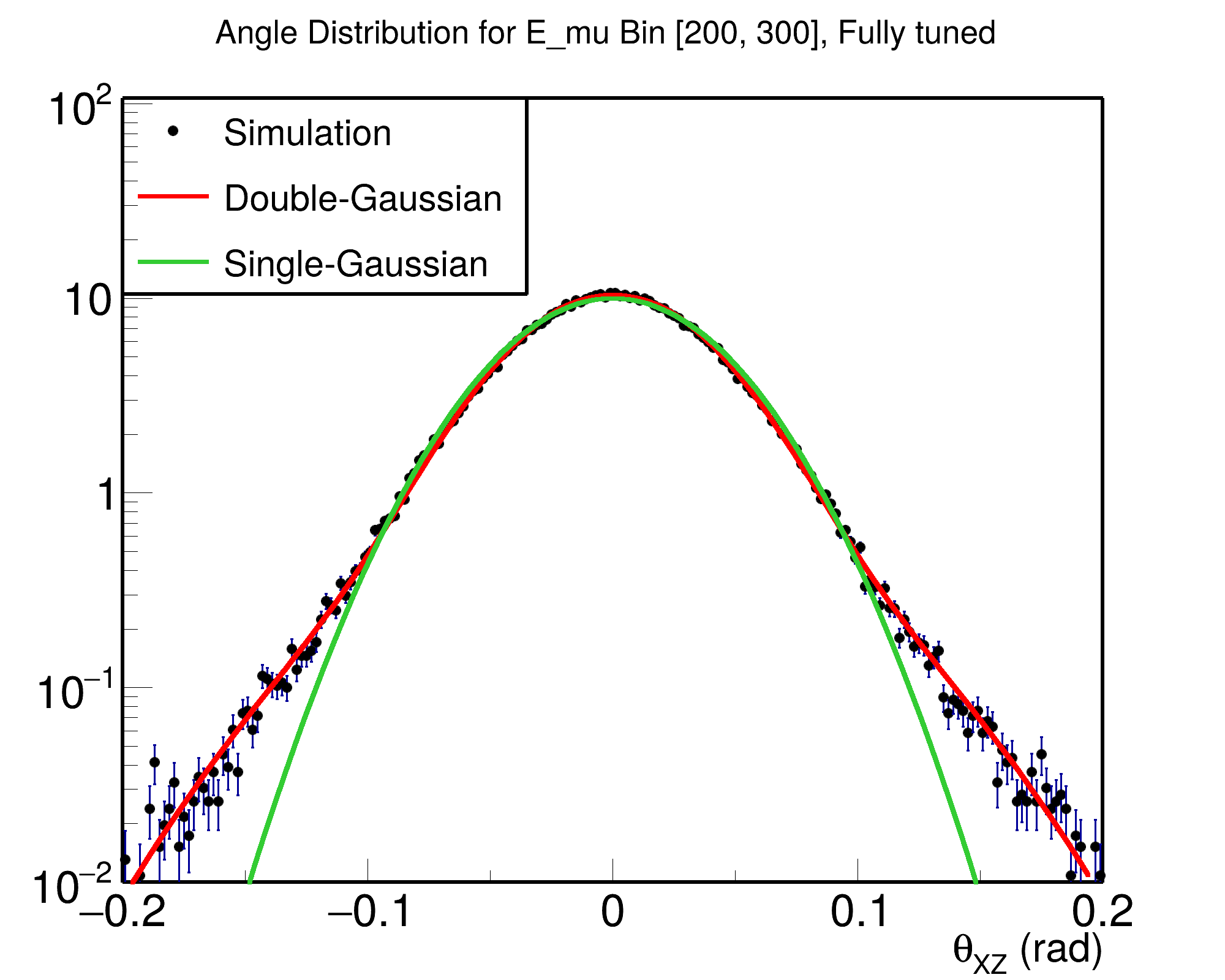}
     \includegraphics[clip,trim={2.0cm 0.0cm 5.3cm 4.9cm},width=0.32\textwidth]{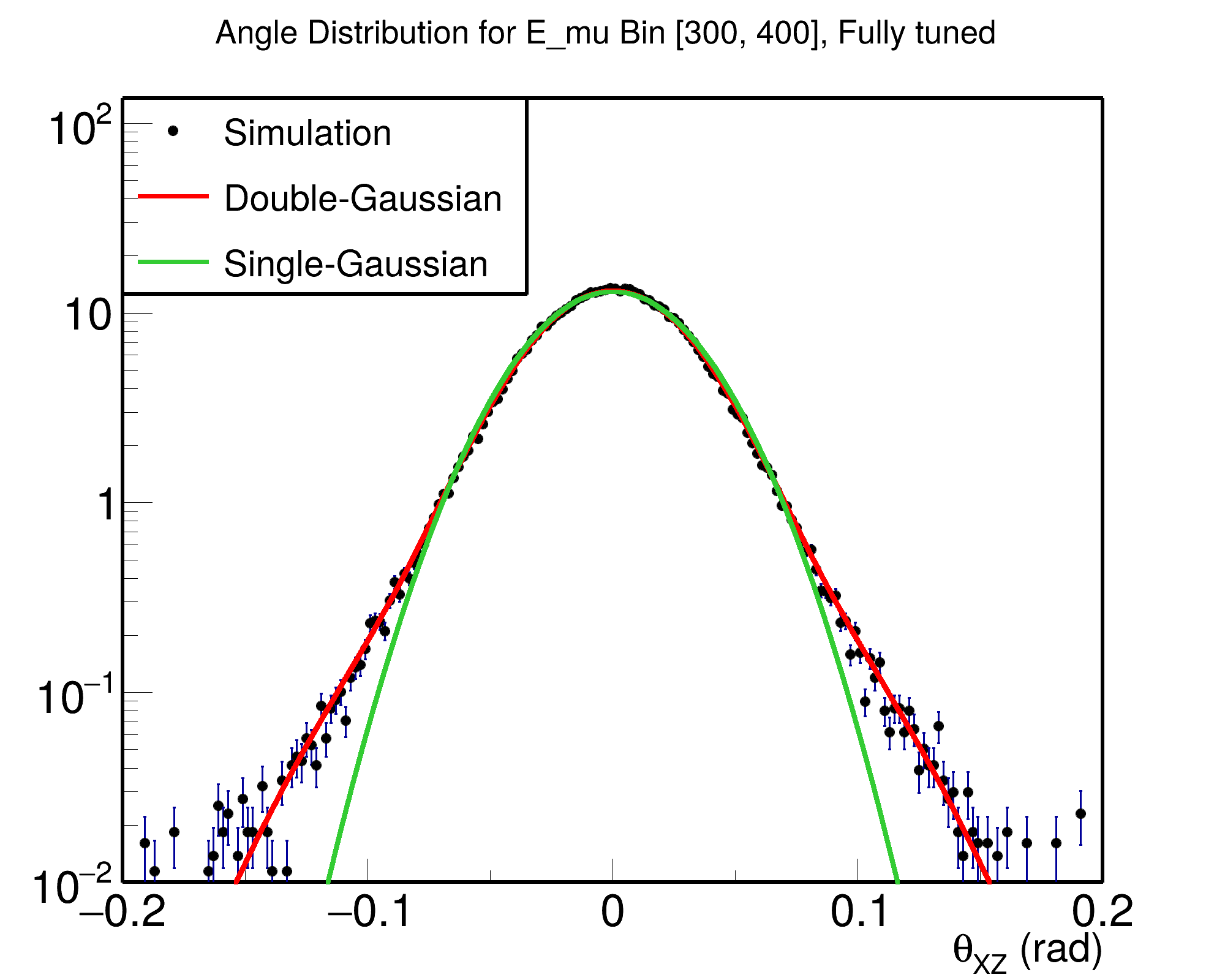}
     \includegraphics[clip,trim={2.0cm 0.0cm 5.3cm 4.9cm},width=0.32\textwidth]{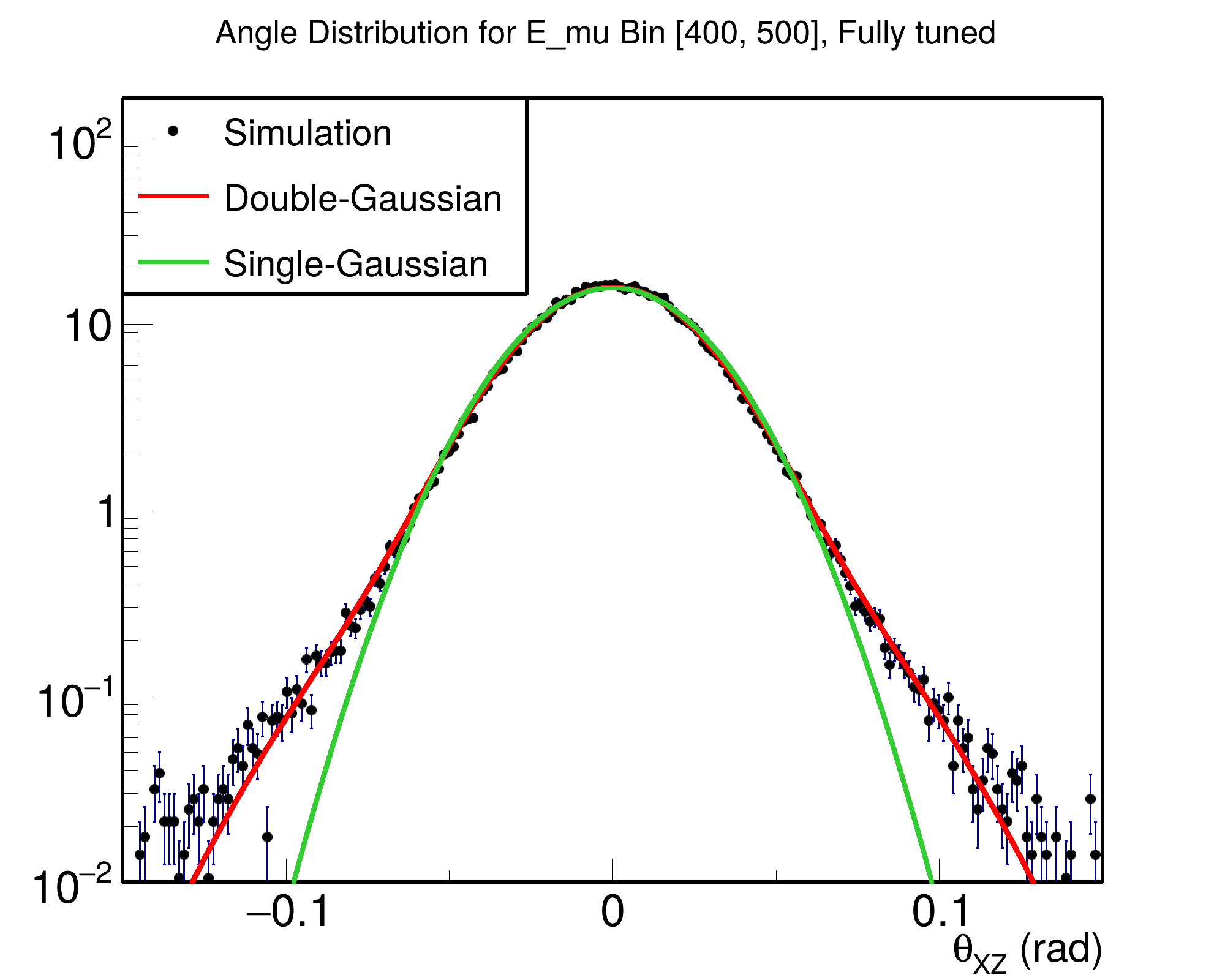}
     \vspace{0.3cm}
     \put(-257,116){\footnotesize Local $KE_{\mu} \in [0.3,0.4]$\,GeV}
     \put(-120,116){\footnotesize Local $KE_{\mu} \in [0.4,0.5]$\,GeV}
     \put(  20,116){\footnotesize Local $KE_{\mu} \in [0.5,0.6]$\,GeV}
     \put(-298,23){\rotatebox{90}{\small Probability Density}}
     \includegraphics[clip,trim={2.0cm 0.0cm 5.3cm 4.9cm},width=0.32\textwidth]{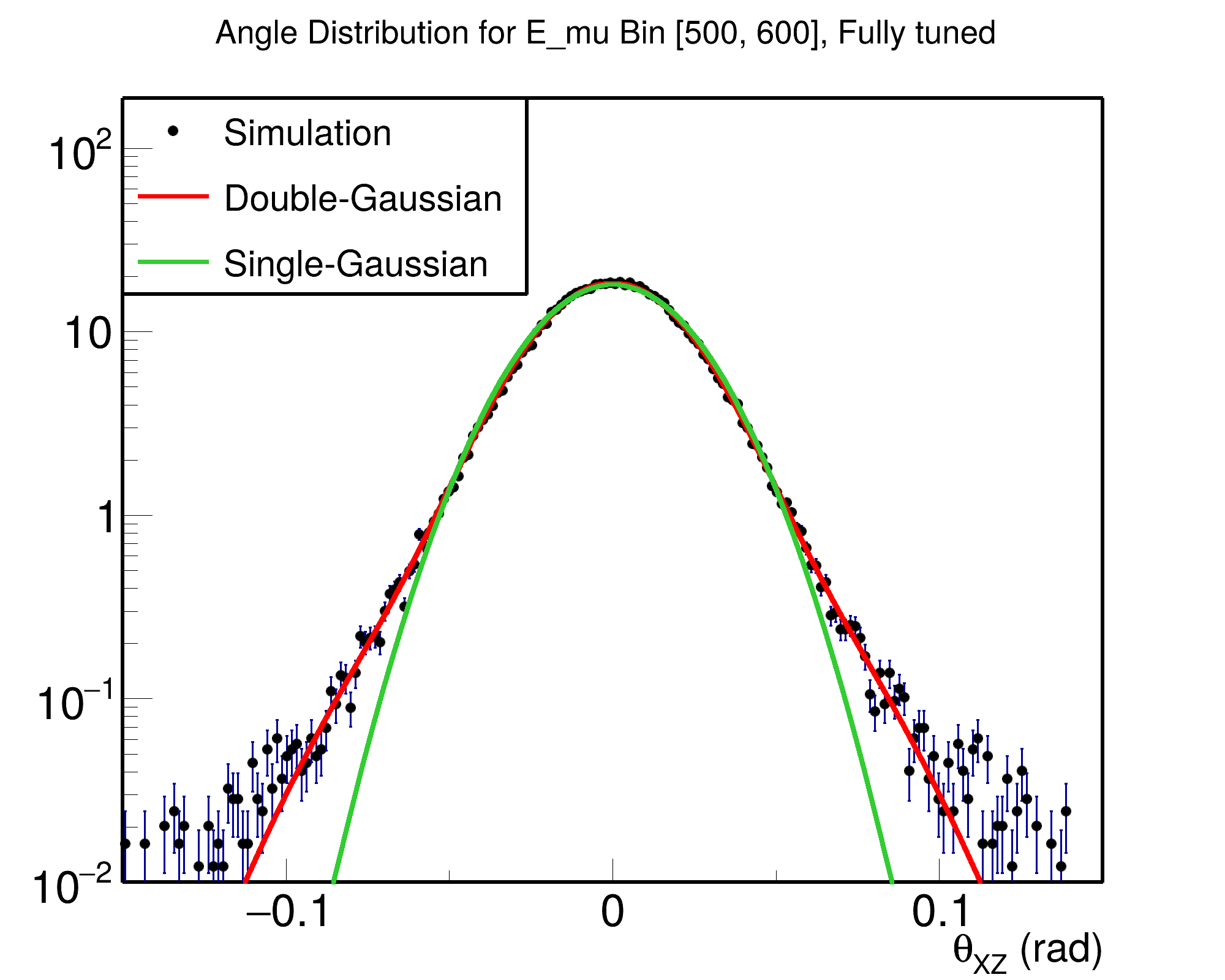}
     \includegraphics[clip,trim={2.0cm 0.0cm 5.3cm 4.9cm},width=0.32\textwidth]{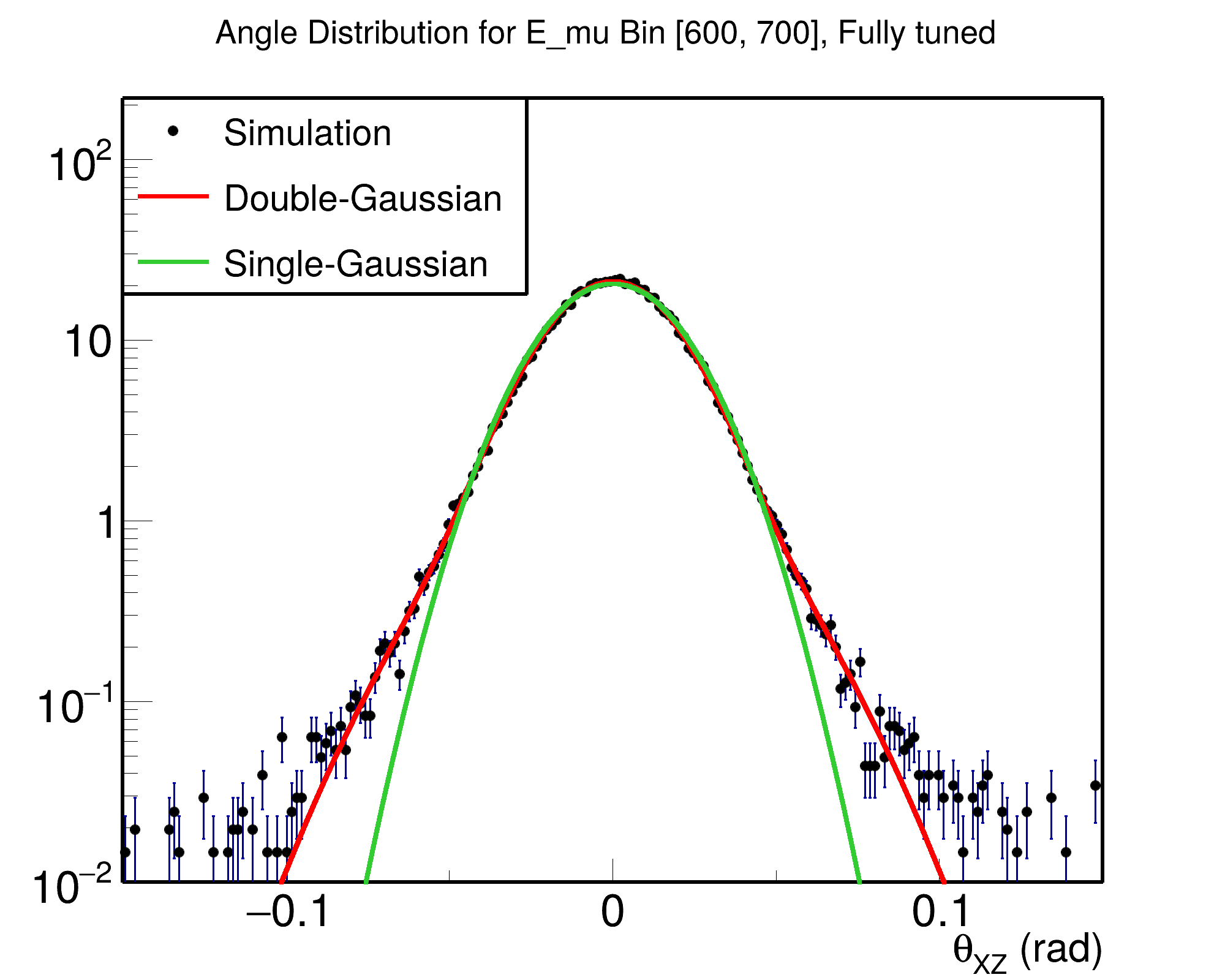}
     \includegraphics[clip,trim={2.0cm 0.0cm 5.3cm 4.9cm},width=0.32\textwidth]{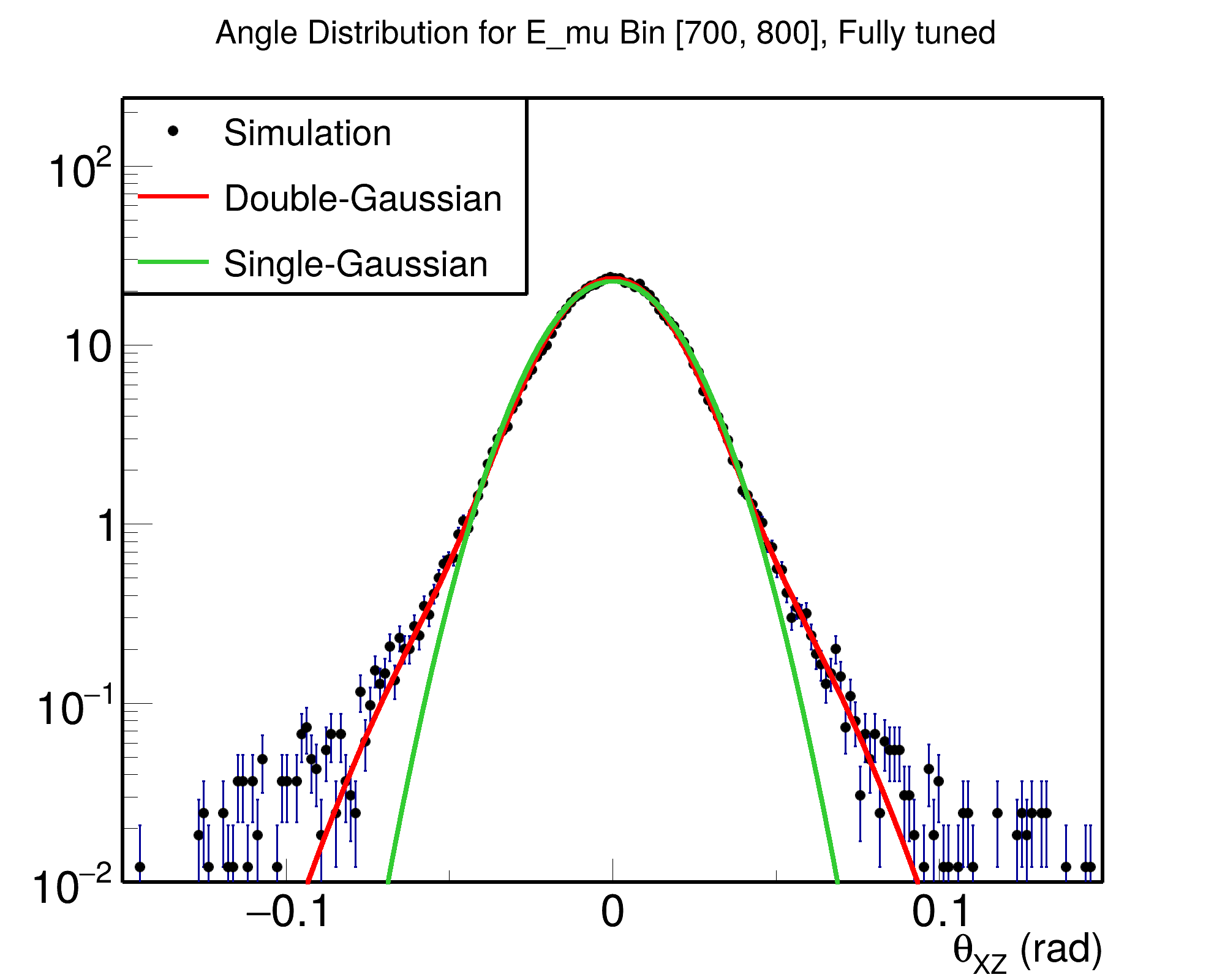}
     \put(-257,116){\footnotesize Local $KE_{\mu} \in [0.6,0.7]$\,GeV}
     \put(-118,116){\footnotesize Local $KE_{\mu} \in [0.7,0.8]$\,GeV}
     \put(  20,116){\footnotesize Local $KE_{\mu} \in [0.8  ,0.9]$\,GeV}
     \includegraphics[clip,trim={2.0cm 0.0cm 5.3cm 4.9cm},width=0.32\textwidth]{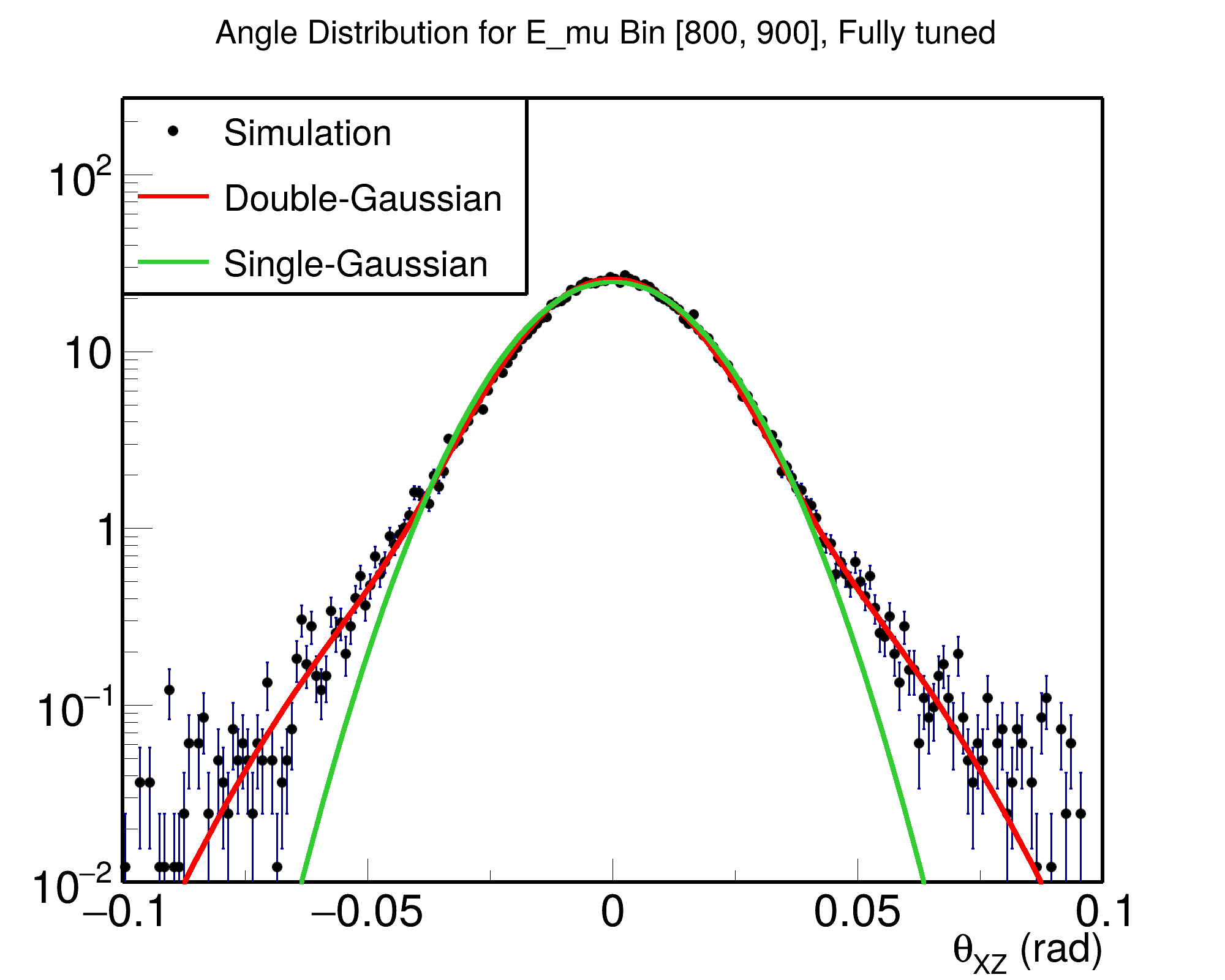}
     \caption{Comparison of single- and double-Gaussian PDFs to the reconstructed $\theta_{xz}$ distribution in simulation. Probability density is plotted on a log scale so differences at large angles can be seen. The average muon kinetic energy of the reconstructed segments adjacent to each angle measurement is in bins from 0 to 0.9\,GeV.}
    \label{fig:theta_xz_low}
\end{figure}

\begin{figure}[hbtp!]
     \centering
     \includegraphics[clip,trim={2.0cm 0.0cm 5.3cm 5.0cm},width=0.32\textwidth]{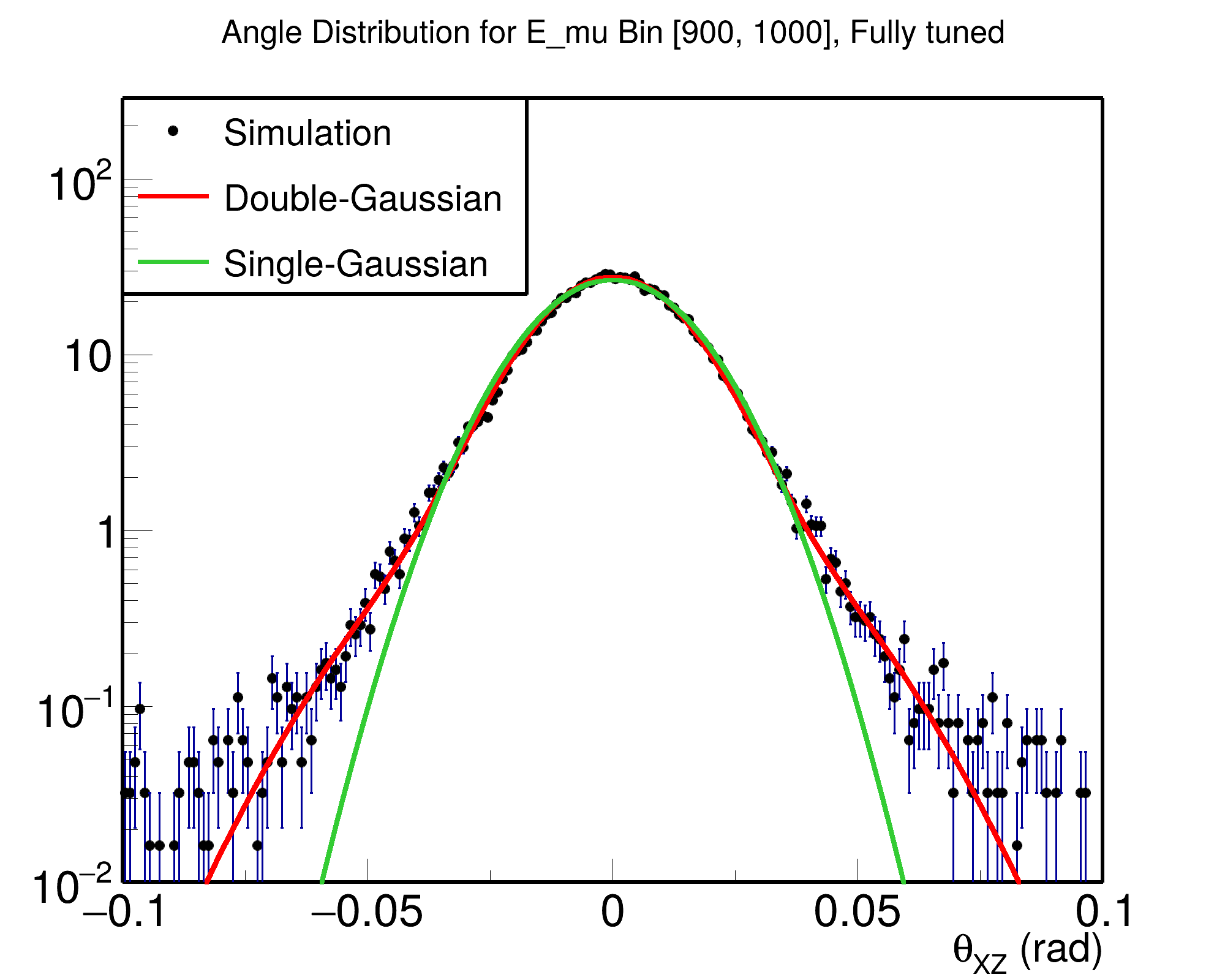}
     \includegraphics[clip,trim={2.0cm 0.0cm 5.3cm 5.0cm},width=0.32\textwidth]{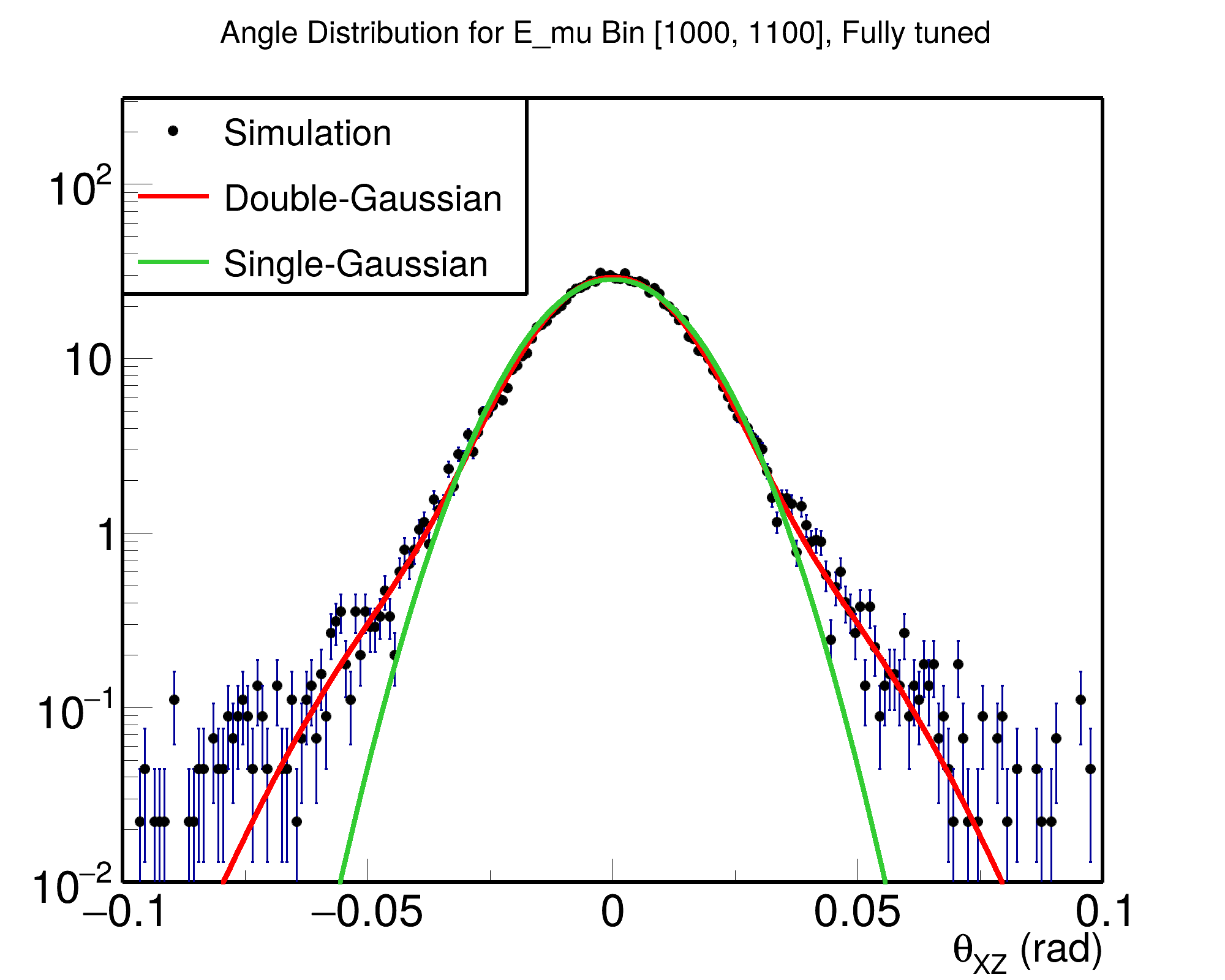}
     \vspace{0.3cm}
     \put(-255,116){\footnotesize Local $KE_{\mu} \in [0.9,1.0]$\,GeV}
     \put(-120,116){\footnotesize Local $KE_{\mu} \in [1.0,1.1]$\,GeV}
     \put(  20,116){\footnotesize Local $KE_{\mu} \in [1.1,1.2]$\,GeV}
     \put(-60,102){\footnotesize MicroBooNE}
     \put(-53, 91){\footnotesize Simulation}
     \includegraphics[clip,trim={2.0cm 0.0cm 5.3cm 4.9cm},width=0.32\textwidth]{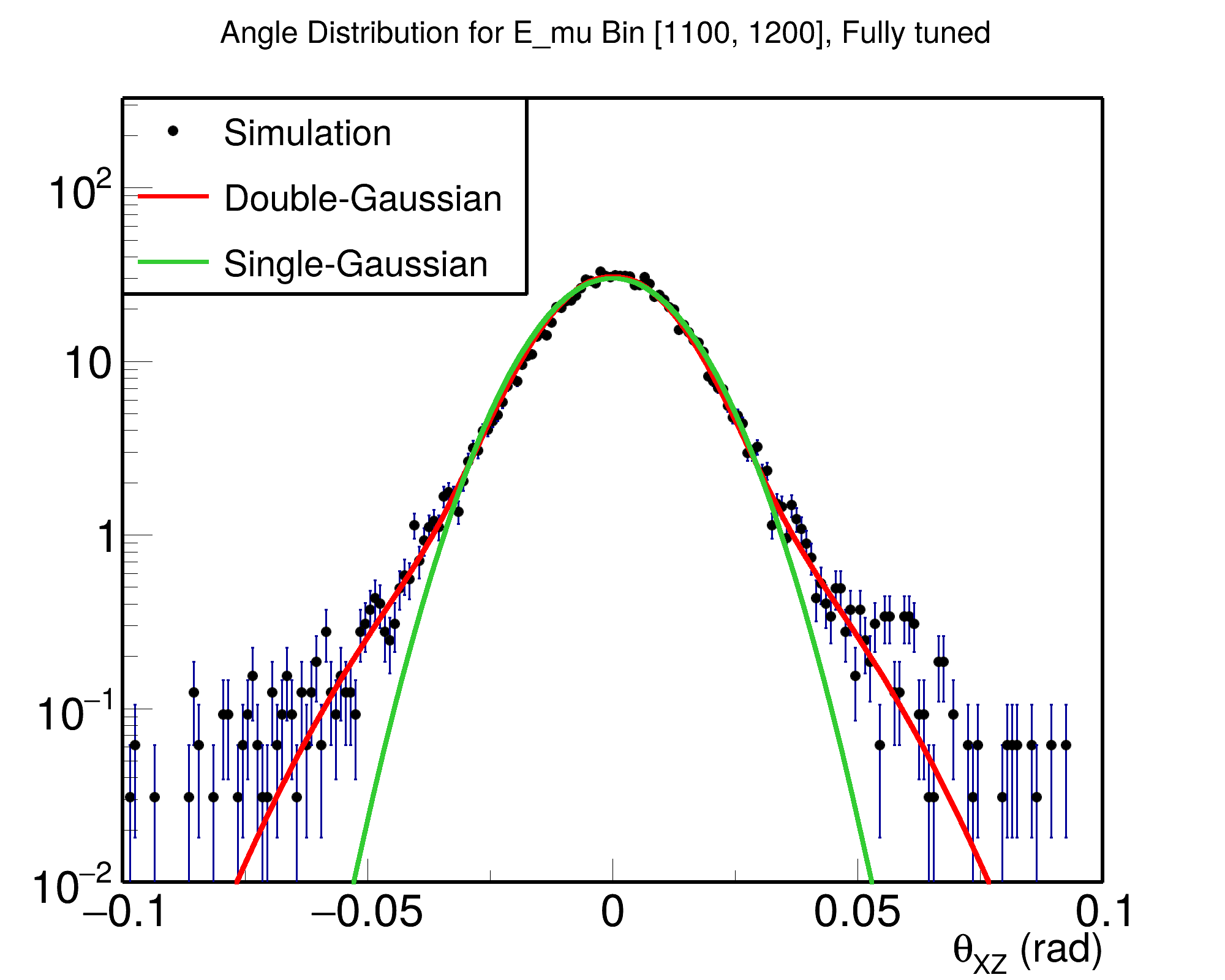}
     \includegraphics[clip,trim={2.0cm 0.0cm 5.3cm 4.9cm},width=0.32\textwidth]{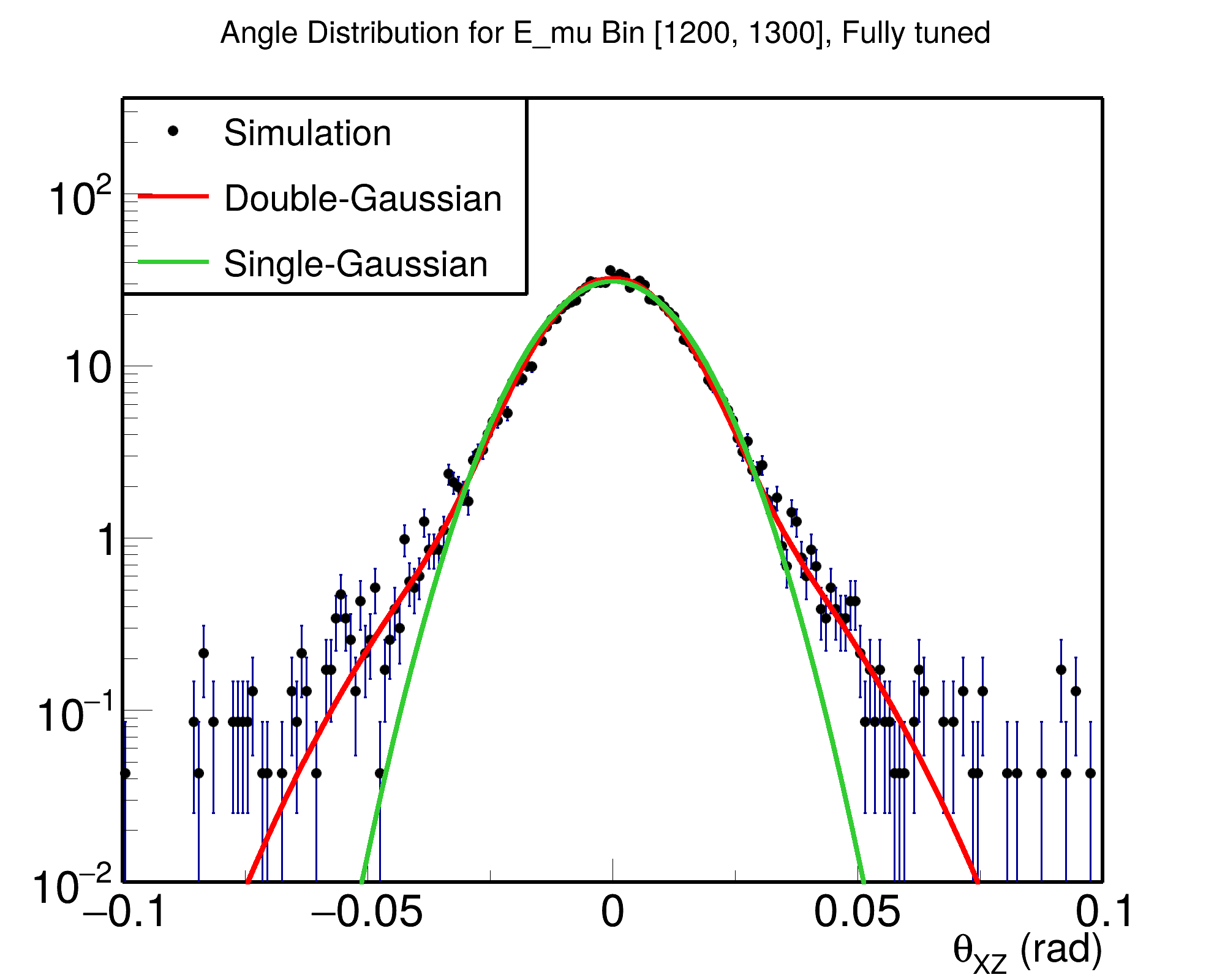}
     \includegraphics[clip,trim={2.0cm 0.0cm 5.3cm 4.9cm},width=0.32\textwidth]{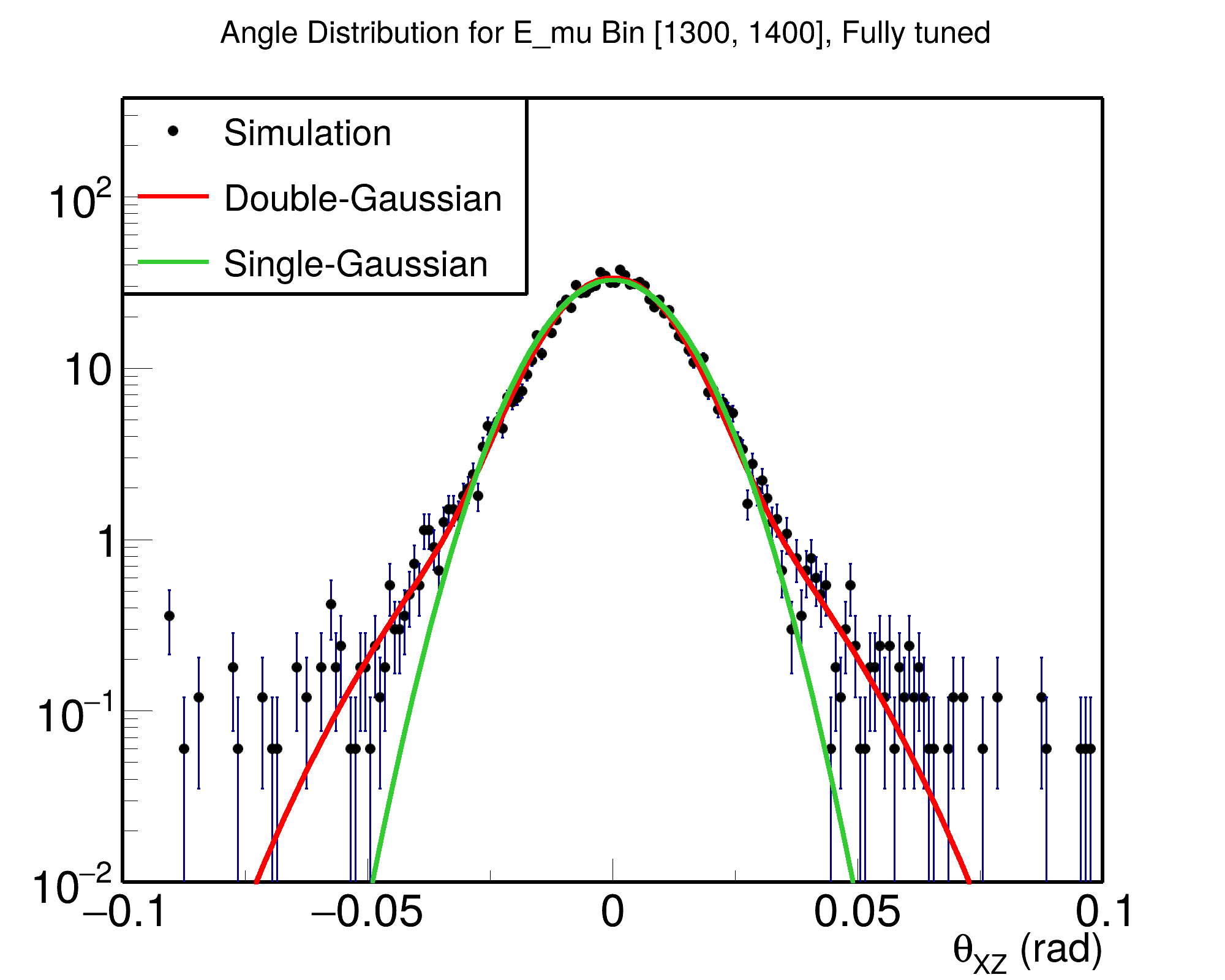}
     \vspace{0.3cm}
     \put(-257,116){\footnotesize Local $KE_{\mu} \in [1.2,1.3]$\,GeV}
     \put(-120,116){\footnotesize Local $KE_{\mu} \in [1.3,1.4]$\,GeV}
     \put(  20,116){\footnotesize Local $KE_{\mu} \in [1.4,1.5]$\,GeV}
     \put(-298,23){\rotatebox{90}{\small Probability Density}}
     \includegraphics[clip,trim={2.0cm 0.0cm 5.3cm 4.9cm},width=0.32\textwidth]{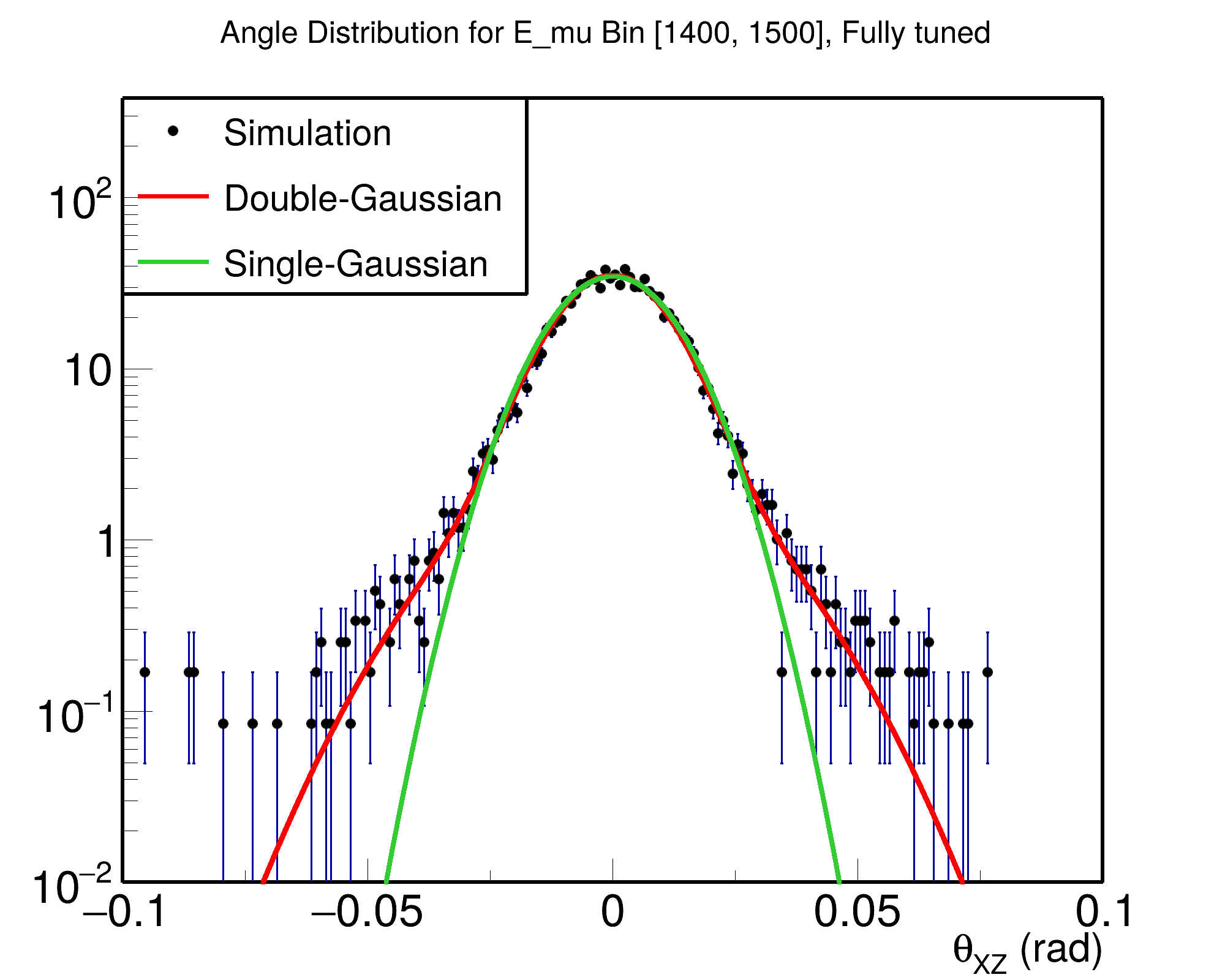}
     \includegraphics[clip,trim={2.0cm 0.0cm 5.3cm 4.9cm},width=0.32\textwidth]{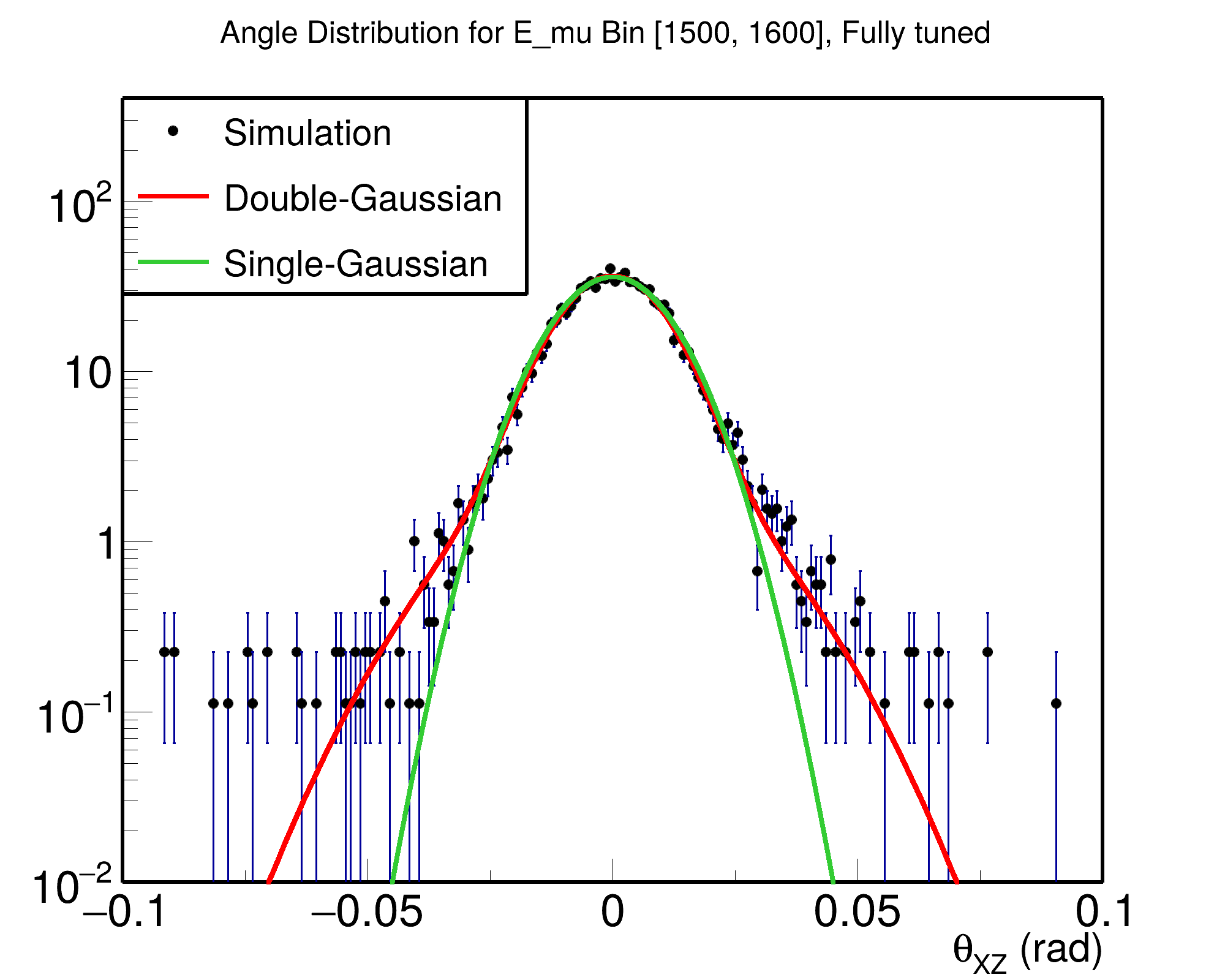}
     \includegraphics[clip,trim={2.0cm 0.0cm 5.3cm 4.9cm},width=0.32\textwidth]{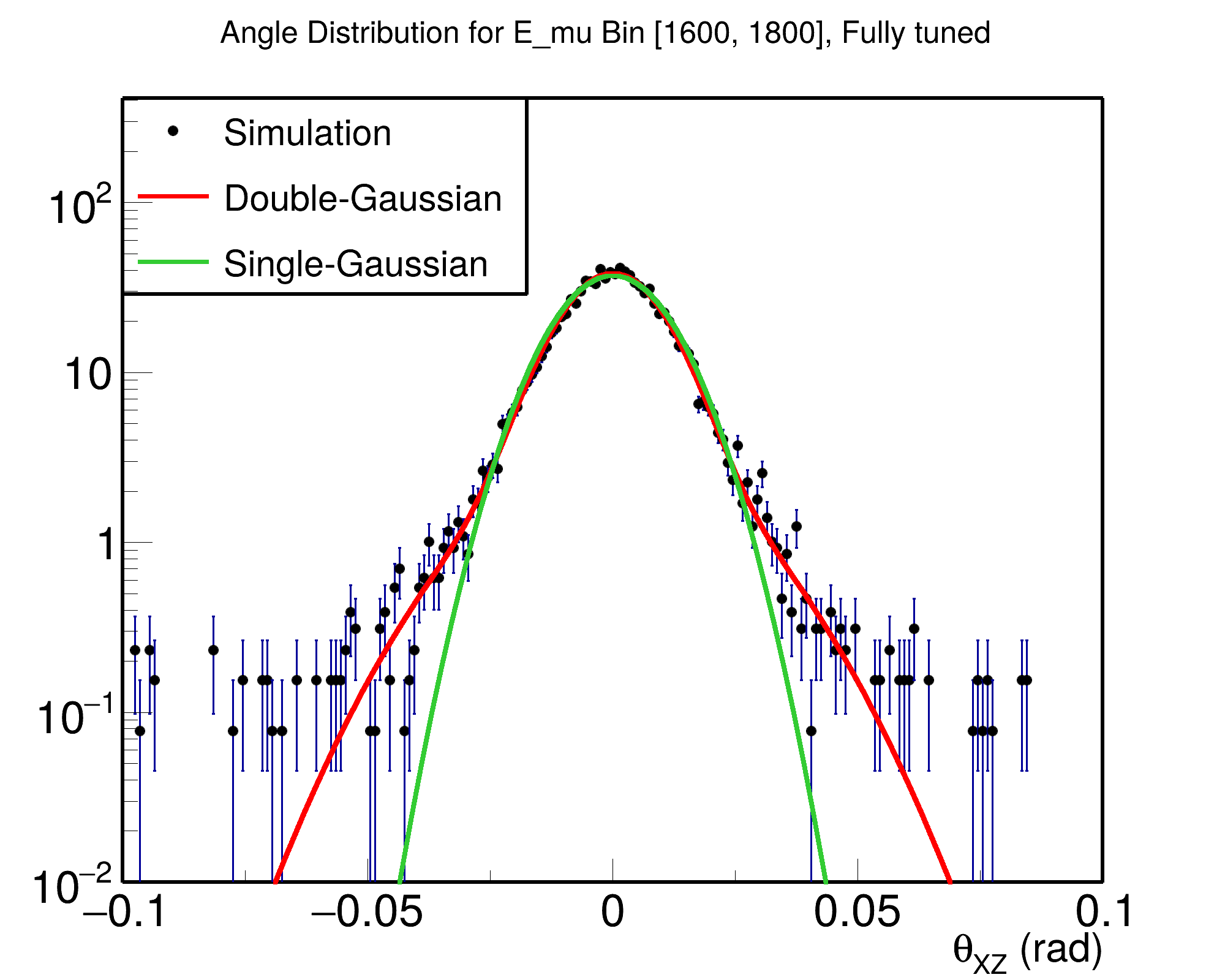}
     \vspace{0.3cm}
     \put(-260,116){\footnotesize Local $KE_{\mu} \in [1.5,1.6]$\,GeV}
     \put(-120,116){\footnotesize Local $KE_{\mu} \in [1.6,1.8]$\,GeV}
     \put(  20,116){\footnotesize Local $KE_{\mu} \in [1.8,2.0]$\,GeV}
     \includegraphics[clip,trim={2.0cm 0.0cm 5.3cm 4.9cm},width=0.32\textwidth]{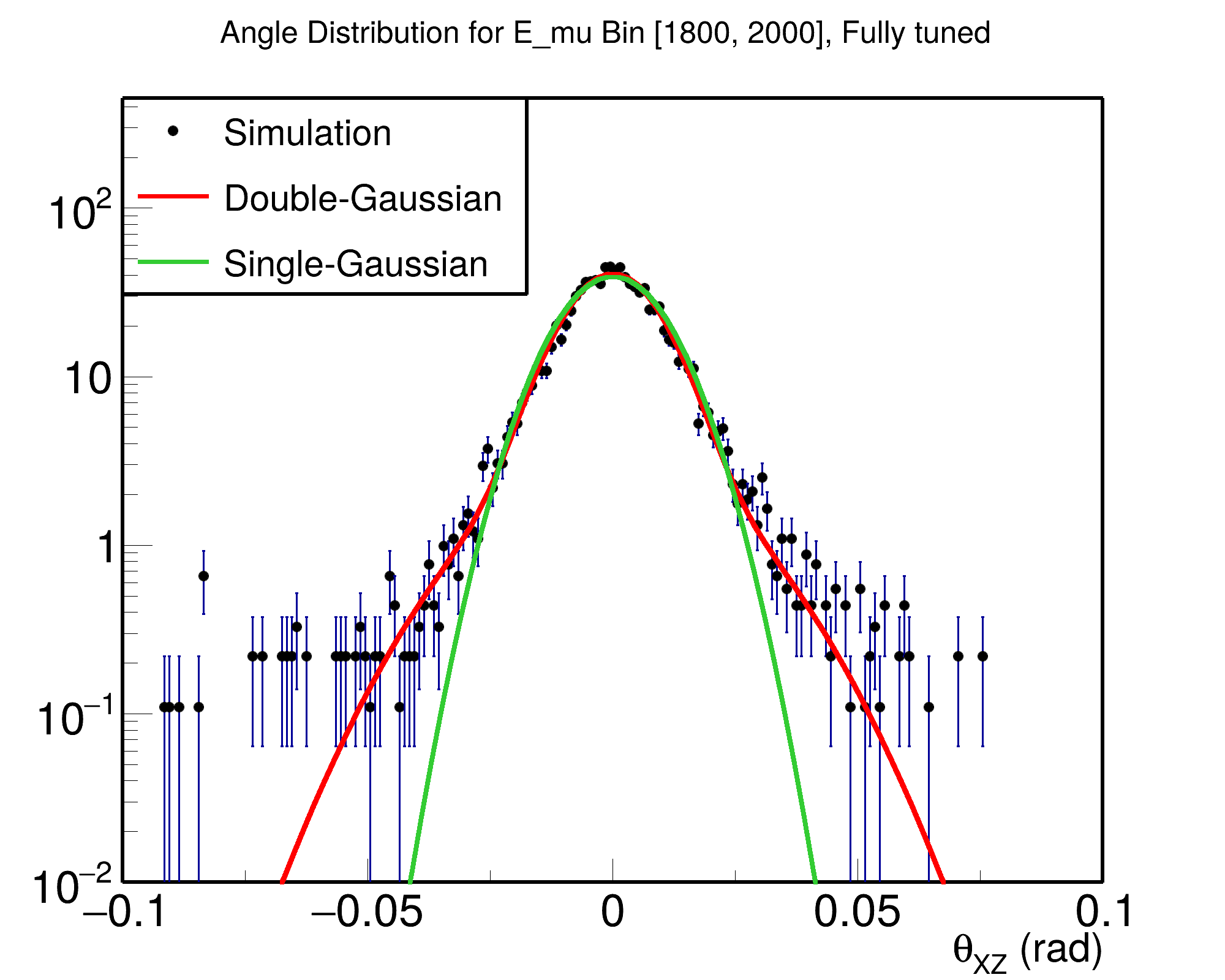}
     \includegraphics[clip,trim={2.0cm 0.0cm 5.3cm 4.9cm},width=0.32\textwidth]{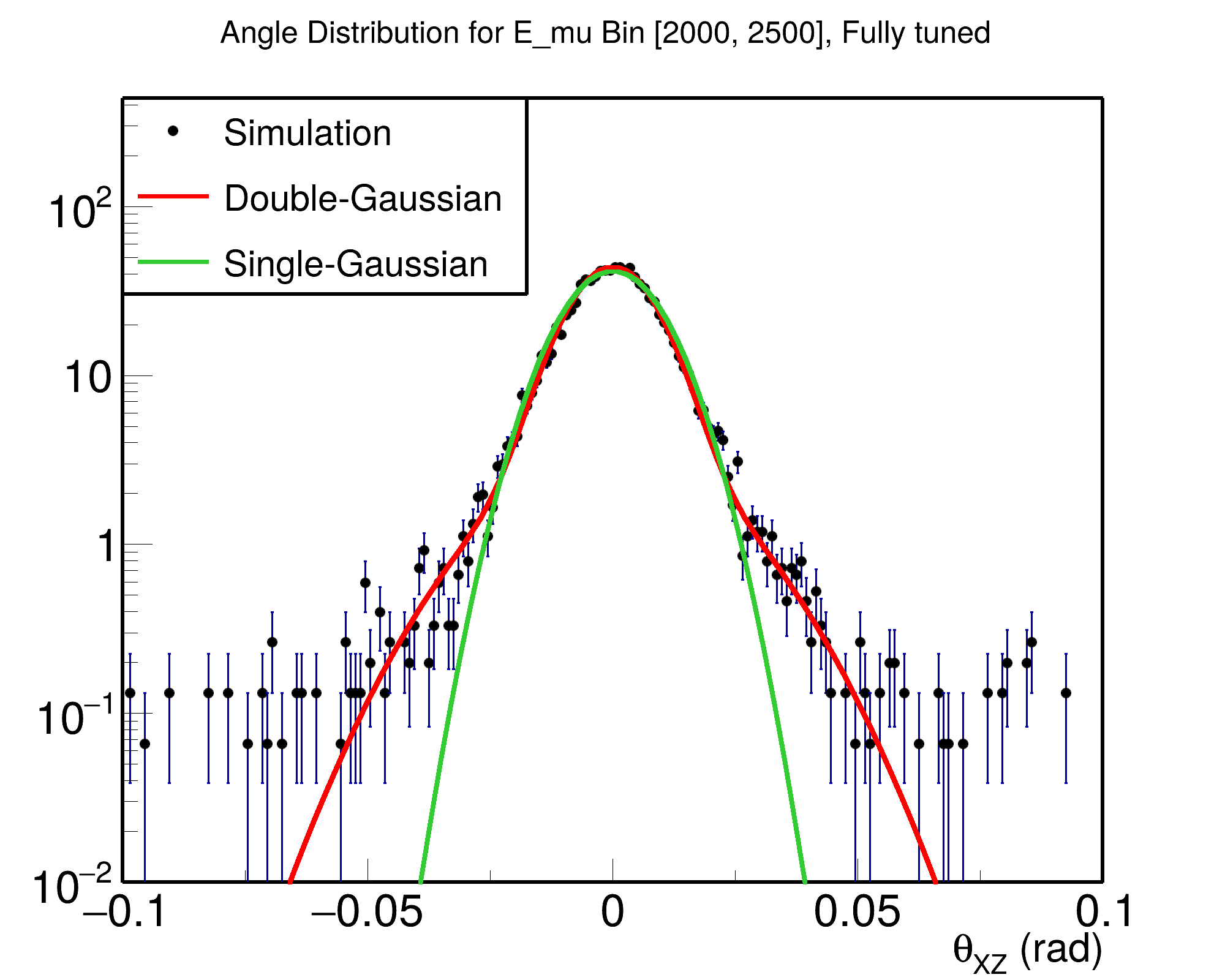}
     \includegraphics[clip,trim={2.0cm 0.0cm 5.3cm 4.9cm},width=0.32\textwidth]{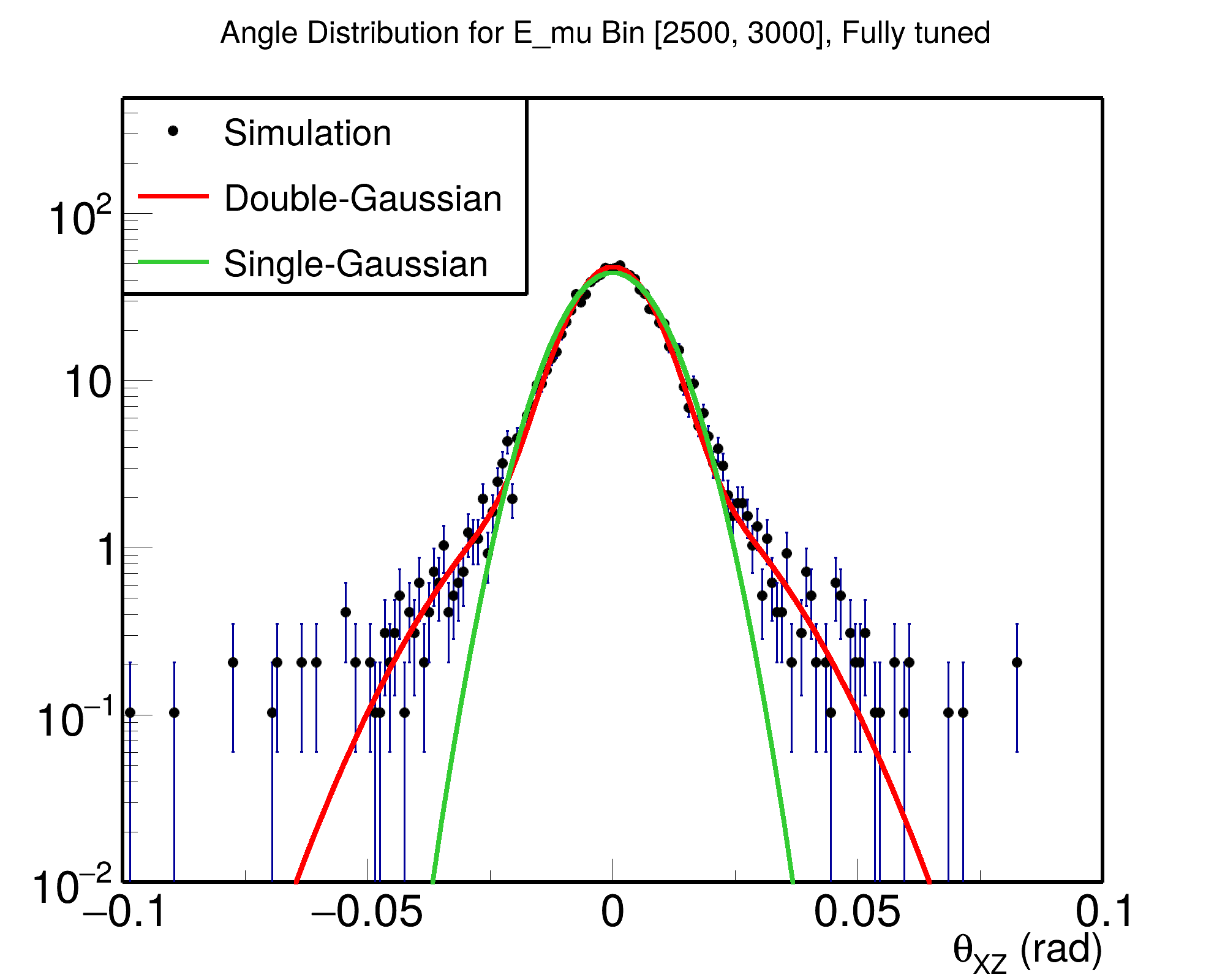}
     \vspace{0.3cm}
     \put(-260,117){\footnotesize Local $KE_{\mu} \in [2.0,2.5]$\,GeV}
     \put(-120,117){\footnotesize Local $KE_{\mu} \in [2.5,3.0]$\,GeV}
     \put(  20,117){\footnotesize Local $KE_{\mu} \in [3.0,4.0]$\,GeV}
     \includegraphics[clip,trim={2.0cm 0.0cm 5.3cm 4.9cm},width=0.32\textwidth]{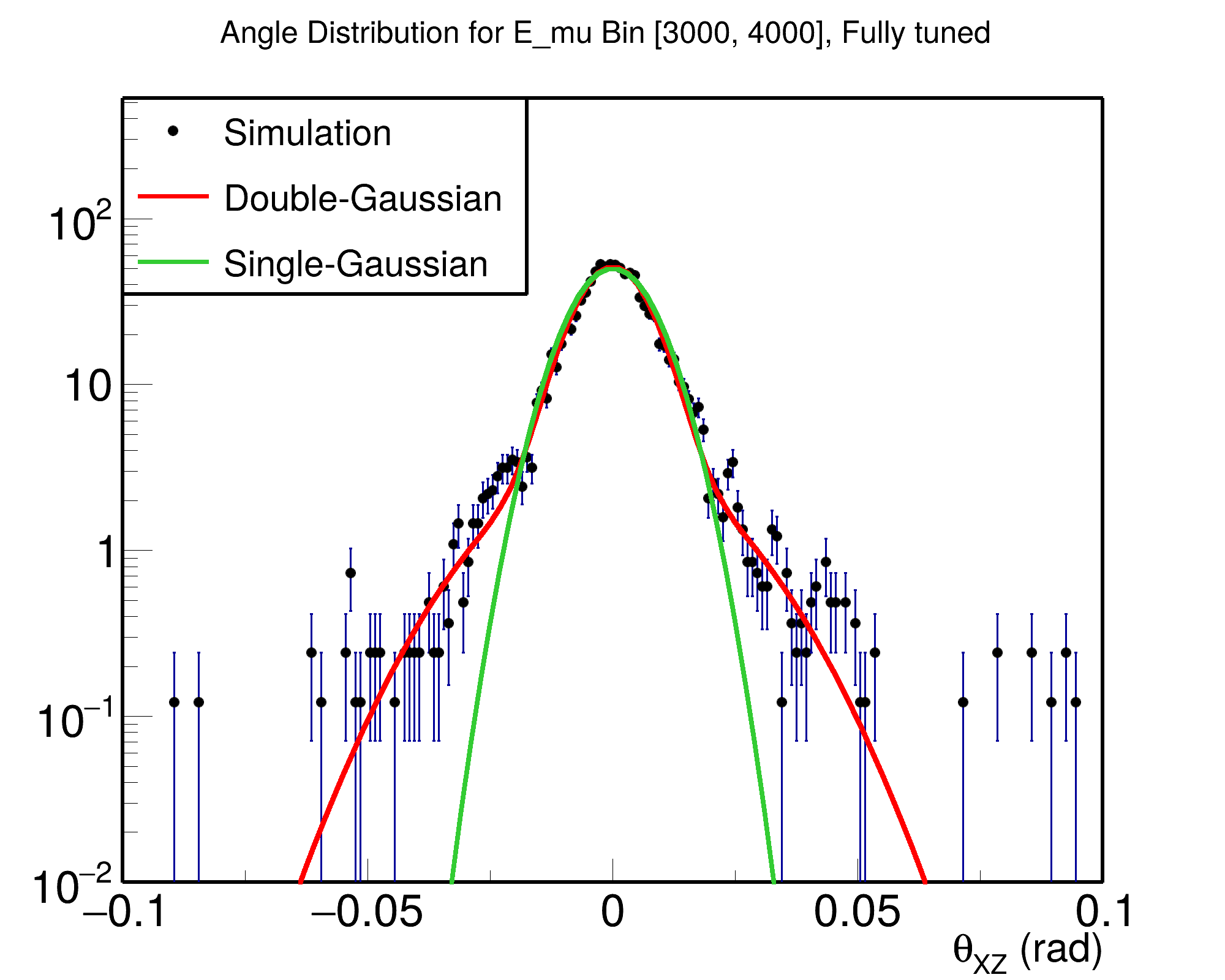}
     \caption{Comparison of single- and double-Gaussian PDFs to the reconstructed $\theta_{xz}$ distribution in simulation. Probability density is plotted on a log scale so differences at large angles can be seen. The average muon kinetic energy of the reconstructed segments adjacent to each angle measurement is in bins from 0.9 to 4.0\,GeV.}
    \label{fig:theta_xz_high}
\end{figure}

\begin{figure}[hbtp!]
     \centering
     \includegraphics[clip,trim={2.0cm 0.0cm 5.3cm 4.9cm},width=0.32\textwidth]{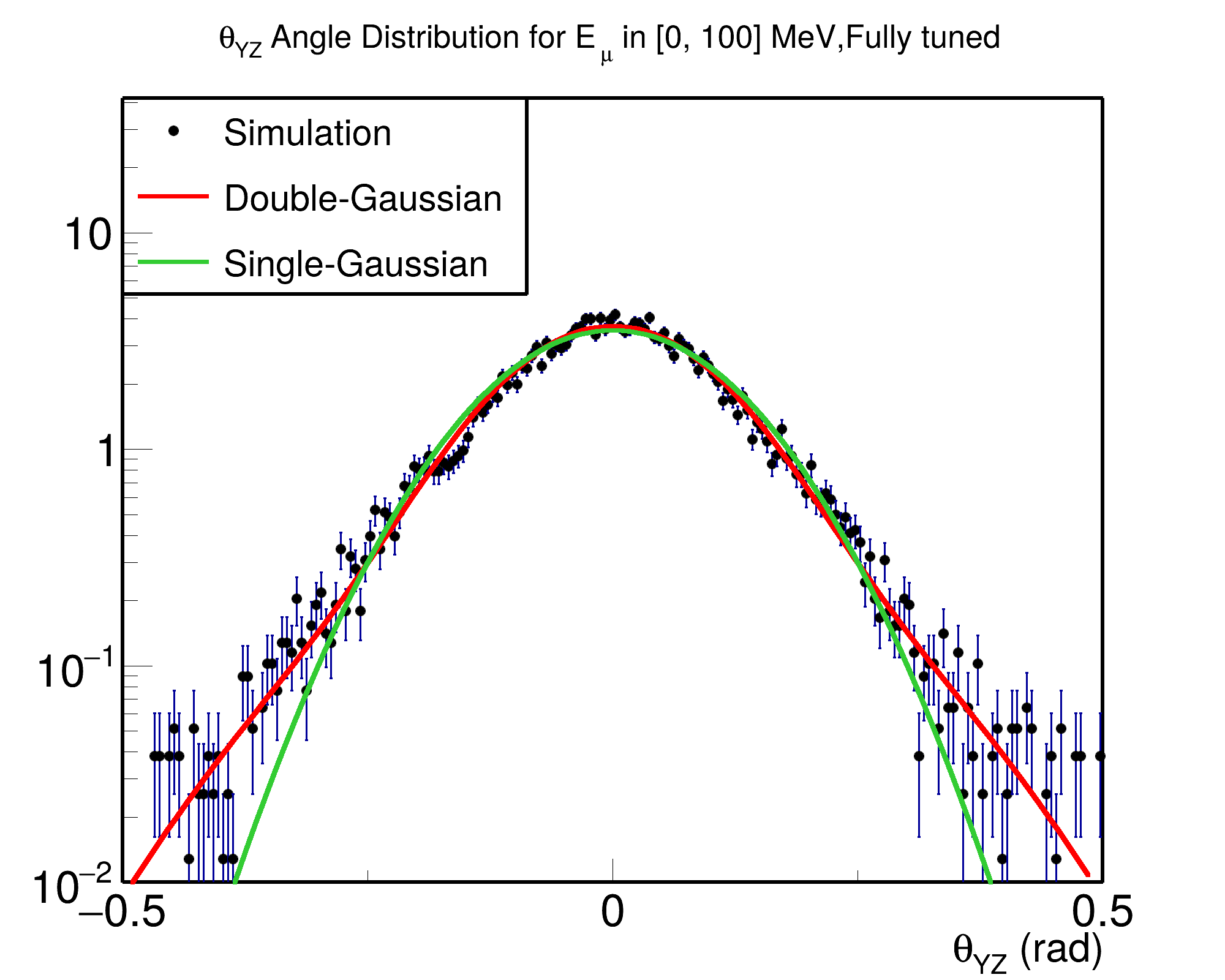}
     \includegraphics[clip,trim={2.0cm 0.0cm 5.3cm 4.9cm},width=0.32\textwidth]{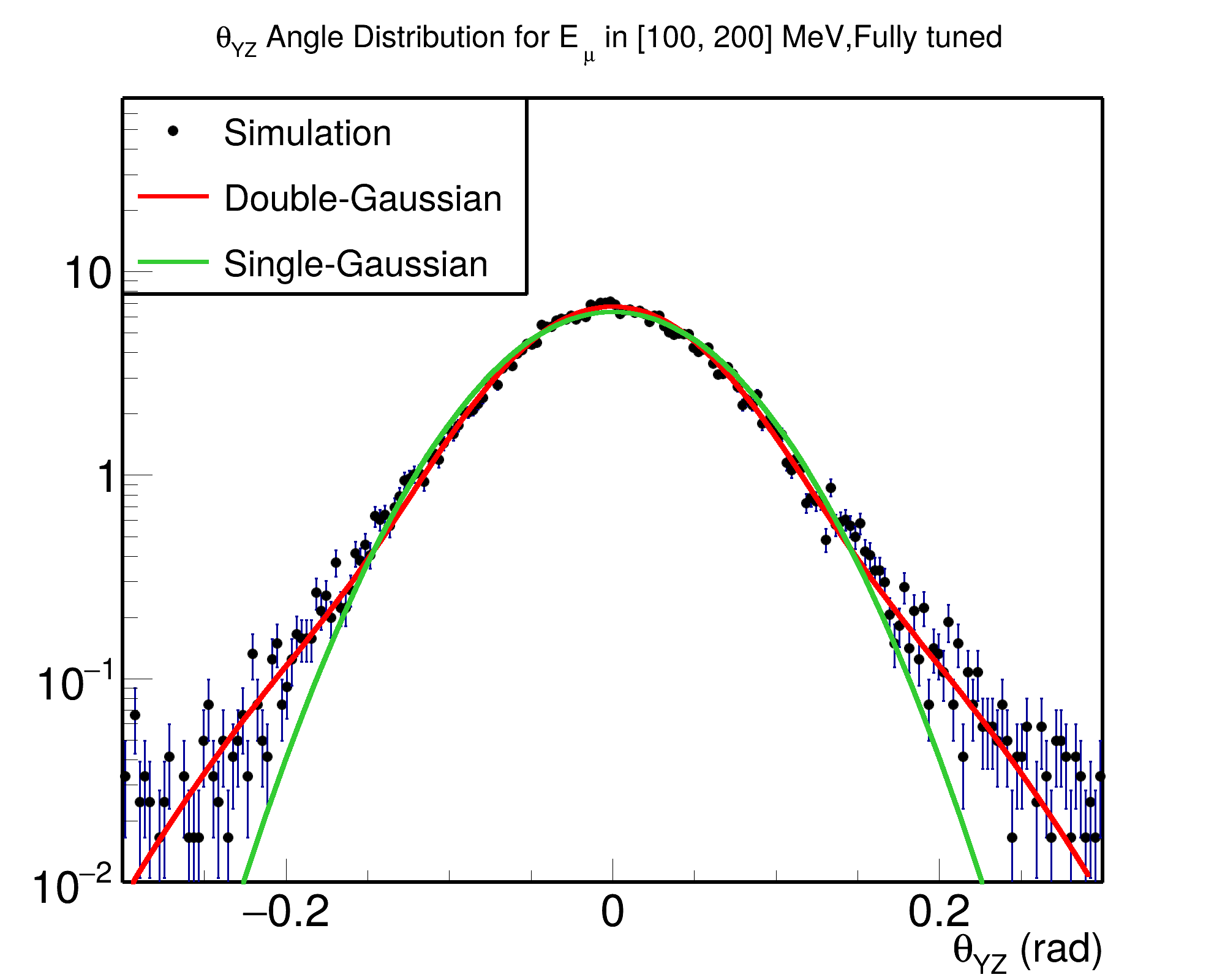}
     \vspace{0.3cm}
     \put(-255,116){\footnotesize Local $KE_{\mu} \in [0,0.1]$\,GeV}
     \put(-120,116){\footnotesize Local $KE_{\mu} \in [0.1,0.2]$\,GeV}
     \put(  20,116){\footnotesize Local $KE_{\mu} \in [0.2,0.3]$\,GeV}
     \put(-60,102){\footnotesize MicroBooNE}
     \put(-53, 91){\footnotesize Simulation}
     \includegraphics[clip,trim={2.0cm 0.0cm 5.3cm 4.9cm},width=0.32\textwidth]{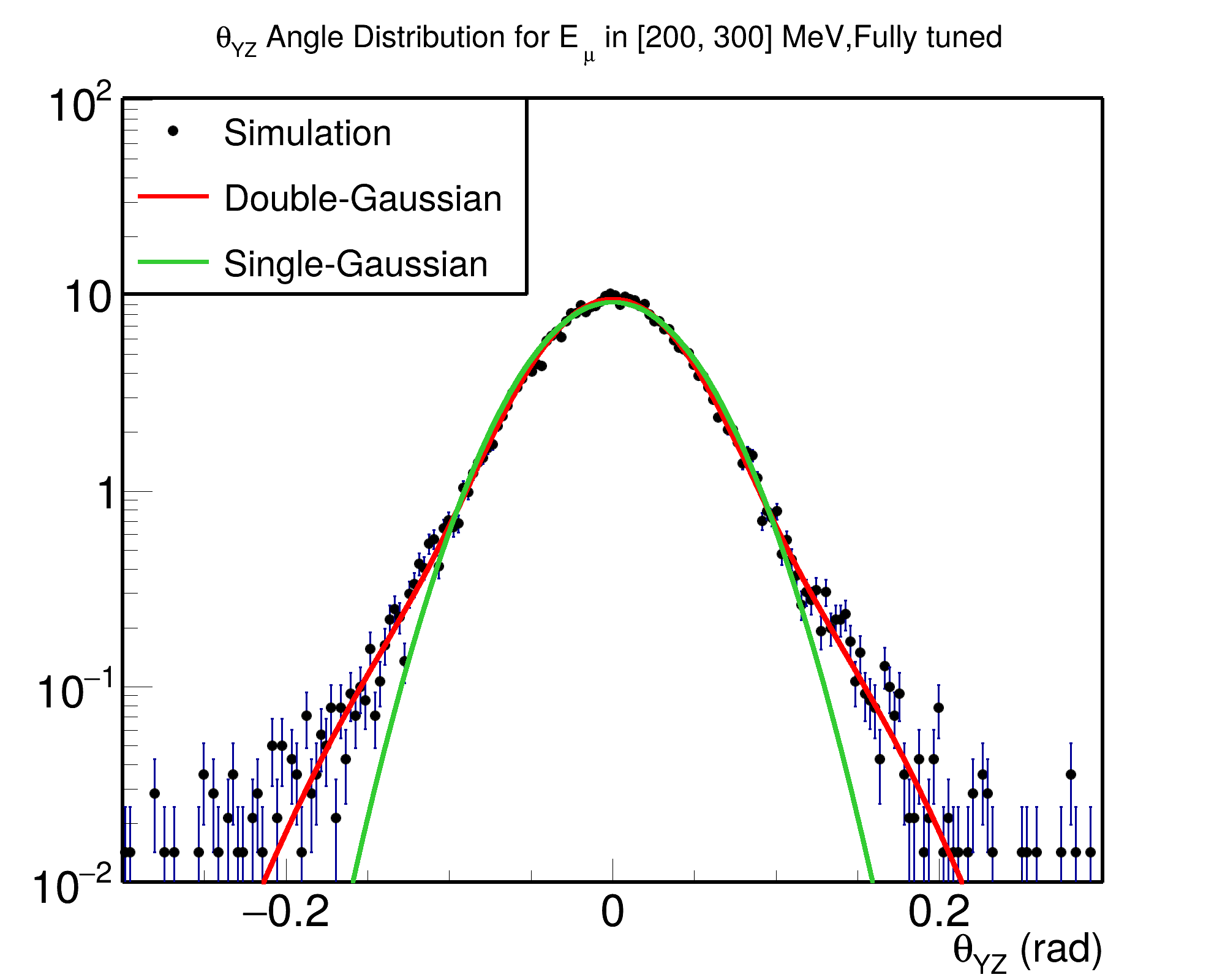}
     \includegraphics[clip,trim={2.0cm 0.0cm 5.3cm 4.9cm},width=0.32\textwidth]{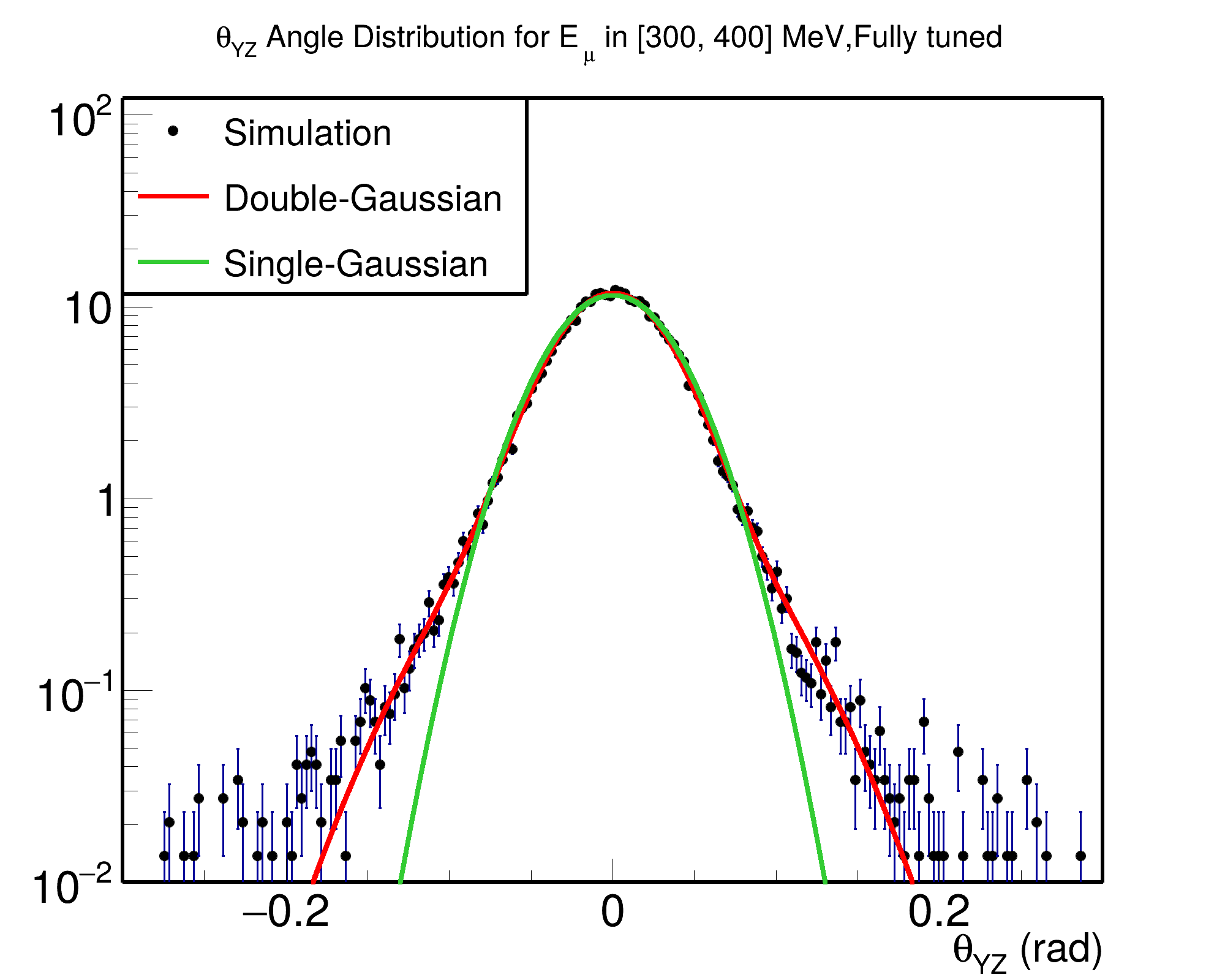}
     \includegraphics[clip,trim={2.0cm 0.0cm 5.3cm 4.9cm},width=0.32\textwidth]{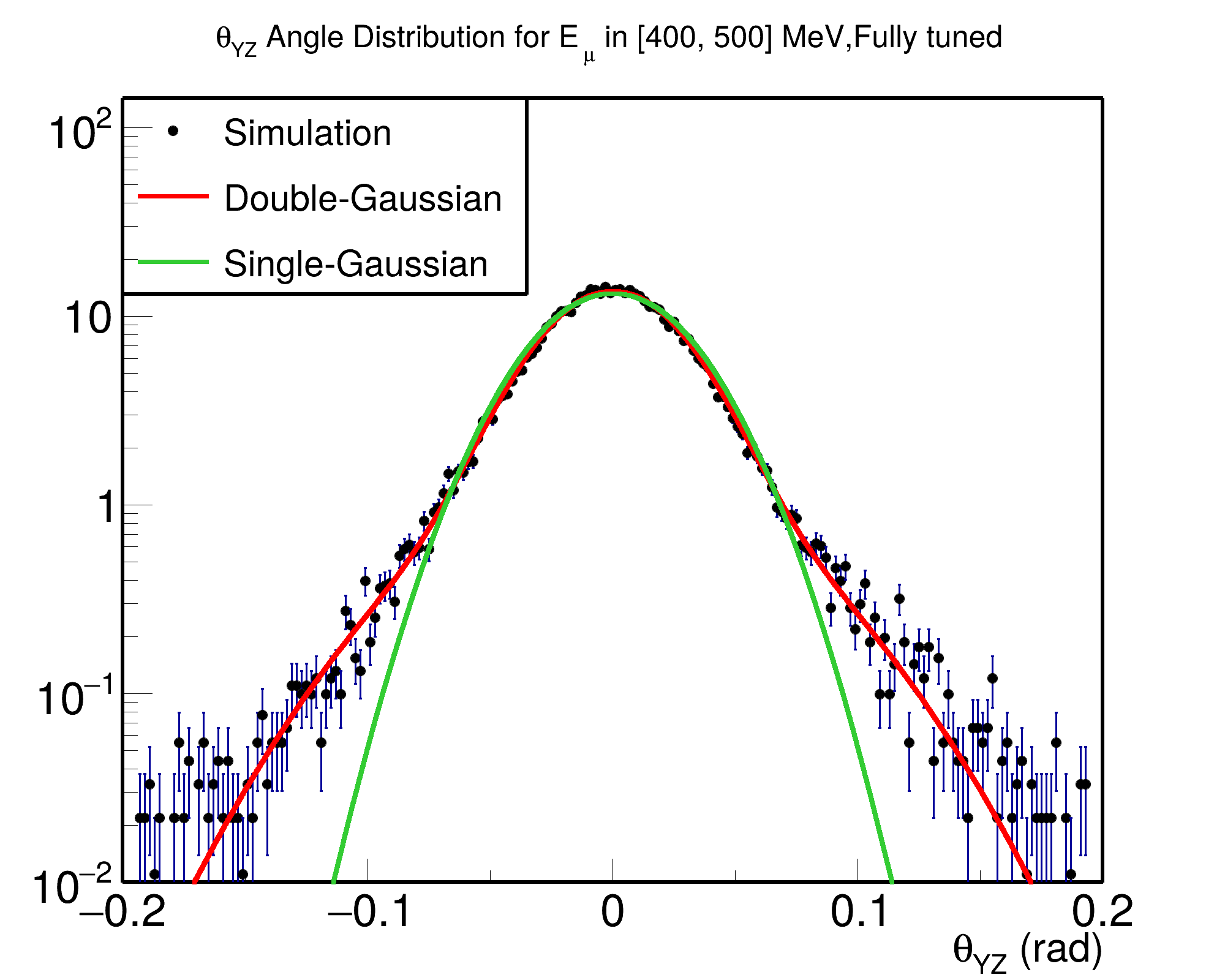}
     \vspace{0.3cm}
     \put(-257,116){\footnotesize Local $KE_{\mu} \in [0.3,0.4]$\,GeV}
     \put(-120,116){\footnotesize Local $KE_{\mu} \in [0.4,0.5]$\,GeV}
     \put(  20,116){\footnotesize Local $KE_{\mu} \in [0.5,0.6]$\,GeV}
     \put(-298,23){\rotatebox{90}{\small Probability Density}}
     \includegraphics[clip,trim={2.0cm 0.0cm 5.3cm 4.9cm},width=0.32\textwidth]{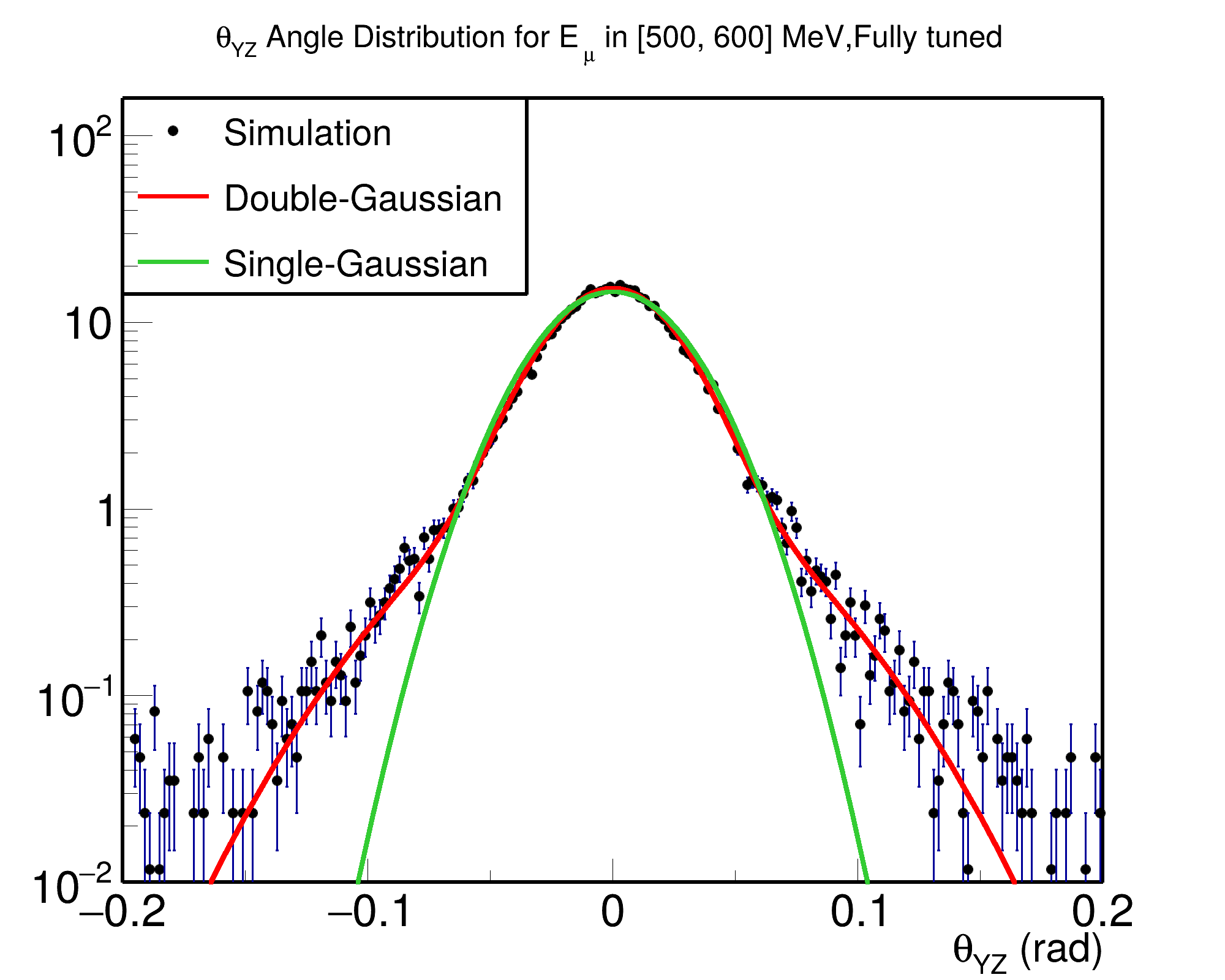}
     \includegraphics[clip,trim={2.0cm 0.0cm 5.3cm 4.9cm},width=0.32\textwidth]{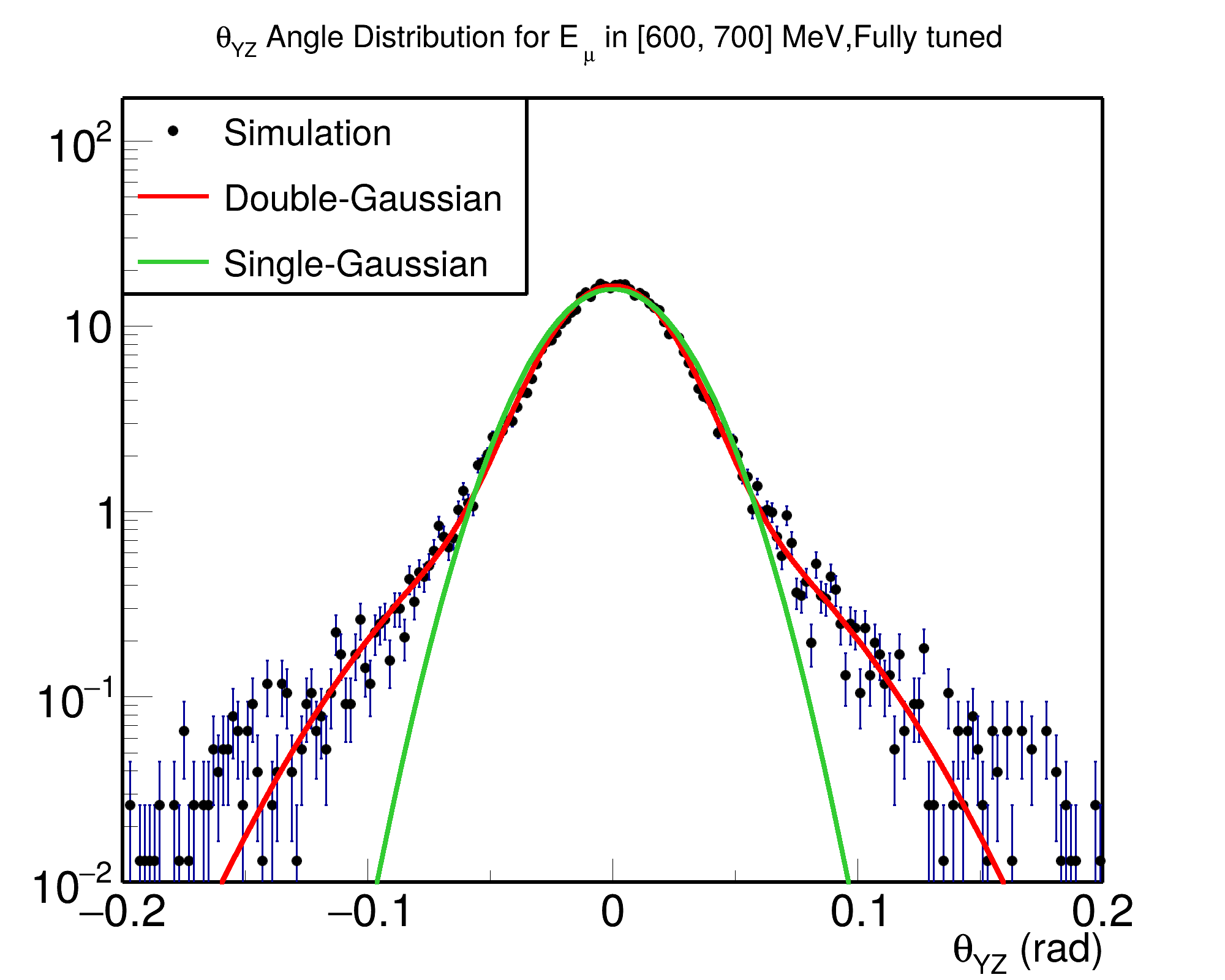}
     \includegraphics[clip,trim={2.0cm 0.0cm 5.3cm 4.9cm},width=0.32\textwidth]{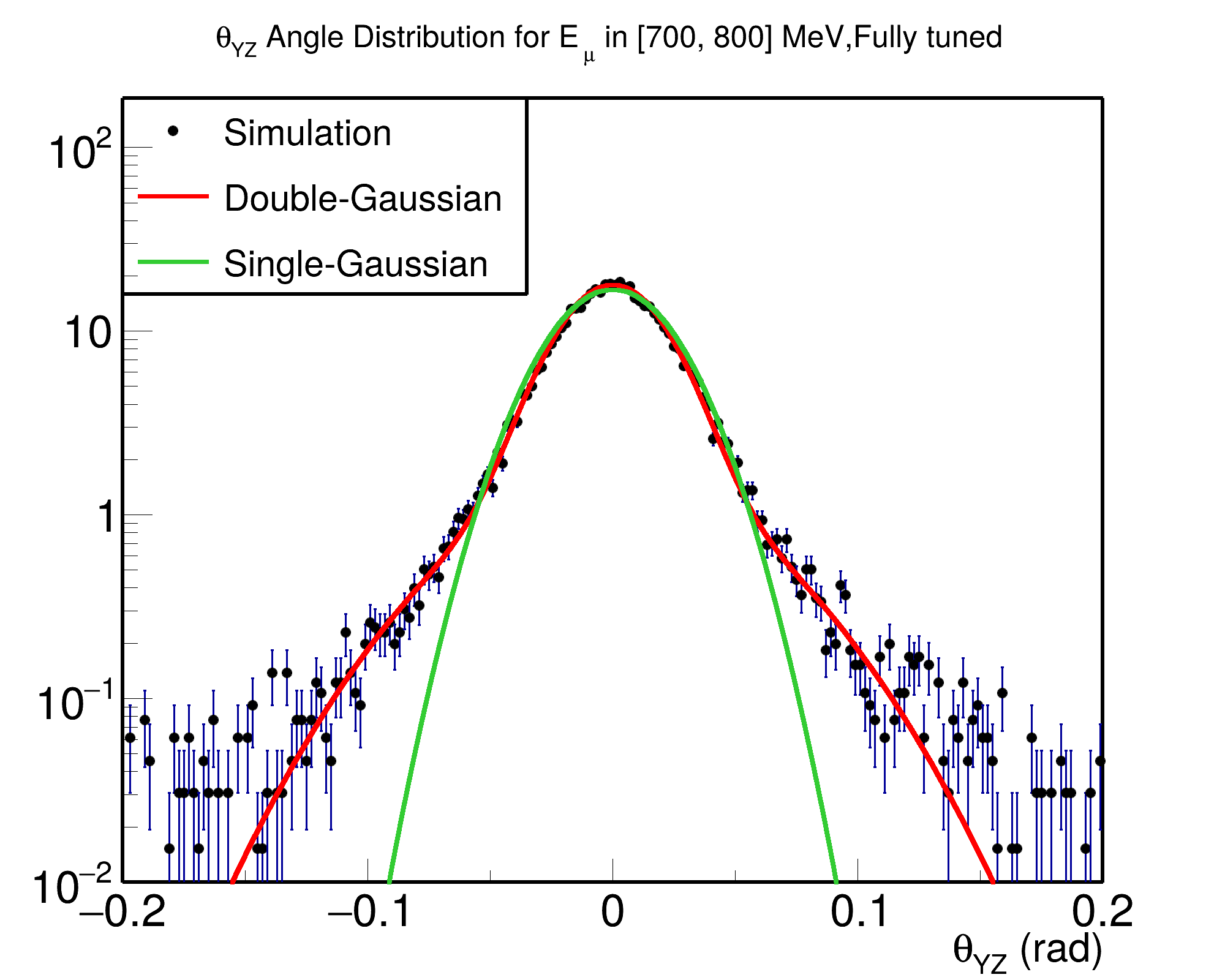}
     \put(-257,116){\footnotesize Local $KE_{\mu} \in [0.6,0.7]$\,GeV}
     \put(-118,116){\footnotesize Local $KE_{\mu} \in [0.7,0.8]$\,GeV}
     \put(  20,116){\footnotesize Local $KE_{\mu} \in [0.8,0.9]$\,GeV}
     \includegraphics[clip,trim={2.0cm 0.0cm 5.3cm 4.9cm},width=0.32\textwidth]{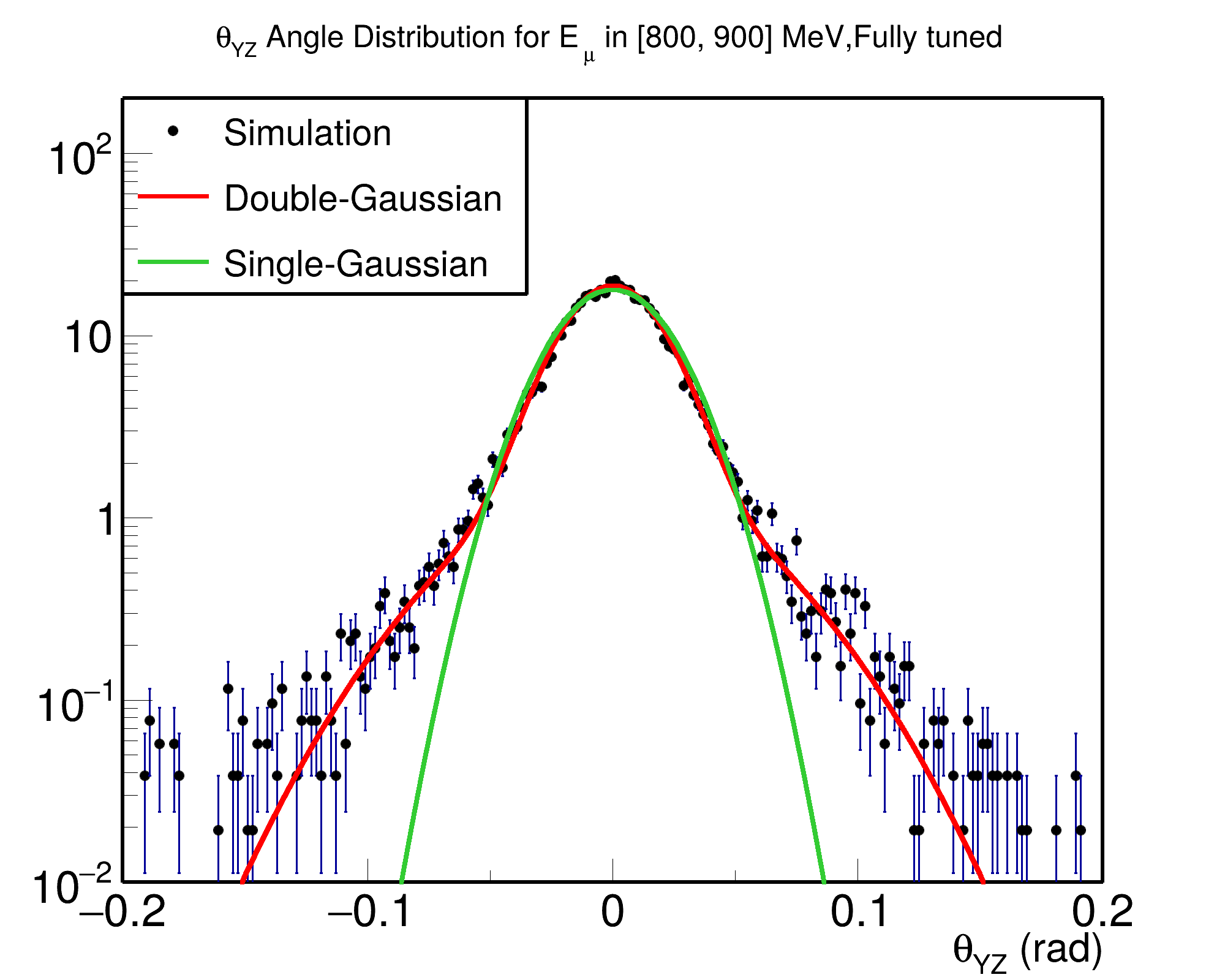}
     \caption{Comparison of single- and double-Gaussian PDFs to the reconstructed $\theta_{yz}$ distribution in simulation. Probability density is plotted on a log scale so differences at large angles can be seen. Both the tune of the double-Gaussian PDF and the simulation only correspond to segments with $|v_{x}| \in [0.2,0.35]$. The average muon kinetic energy of the reconstructed segments adjacent to each angle measurement is in bins from 0 to 0.9\,GeV.}
    \label{fig:theta_yz_vx2_low}
\end{figure}

\begin{figure}[hbtp!]
     \centering
     \includegraphics[clip,trim={2.0cm 0.0cm 5.3cm 4.9cm},width=0.32\textwidth]{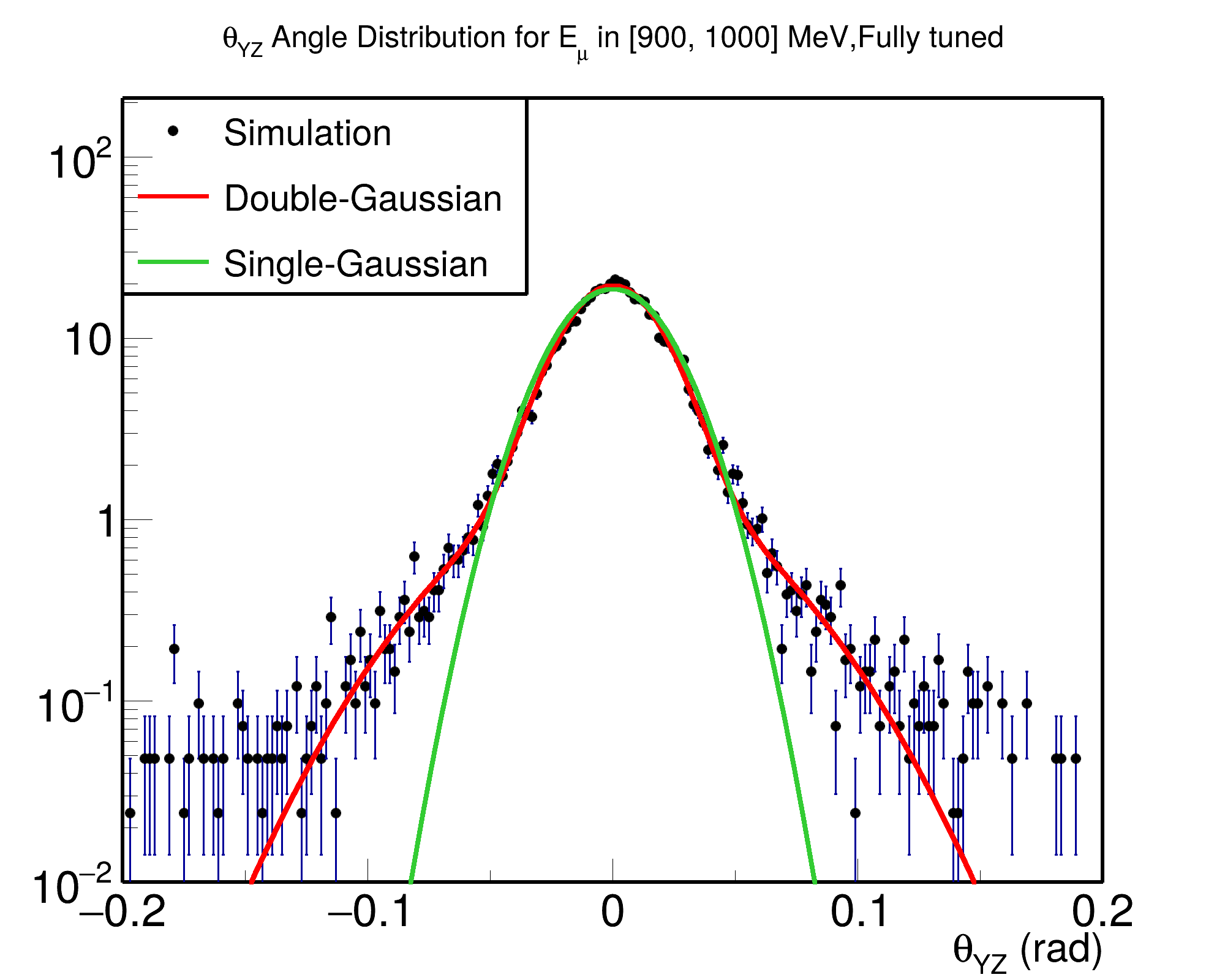}
     \includegraphics[clip,trim={2.0cm 0.0cm 5.3cm 4.9cm},width=0.32\textwidth]{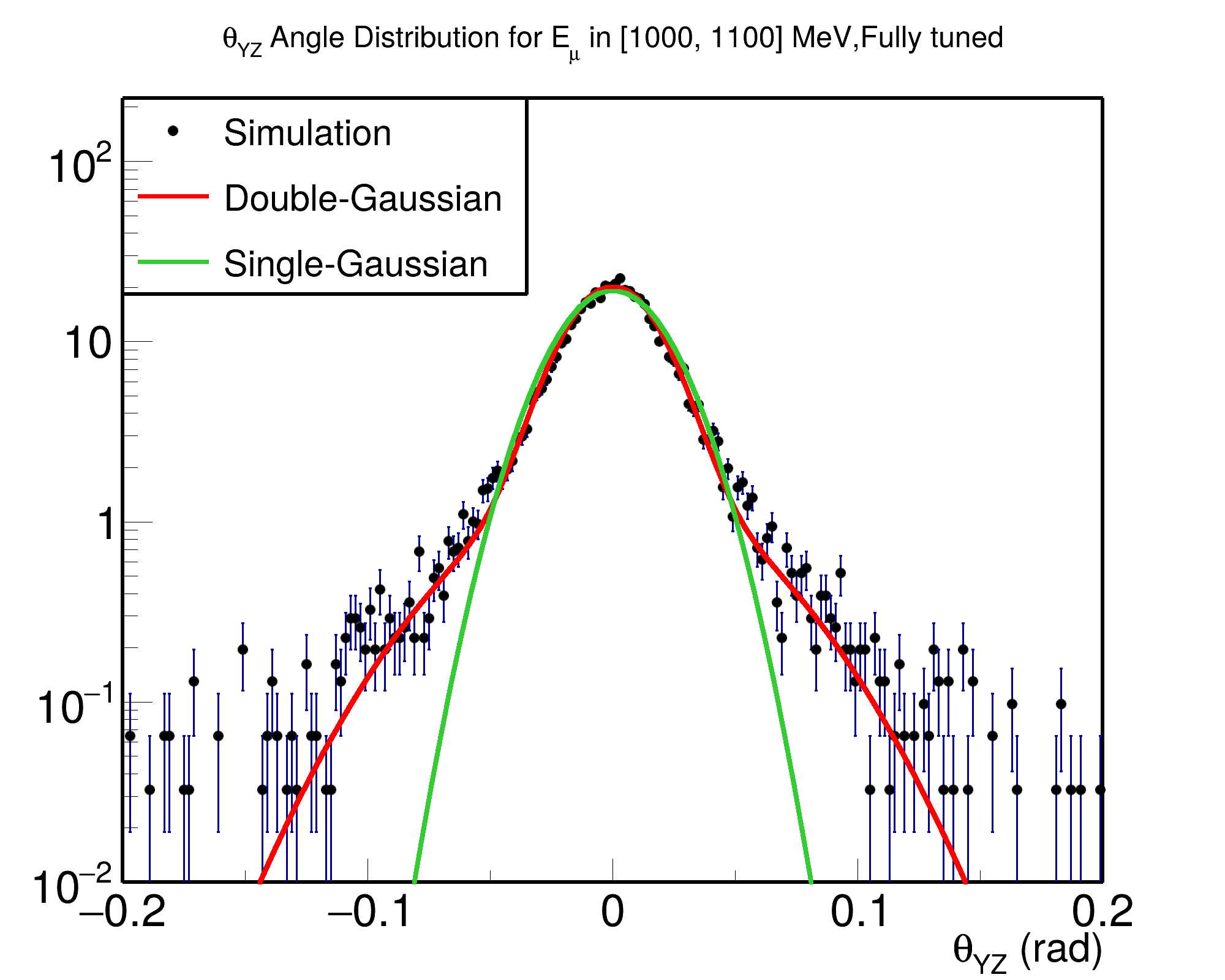}
     \vspace{0.3cm}
     \put(-255,116){\footnotesize Local $KE_{\mu} \in [0.9,1.0]$\,GeV}
     \put(-120,116){\footnotesize Local $KE_{\mu} \in [1.0,1.1]$\,GeV}
     \put(  20,116){\footnotesize Local $KE_{\mu} \in [1.1,1.2]$\,GeV}
     \put(-60,102){\footnotesize MicroBooNE}
     \put(-53, 91){\footnotesize Simulation}
     \includegraphics[clip,trim={2.0cm 0.0cm 5.3cm 4.9cm},width=0.32\textwidth]{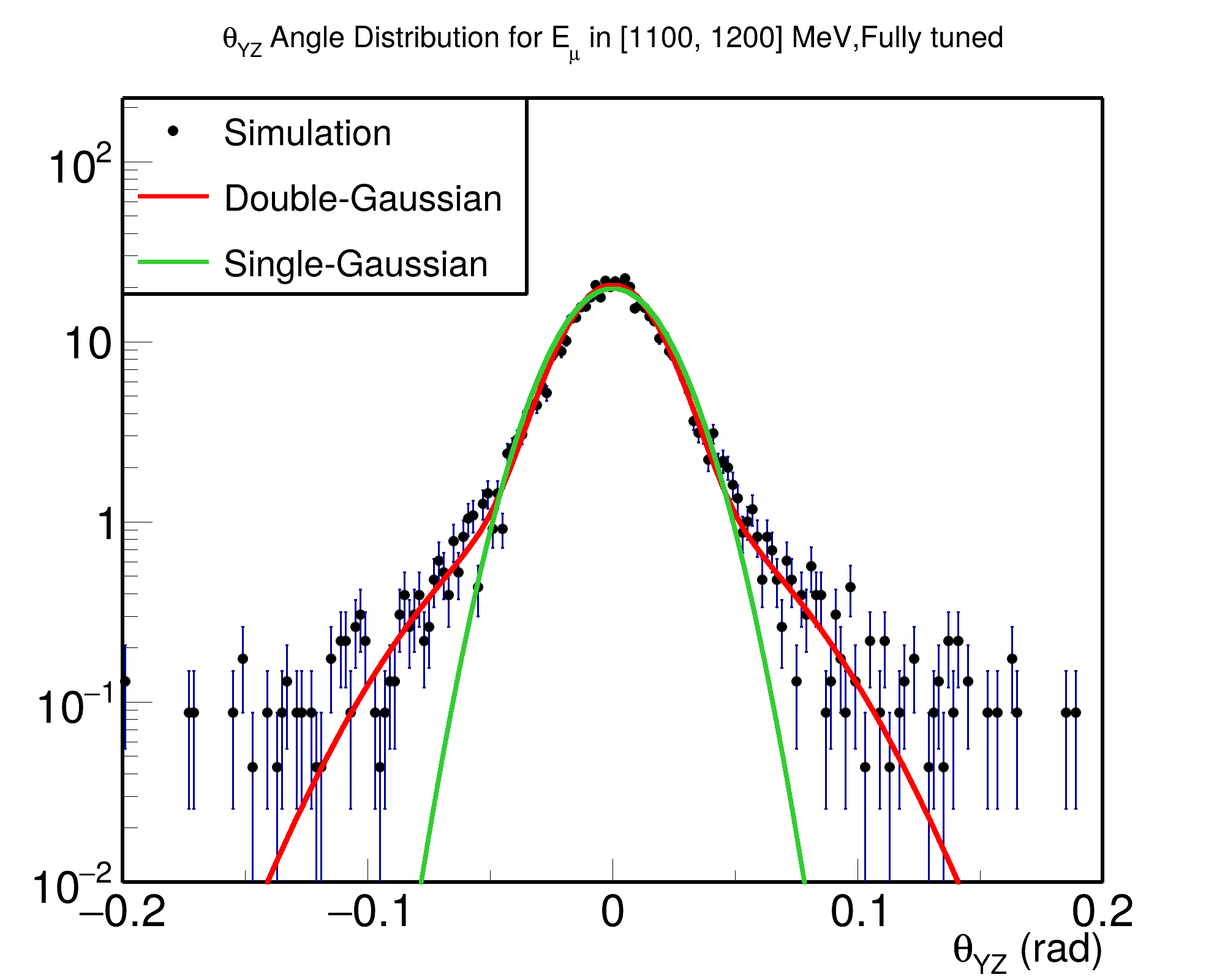}
     \includegraphics[clip,trim={2.0cm 0.0cm 5.3cm 4.9cm},width=0.32\textwidth]{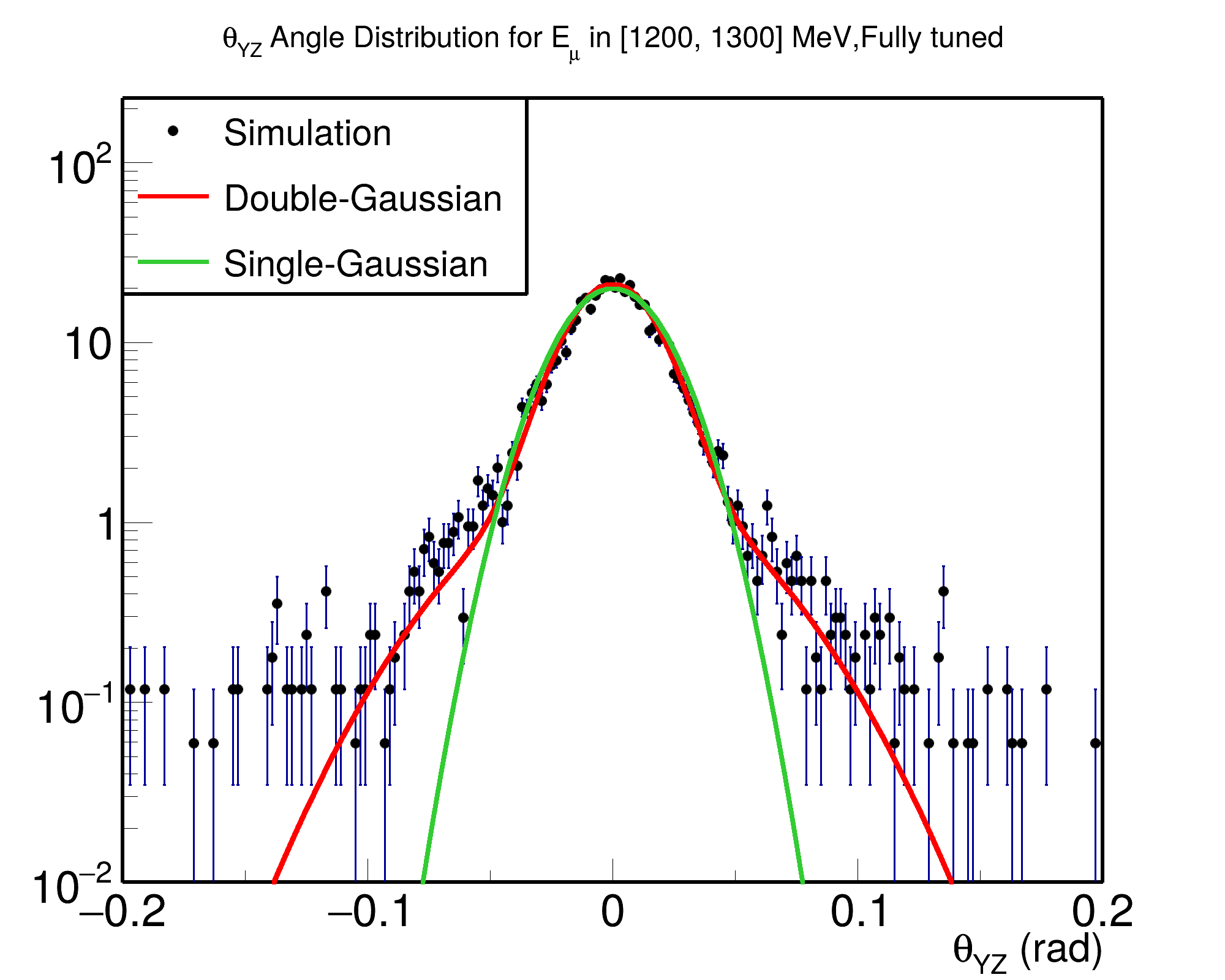}
     \includegraphics[clip,trim={2.0cm 0.0cm 5.3cm 4.9cm},width=0.32\textwidth]{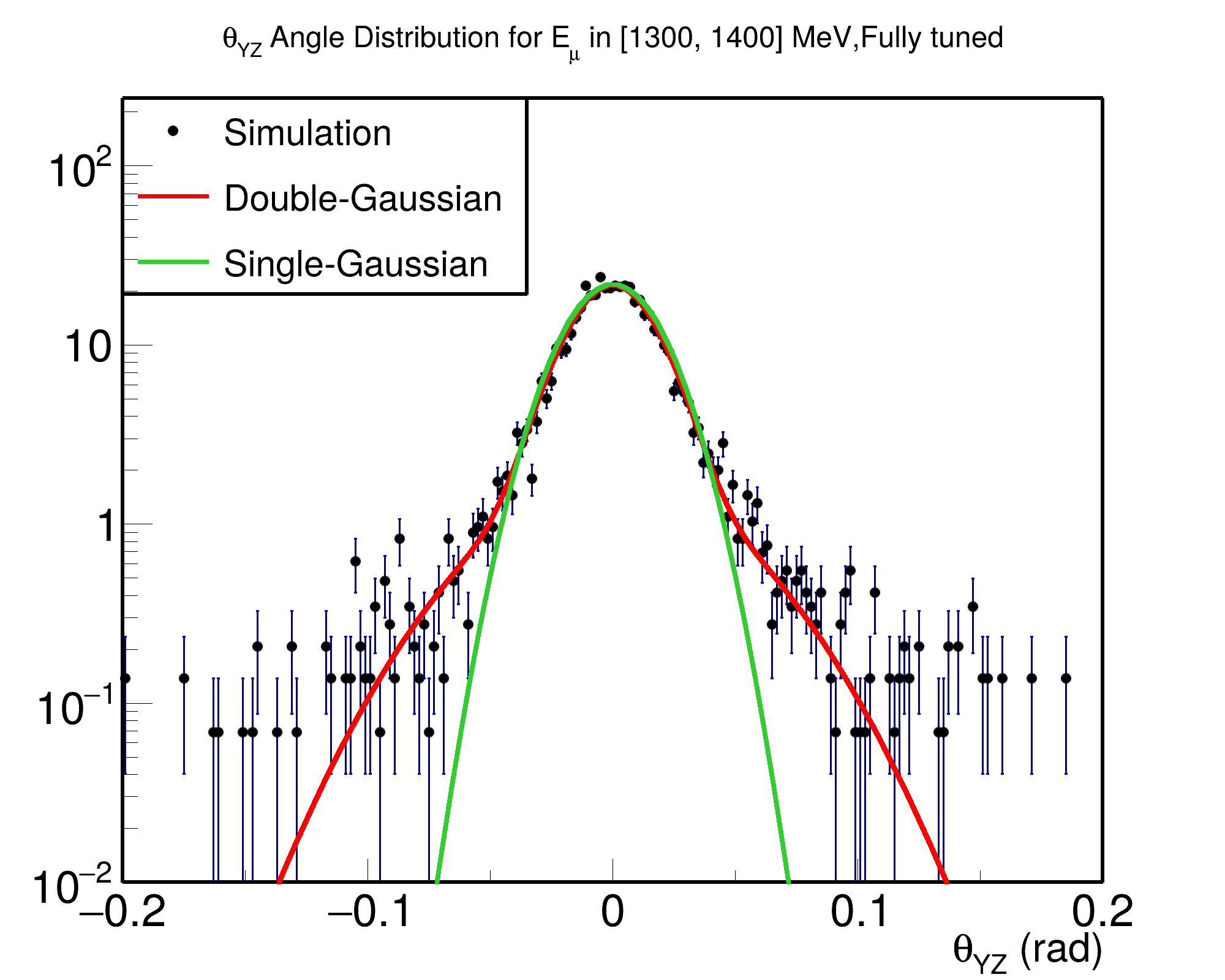}
     \vspace{0.3cm}
     \put(-257,116){\footnotesize Local $KE_{\mu} \in [1.2,1.3]$\,GeV}
     \put(-120,116){\footnotesize Local $KE_{\mu} \in [1.3,1.4]$\,GeV}
     \put(  20,116){\footnotesize Local $KE_{\mu} \in [1.4,1.5]$\,GeV}
     \put(-298,23){\rotatebox{90}{\small Probability Density}}
     \includegraphics[clip,trim={2.0cm 0.0cm 5.3cm 4.9cm},width=0.32\textwidth]{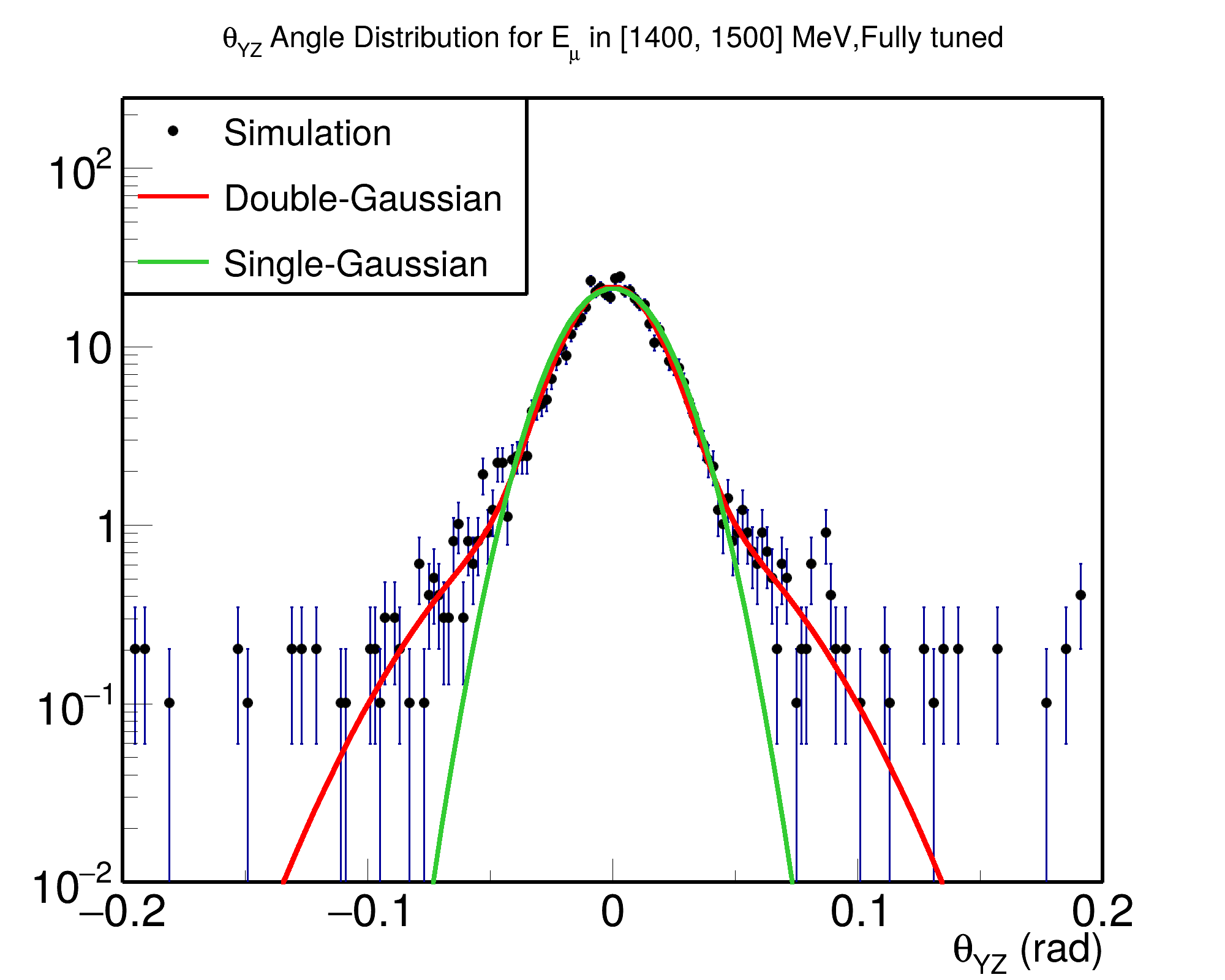}
     \includegraphics[clip,trim={2.0cm 0.0cm 5.3cm 4.9cm},width=0.32\textwidth]{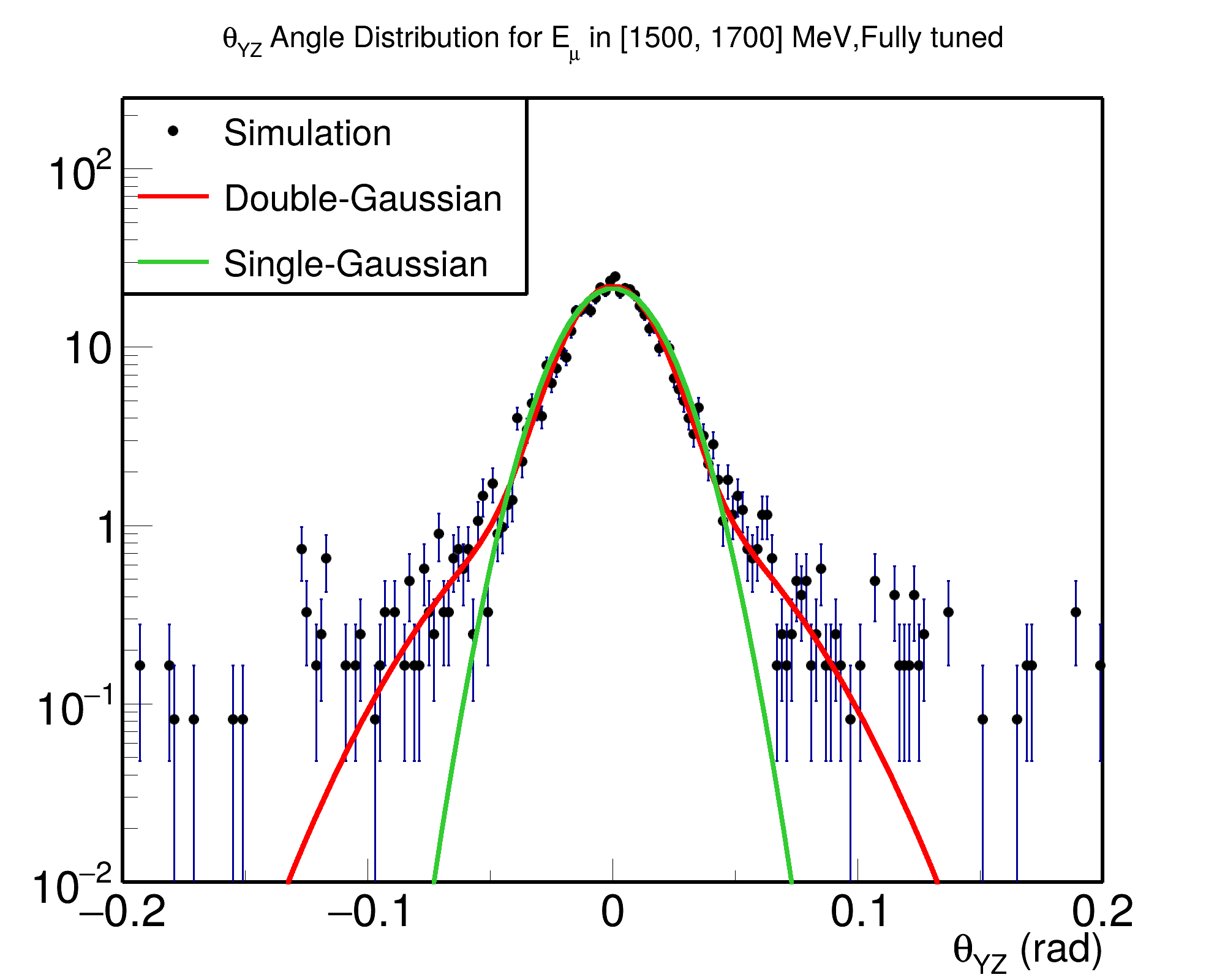}
     \includegraphics[clip,trim={2.0cm 0.0cm 5.3cm 4.9cm},width=0.32\textwidth]{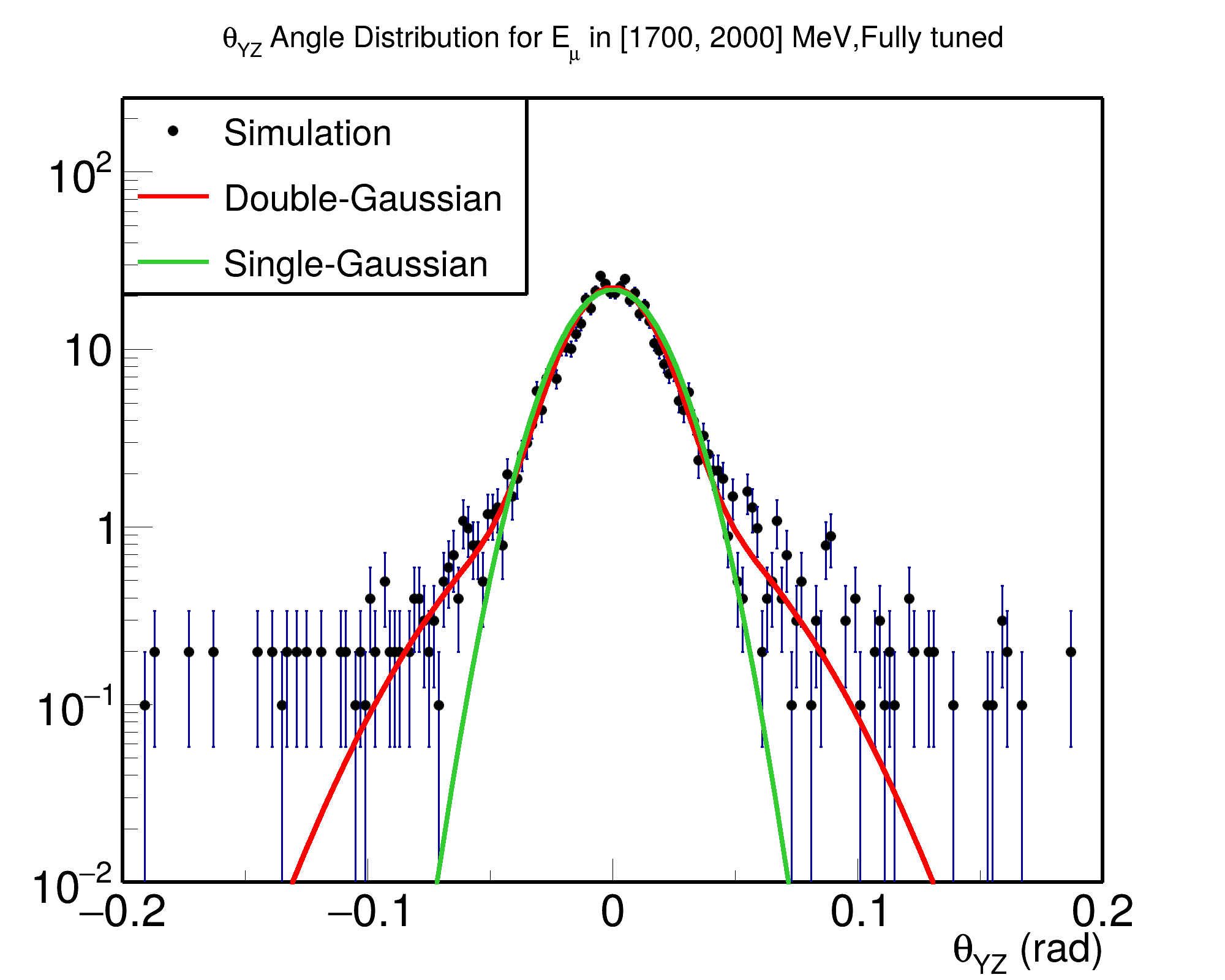}
     \put(-257,116){\footnotesize Local $KE_{\mu} \in [1.5,1.7]$\,GeV}
     \put(-118,116){\footnotesize Local $KE_{\mu} \in [1.7,2.0]$\,GeV}
     \put(  20,116){\footnotesize Local $KE_{\mu} \in [2.0,4.0]$\,GeV}
     \includegraphics[clip,trim={2.0cm 0.0cm 5.3cm 4.9cm},width=0.32\textwidth]{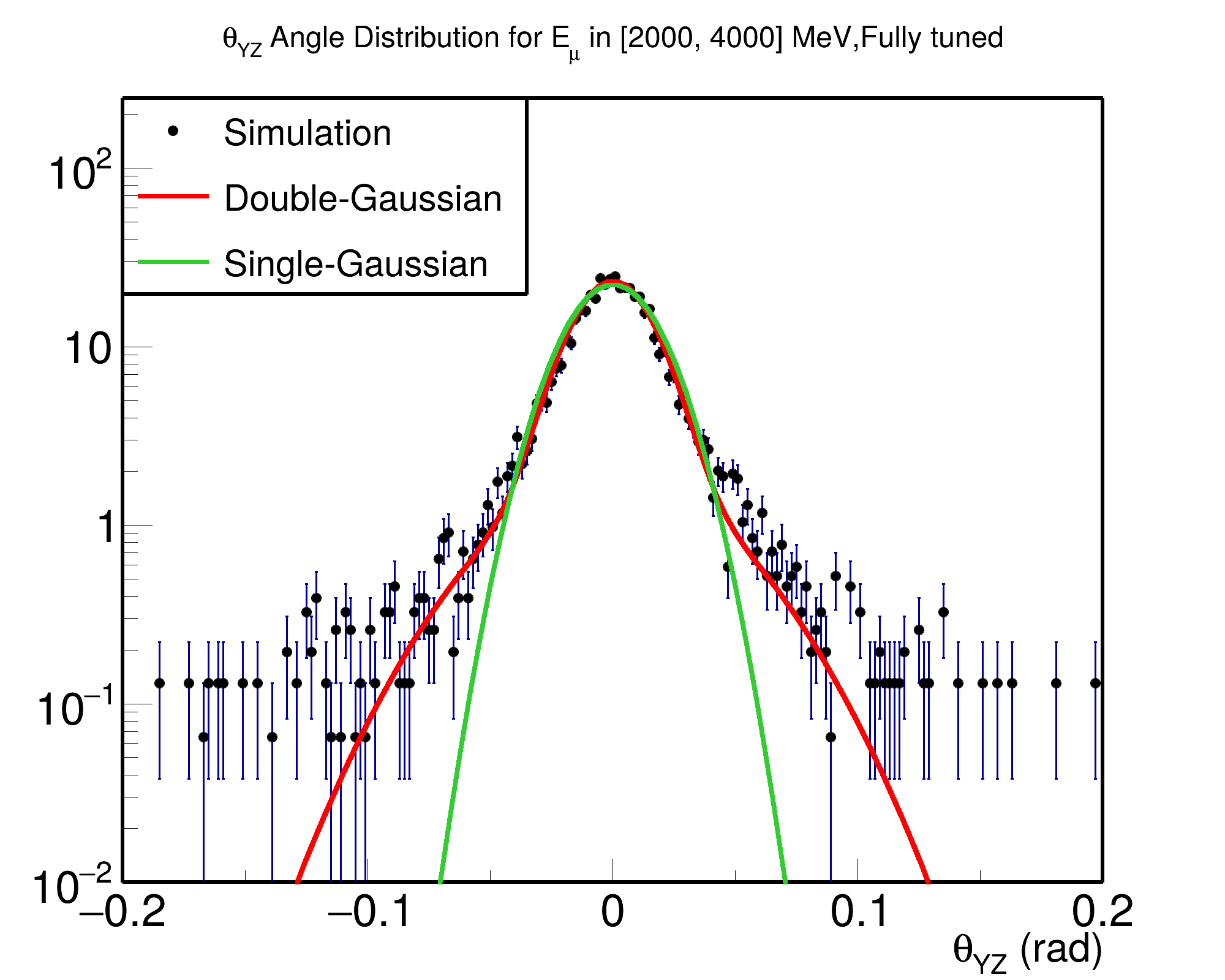}
     \caption{Comparison of single- and double-Gaussian PDFs to the reconstructed $\theta_{yz}$ distribution in simulation. Probability density is plotted on a log scale so differences at large angles can be seen. Both the tune of the double-Gaussian PDF and the simulation only correspond to segments with $|v_{x}| \in [0.2,0.35]$. The average muon kinetic energy of the reconstructed segments adjacent to each angle measurement is in bins from 0.9 to 4.0\,GeV.}
    \label{fig:theta_yz_vx2_high}
\end{figure}

To support visual comparisons of scattering angle PDFs with simulated distributions, a quantitative metric is used. Normally, a $\chi^{2}$ test statistic would be constructed from the summed squares of deviations in each bin between the prediction from the PDF and the estimated probability in simulation, and a $p$-value computed from this test statistic. However, in this situation such an approach fails to align with the overall goal of producing an effective likelihood function. Since likelihood maximization can be thought of as sampling from bins in accordance with their integrated probability density, different bins will have differing significances to the overall likelihood estimation in accordance with their integrated probability densities, which a simple $\chi^{2}$ metric does not capture.

Instead, we define the average likelihood score L:
\begin{equation}
    L = \left( \prod_{i=1}^{N} (w_{i}f(\theta_{i};E_{\mu}) )^{m_{i}} \right)^{\frac{1}{N}},
    \label{eqn:likelihood}
\end{equation}
where $f(\theta_{i} ; E_{\mu})$ is a given PDF evaluated at an angle bin $\theta_{i}$ and energy $E_{\mu}$, $w_{i}$ is the width of bin $i$, $m_{i}$ is the fraction of all simulated angles that correspond to bin $i$, and $N$ is the total number of bins. This quantity samples likelihoods from a given PDF from each angle bin in proportion to the number of measured angles in that bin, and then geometrically averages these likelihoods. Therefore, the maximum possible likelihood is achieved by a PDF constructed from the reconstructed angle histogram itself. We report the average likelihood scores, denoted here as the $L$-score, for the single- and double-Gaussian models as a fraction of this maximal likelihood, so that values closer to 1 can be interpreted as better descriptions of the reconstructed distribution. The $L$-score provides a way of mathematically determining how much a given PDF differs, in terms of its average impact on the total likelihood, from a hypothetical optimal PDF in simulation. In most $E_{\mu}$ bins the statistical fluctuations on the simulation-derived PDF are small, but at high energies where the sample size in a given bin is small, these fluctuations are non-negligible. As a result, it is not practical or desired for a model to always approach an $L$-score of 1, as this would require over-fitting to the non-physical statistical fluctuations present in the simulation.

\begin{table}[h!]
\centering
\caption{Computed $L$-scores for double- and single-Gaussian PDFs, labeled as $L_{2G}$ and $L_{1G}$ respectively, of $\theta_{xz}$ across $E_{\mu}$ bins.}
\begin{tabular}{|c|c|c|}
\hline
$E_{\mu}$ Bin (GeV) & $L_{2G}$ & $L_{1G}$ \\
\hline
$[0, 0.1]$   & $0.996 \pm 0.0003$  & $0.985 \pm 0.0007$ \\
$[0.1, 0.2]$ & $0.995 \pm 0.0003$  & $0.974 \pm 0.0008$ \\
$[0.2, 0.3]$ & $0.997 \pm 0.0002$ & $0.970 \pm 0.0011$ \\
$[0.3, 0.4]$ & $0.996 \pm 0.0003$  & $0.958 \pm 0.0016$ \\
$[0.4, 0.5]$ & $0.995 \pm 0.0005$ & $0.941 \pm 0.0022$ \\
$[0.5, 0.6]$ & $0.992 \pm 0.0007$ & $0.923 \pm 0.0030$ \\
$[0.6, 0.7]$ & $0.987 \pm 0.0011$  & $0.901 \pm 0.0043$ \\
$[0.7, 0.8]$ & $0.983 \pm 0.0015$  & $0.879 \pm 0.006$ \\
$[0.8, 0.9]$ & $0.983 \pm 0.0017$  & $0.873 \pm 0.006$ \\
$[0.9, 1.0]$ & $0.980 \pm 0.0022$  & $0.844 \pm 0.008$ \\
$[1.0, 1.1]$ & $0.970 \pm 0.0030$  & $0.791 \pm 0.011$ \\
$[1.1, 1.2]$ & $0.961 \pm 0.0039$  & $0.750 \pm 0.013$ \\
$[1.2, 1.3]$ & $0.948 \pm 0.006$  & $0.734 \pm 0.018$ \\
$[1.3, 1.4]$ & $0.932 \pm 0.008$  & $0.706 \pm 0.023$ \\
$[1.4, 1.5]$ & $0.937 \pm 0.008$  & $0.675 \pm 0.025$ \\
$[1.5, 1.6]$ & $0.920 \pm 0.011$   & $0.650 \pm 0.031$ \\
$[1.6, 1.8]$ & $0.908 \pm 0.011$   & $0.556 \pm 0.031$ \\
$[1.8, 2.0]$ & $0.881 \pm 0.014$   & $0.456 \pm 0.033$ \\
$[2.0, 2.5]$ & $0.903 \pm 0.011$   & $0.476 \pm 0.031$ \\
$[2.5, 3.0]$ & $0.918 \pm 0.011$   & $0.494 \pm 0.034$ \\
$[3.0, 4.0]$ & $0.878 \pm 0.015$   & $0.299 \pm 0.034$ \\
\hline
\end{tabular}
\label{table:theta_xz_lscores}
\end{table}

\begin{table}[h!]
\centering
\caption{Computed $L$-scores for double- and single-Gaussian PDFs, labeled as $L_{2G}$ and $L_{1G}$ respectively, of $\theta_{yz}$ in each $|v_{x}|$ tune across $E_{\mu}$ bins.}
\resizebox{0.98\textwidth}{!}{\begin{tabular}{|c|cc|cc|cc|}
\hline
$E_{\mu}$ Bin (GeV) & \multicolumn{2}{c|}{$|v_{x}| \in [0, 0.1]$} & \multicolumn{2}{c|}{$|v_{x}| \in [0.1, 0.2]$} & \multicolumn{2}{c|}{$|v_{x}| \in [0.2, 0.35]$}  \\
     & $L_{\rm 2G}$ & $L_{\rm 1G}$ & $L_{\rm 2G}$ & $L_{\rm 1G}$ & $L_{\rm 2G}$ & $L_{\rm 1G}$ \\
\hline
$[0,0.1]$     
& 0.978 $\pm$ 0.002 & 0.976 $\pm$ 0.002 
& 0.980 $\pm$ 0.002 & 0.954 $\pm$ 0.004 
& 0.988 $\pm$ 0.001 & 0.975 $\pm$ 0.002\\

$[0.1,0.2]$   
& 0.986 $\pm$ 0.001 & 0.938 $\pm$ 0.003 
& 0.979 $\pm$ 0.002 & 0.934 $\pm$ 0.004 
& 0.991 $\pm$ 0.001 & 0.960 $\pm$ 0.002\\

$[0.2,0.3]$   
& 0.984 $\pm$ 0.001 & 0.870 $\pm$ 0.005 
& 0.988 $\pm$ 0.001 & 0.883 $\pm$ 0.006 
& 0.990 $\pm$ 0.001 & 0.932 $\pm$ 0.004\\

$[0.3,0.4]$   
& 0.983 $\pm$ 0.001 & 0.782 $\pm$ 0.006 
& 0.981 $\pm$ 0.002 & 0.786 $\pm$ 0.009 
& 0.986 $\pm$ 0.001 & 0.892 $\pm$ 0.006\\

$[0.4,0.5]$   
& 0.983 $\pm$ 0.001 & 0.747 $\pm$ 0.008
& 0.976 $\pm$ 0.002 & 0.785 $\pm$ 0.009
& 0.989 $\pm$ 0.001 & 0.852 $\pm$ 0.006\\

$[0.5,0.6]$   
& 0.982 $\pm$ 0.001 & 0.696 $\pm$ 0.009 
& 0.968 $\pm$ 0.003 & 0.730 $\pm$ 0.010 
& 0.983 $\pm$ 0.002 & 0.780 $\pm$ 0.009\\

$[0.6,0.7]$   
& 0.983 $\pm$ 0.002 & 0.627 $\pm$ 0.010 
& 0.963 $\pm$ 0.003 & 0.619 $\pm$ 0.013 
& 0.981 $\pm$ 0.002 & 0.745 $\pm$ 0.010\\

$[0.7,0.8]$   
& 0.982 $\pm$ 0.002 & 0.661 $\pm$ 0.011 
& 0.970 $\pm$ 0.003 & 0.615 $\pm$ 0.014 
& 0.975 $\pm$ 0.002 & 0.669 $\pm$ 0.012\\

$[0.8,0.9]$   
& 0.975 $\pm$ 0.002 & 0.655 $\pm$ 0.012 
& 0.959 $\pm$ 0.004 & 0.530 $\pm$ 0.016 
& 0.974 $\pm$ 0.003 & 0.660 $\pm$ 0.014\\

$[0.9,1.0]$  
& 0.976 $\pm$ 0.003 & 0.692 $\pm$ 0.012 
& 0.950 $\pm$ 0.004 & 0.557 $\pm$ 0.017 
& 0.959 $\pm$ 0.004 & 0.570 $\pm$ 0.017\\

$[1.0,1.1]$ 
& 0.977 $\pm$ 0.003 & 0.710 $\pm$ 0.013 
& 0.944 $\pm$ 0.006 & 0.506 $\pm$ 0.020 
& 0.950 $\pm$ 0.005 & 0.537 $\pm$ 0.019\\

$[1.1,1.2]$ 
& 0.971 $\pm$ 0.004 & 0.688 $\pm$ 0.017 
& 0.926 $\pm$ 0.007 & 0.471 $\pm$ 0.022 
& 0.941 $\pm$ 0.006 & 0.534 $\pm$ 0.024\\

$[1.2,1.3]$ 
& 0.966 $\pm$ 0.004 & 0.639 $\pm$ 0.019 
& 0.915 $\pm$ 0.008 & 0.524 $\pm$ 0.024 
& 0.914 $\pm$ 0.009 & 0.459 $\pm$ 0.026\\

\hline
$E_{\mu}$ Bin (GeV) & \multicolumn{2}{c|}{$|v_{x}| \in [0.35, 0.75]$} & \multicolumn{2}{c|}{$|v_{x}| \in [0.75, 1]$} & & \\
     & $L_{\rm 2G}$ & $L_{\rm 1G}$ & $L_{\rm 2G}$ & $L_{\rm 1G}$ & & \\
\hline
$[0,0.1]$     
& 0.995 $\pm$ 0.001 & 0.982 $\pm$ 0.001 
& 0.988 $\pm$ 0.001 & 0.982 $\pm$ 0.002
& -- & -- \\

$[0.1,0.2]$    
& 0.996 $\pm$ 0.001 & 0.973 $\pm$ 0.001 
& 0.993 $\pm$ 0.001 & 0.971 $\pm$ 0.003
& -- & -- \\

$[0.2,0.3]$   
& 0.995 $\pm$ 0.001 & 0.953 $\pm$ 0.002 
& 0.993 $\pm$ 0.001 & 0.943 $\pm$ 0.005
& -- & -- \\

$[0.3,0.4]$   
& 0.993 $\pm$ 0.001 & 0.926 $\pm$ 0.003 
& 0.992 $\pm$ 0.001 & 0.875 $\pm$ 0.010
& -- & -- \\

$[0.4,0.5]$    
& 0.989 $\pm$ 0.001 & 0.848 $\pm$ 0.005 
& 0.973 $\pm$ 0.003 & 0.837 $\pm$ 0.014
& -- & -- \\

$[0.5,0.6]$   
& 0.989 $\pm$ 0.001 & 0.815 $\pm$ 0.006 
& 0.956 $\pm$ 0.005 & 0.753 $\pm$ 0.027
& -- & -- \\

$[0.6,0.7]$   
& 0.987 $\pm$ 0.001 & 0.777 $\pm$ 0.008 
& -- & -- 
& -- & -- \\

$[0.7,0.8]$
& 0.981 $\pm$ 0.002 & 0.735 $\pm$ 0.011 
& -- & -- 
& -- & -- \\

$[0.8,0.9]$   
& 0.977 $\pm$ 0.003 & 0.707 $\pm$ 0.014 
& -- & -- 
& -- & -- \\

$[0.9,1.0]$  
& 0.966 $\pm$ 0.004 & 0.620 $\pm$ 0.019 
& -- & -- 
& -- & -- \\

$[1.0,1.1]$ 
& 0.948 $\pm$ 0.006 & 0.573 $\pm$ 0.024 
& -- & -- 
& -- & -- \\

$[1.1,1.2]$ 
& 0.926 $\pm$ 0.009 & 0.504 $\pm$ 0.031 
& -- & -- 
& -- & -- \\

$[1.2,1.3]$ 
& 0.897 $\pm$ 0.013 & 0.478 $\pm$ 0.038 
& -- & -- 
& -- & -- \\

\hline
\end{tabular} }
\label{table:theta_yz_lscores}
\end{table}

Comparisons of predicted and reconstructed angle scattering distributions are shown in figure~\ref{fig:theta_xz_low} and figure~\ref{fig:theta_xz_high} for $\theta_{xz}$ and figure~\ref{fig:theta_yz_vx2_low} and figure~\ref{fig:theta_yz_vx2_high} for the $|v_{x}| \in [0.2,0.35]$ tune of $\theta_{yz}$, as it contains a large number of reconstructed angles across $E_{\mu}$. In each plot, the fully tuned double-Gaussian model, as well as a single-Gaussian fit, are compared to the reconstructed distribution in a given bin of muon kinetic energy, $KE_{\mu}$, so that the differences between these PDFs can be seen. Table~\ref{table:theta_xz_lscores} and Table~\ref{table:theta_yz_lscores} summarize the $L$-scores for the single- and double-Gaussian models for $\theta_{xz}$ and $\theta_{yz}$, respectively.

Both single- and double-Gaussian models provide good descriptions of the reconstructed distributions at low energies. Below a GeV the double-Gaussian model does extremely well, with an $L$-score often above 0.98, and the single Gaussian model also performs reasonably well with an $L$-score often above 0.8. This is partially because of the reduced importance of the detector resolution and distribution tails, and partially because of increased sample sizes in each bin, which reduces statistical fluctuations.

At higher energies the descriptive power of the double-Gaussian model degrades slightly but still typically maintains an $L$-score above 0.9. By comparison, the single-Gaussian model often struggles at higher energies with $L$-scores frequently around 0.5. This factor of two difference in $L$-scores demonstrates the improvements gained through the use of the double-Gaussian model, as poor $L$-scores, particularly with non-flat distributions as seen in the single-Gaussian model, worsen the bias and resolution of the maximum likelihood estimate. Furthermore, the consistently high $L$-scores of the double-Gaussian model demonstrate the success of the tuning procedure outlined above.

\subsection{Maximum likelihood energy estimation}

Maximum likelihood estimation uses a series of measurements $\theta_{i}$ and corresponding PDFs $f_{i}(\theta_{i})$ to estimate the values of the parameters of the PDFs. This is achieved by computing the total likelihood $\mathcal{L}$ as the product of $N$ probability densities, or likelihoods, and finding the optimal parameter values that maximize $\mathcal{L}$:
\begin{equation}
    \mathcal{L} = \prod_{i=1}^{N} f_{i}(\theta_{i}).
    \label{eqn:likelihood_fn}
\end{equation}
We use the previously tuned double-Gaussian PDFs as the $f_{i}(\theta_{i}) = f(\theta_{i};E_{\mu \text{ }i}^{\text{MCS}})$, and vary the estimate of the total muon energy, $E_{\mu}^{\text{MCS}}$, and therefore the estimates $E_{\mu \text{ }i}^{\text{MCS}}$ propagated to each angle measurement, to maximize the likelihood. The total energy estimate is propagated to the center-point of each segment along the muon track using the length of reconstructed track and the mean energy loss predicted by the Bethe-Bloch formula~\cite{bethe-bloch}. Each measured angle $\theta_{i}$ is adjoined by two segments whose average estimated energy is used as the value for $E_{\mu \text{ }i}^{\text{MCS}}$. The predicted likelihood is then calculated using $f(\theta_{i};E_{\mu \text{ }i}^{\text{MCS}})$ from equation~\ref{eqn:double_gaussian}, and the total likelihood is maximized to determine the optimal starting muon energy estimate $E_{\mu}^{\text{MCS}}$. For computational convenience, in practice the natural logarithm of the likelihood is computed and maximized, which is equivalent to maximizing $\mathcal{L}$ as the logarithm function is strictly increasing.

It has already been noted that using accurate PDFs is an important aspect of MLE. The previously detailed model improvements and tuning effort all contribute to accurate PDF models, but it is worth giving particular attention to the estimation of the detector resolution. If the estimated resolution is too low, then measured angles will appear consistent with a larger MCS effect, biasing the estimate to lower energies. If the estimated resolution is too high, then angle measurements for some tracks may not be consistent with any energy estimate if measured values are frequently smaller than the estimated resolution. In this case, the likelihood will be maximized when the energy estimate is infinity, rendering MLE nearly useless for that particle. These effects are most relevant at higher energies where the detector resolution term dominates the angle distribution prediction. Given the wide variation of detector resolutions observed in this work, simpler models, such as those used in previous works, may lack the capacity to describe detector variations and may be susceptible to bias in their energy estimate.

\section{Estimated resolution and bias}\label{sec:resolution}

\begin{figure}[hbtp!]
     \centering
     \includegraphics[clip,trim={0.0cm 0.0cm 0.0cm 2.5cm},width=0.95\textwidth]{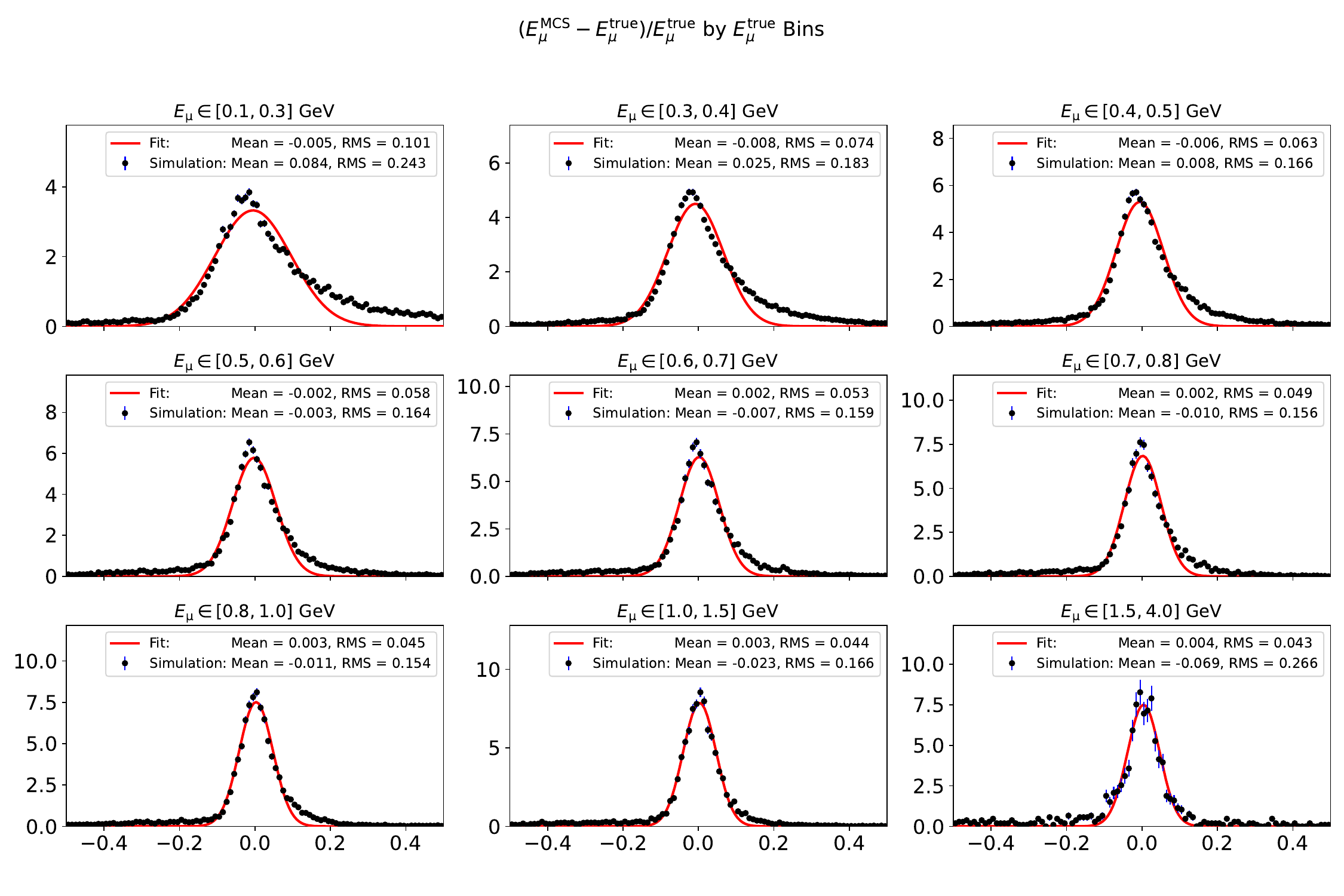}
     \put(-223,-4){\small $\frac{E_{\mu}^{MCS}-E_{\mu}}{E_{\mu}}$}
     \put(-412,95){\rotatebox{90}{\scriptsize Probability Density}}
     \put(-383,210){\scalebox{.8}{\footnotesize Fit Gaussian}}
     \put(-382,200){\scalebox{.8}{\footnotesize Area = 0.85}}
     \put(-247,210){\scalebox{.8}{\footnotesize Fit Gaussian}}
     \put(-246,200){\scalebox{.8}{\footnotesize Area = 0.84}}
     \put(-110,210){\scalebox{.8}{\footnotesize Fit Gaussian}}
     \put(-109,200){\scalebox{.8}{\footnotesize Area = 0.83}}
     \put(-383,134){\scalebox{.8}{\footnotesize Fit Gaussian}}
     \put(-382,124){\scalebox{.8}{\footnotesize Area = 0.84}}
     \put(-247,134){\scalebox{.8}{\footnotesize Fit Gaussian}}
     \put(-246,124){\scalebox{.8}{\footnotesize Area = 0.84}}
     \put(-110,134){\scalebox{.8}{\footnotesize Fit Gaussian}}
     \put(-109,124){\scalebox{.8}{\footnotesize Area = 0.84}}
     \put(-383,57){\scalebox{.8}{\footnotesize Fit Gaussian}}
     \put(-382,47){\scalebox{.8}{\footnotesize Area = 0.84}}
     \put(-247,57){\scalebox{.8}{\footnotesize Fit Gaussian}}
     \put(-246,47){\scalebox{.8}{\footnotesize Area = 0.86}}
     \put(-110,57){\scalebox{.8}{\footnotesize Fit Gaussian}}
     \put(-109,47){\scalebox{.8}{\footnotesize Area = 0.81}}
     \put(-54,209){\scalebox{.8}{\small MicroBooNE}}
     \put(-50,200){\scalebox{.8}{\small Simulation}}
     \caption{Fractional error of $E_{\mu}^{\text{MCS}}$ in bins of true $E_{\mu}$ for FC muon events. In each bin a Gaussian distribution is fit to describe the bias and resolution on the central region of each distribution. The fit is centered on the most probable value and extended by the RMS of the simulated distribution in each direction. The histograms are area-normalized to have a total integrated area of unity, while the fit Gaussians allow their total area to vary.}
    \label{fig:fc_error}
\end{figure}

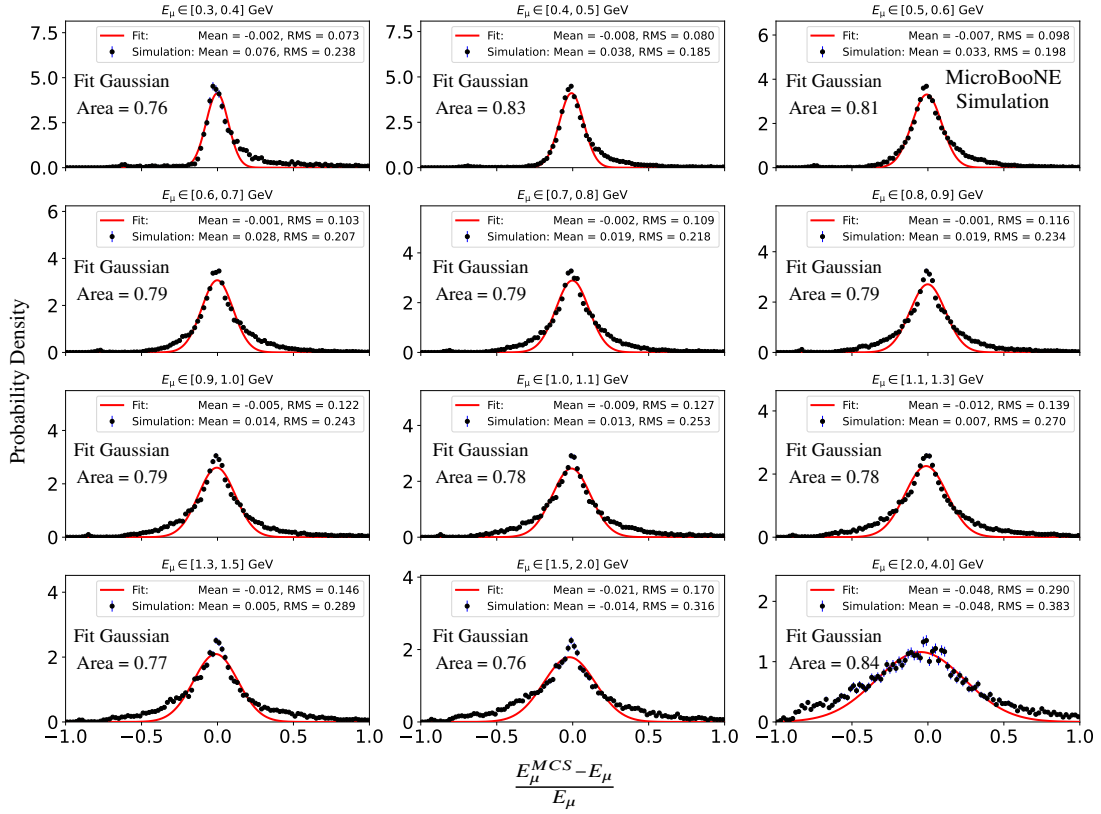
\begin{figure}[hbtp!]
     \centering
     \includegraphics[clip,trim={0.0cm 0.0cm 0.0cm 2.5cm},width=0.95\textwidth]{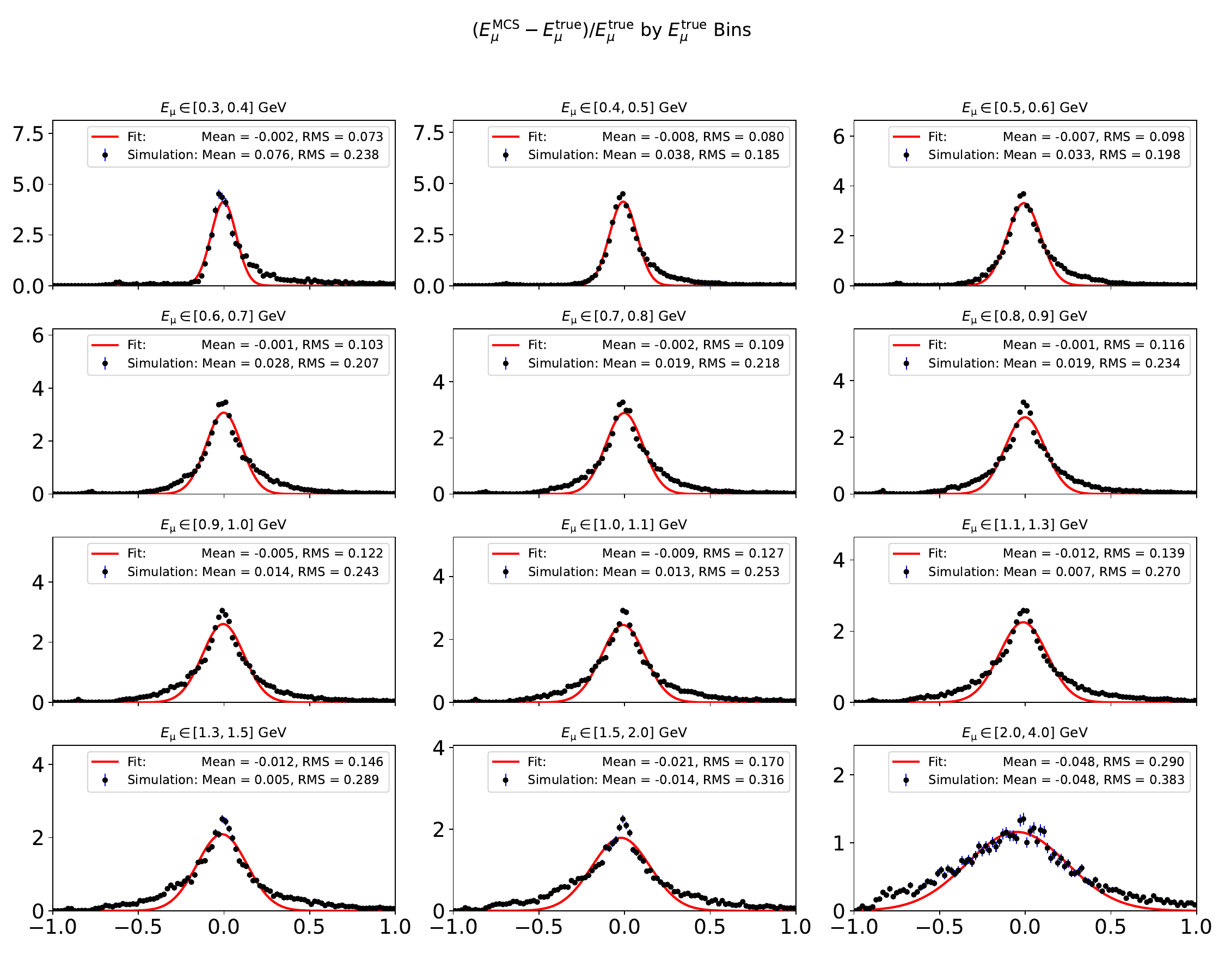}
     \put(-223,-4){\small $\frac{E_{\mu}^{MCS}-E_{\mu}}{E_{\mu}}$}
     \put(-412,120){\rotatebox{90}{\scriptsize Probability Density}}
     \put(-388,262){\scalebox{.8}{\footnotesize Fit Gaussian}}
     \put(-387,252){\scalebox{.8}{\footnotesize Area = 0.76}}
     \put(-253,262){\scalebox{.8}{\footnotesize Fit Gaussian}}
     \put(-252,252){\scalebox{.8}{\footnotesize Area = 0.83}}
     \put(-120,262){\scalebox{.8}{\footnotesize Fit Gaussian}}
     \put(-119,252){\scalebox{.8}{\footnotesize Area = 0.81}}
     \put(-388,192){\scalebox{.8}{\footnotesize Fit Gaussian}}
     \put(-387,182){\scalebox{.8}{\footnotesize Area = 0.79}}
     \put(-253,192){\scalebox{.8}{\footnotesize Fit Gaussian}}
     \put(-252,182){\scalebox{.8}{\footnotesize Area = 0.79}}
     \put(-120,192){\scalebox{.8}{\footnotesize Fit Gaussian}}
     \put(-119,182){\scalebox{.8}{\footnotesize Area = 0.79}}
     \put(-388,123){\scalebox{.8}{\footnotesize Fit Gaussian}}
     \put(-387,113){\scalebox{.8}{\footnotesize Area = 0.79}}
     \put(-253,123){\scalebox{.8}{\footnotesize Fit Gaussian}}
     \put(-252,113){\scalebox{.8}{\footnotesize Area = 0.78}}
     \put(-120,123){\scalebox{.8}{\footnotesize Fit Gaussian}}
     \put(-119,113){\scalebox{.8}{\footnotesize Area = 0.78}}
     \put(-388,53){\scalebox{.8}{\footnotesize Fit Gaussian}}
     \put(-387,43){\scalebox{.8}{\footnotesize Area = 0.77}}
     \put(-253,53){\scalebox{.8}{\footnotesize Fit Gaussian}}
     \put(-252,43){\scalebox{.8}{\footnotesize Area = 0.76}}
     \put(-120,53){\scalebox{.8}{\footnotesize Fit Gaussian}}
     \put(-119,43){\scalebox{.8}{\footnotesize Area = 0.84}}
     \put(-60,263){\scalebox{.8}{\small MicroBooNE}}
     \put(-56,254){\scalebox{.8}{\small Simulation}}
     \caption{Fractional error of $E_{\mu}^{\text{MCS}}$ in bins of true $E_{\mu}$ for PC muon events with at least one meter of reconstructed track. In each bin a Gaussian distribution is fit to describe the bias and resolution on the central region of each distribution. The fit is centered on the most probable value and extended by the RMS of the simulated distribution in each direction. Since each $E_{\mu}$ bin contains a range of reconstructed muon track lengths, each with different resolution performances, the total distribution in each $E_{\mu}$ bin differs from a Gaussian distribution, with a narrow peak from long muon tracks and long tails from short muon tracks. For large fractional errors there is a skew toward positive percentile errors. For example, a factor of two deficit constitutes a -50\% error, while a factor of two excess constitutes a +100\% error. The histograms are area-normalized to have a total integrated area of unity, while the fit Gaussians allow their total area to vary.}
    \label{fig:pc_error}
\end{figure}

In this section we study the performance of $E_{\mu}^{\text{MCS}}$ in simulation, before examining the data-model consistency of $E_{\mu}^{\text{MCS}}$ in section~\ref{sec:validation}. To ensure an independent evaluation of the estimator's performance, separate simulation samples are used for tuning and for validation. Simulation corresponding to runs 1–3 of data taking is used for the algorithm tuning described in section~\ref{sec:tuning}. The performance studies in this section use an independent simulation sample corresponding to runs 4–5 of data taking. Figure~\ref{fig:fc_error} shows the fractional error in bins of true muon energy for fully contained (FC) muons. A Gaussian distribution is fit to each bin to estimate the bias and resolution, which are reported in Table~\ref{table:fc_error}. Each fit is performed on a region centered on the most probable value, extending by the root mean square (RMS) of the simulated distribution in each direction. The total area of the Gaussian fit is allowed to vary, letting it capture the Gaussian-like core of the distribution, leaving a description of the tail to the RMS statistic of the histogram. Below 350\,MeV, the muon track is less than a meter long, allowing at most six measurements of \{$\theta_{xz}$, $\theta_{yz}$\} angle pairs. As a result, the resolution is quite poor and the estimator is biased by a long tail of over-estimated muon energies. At higher energies the performance improves significantly, reaching a resolution of 4.3\%. The fractional error distribution is predominantly Gaussian in shape for $E_{\mu}$ above 0.5\,GeV, and the bias is less than 1\%.

\begin{table}[h!]
\centering
\caption{Estimated bias and resolution for FC muons from fits of Gaussian distributions to $E_{\mu}^{\text{MCS}}$ fractional error in bins of true $E_{\mu}$. The fit is centered on the most probable value and extended by the RMS of the simulated distribution in each direction. Note that there are very few contained muons above 2\,GeV given the detector geometry. }
\begin{tabular}{|c|c|c|}
\hline
$E_{\mu}$ Bin (GeV) & Bias (\%) & Resolution (\%) \\
\hline
$[0.1, 0.3]$   & $-0.51 \pm 0.33$ & $10.15 \pm 0.34$ \\
$[0.3, 0.4]$   & $-0.81 \pm 0.24$ & $7.41 \pm 0.24$ \\
$[0.4, 0.5]$   & $-0.59 \pm 0.21$ & $6.28 \pm 0.22$ \\
$[0.5, 0.6]$   & $-0.19 \pm 0.19$ & $5.77 \pm 0.20$ \\
$[0.6, 0.7]$   & $0.20 \pm 0.19$  & $5.31 \pm 0.19$ \\
$[0.7, 0.8]$   & $0.19 \pm 0.18$  & $4.87 \pm 0.18$ \\
$[0.8, 1.0]$  & $0.29 \pm 0.14$  & $4.48 \pm 0.14$ \\
$[1.0, 1.5]$ & $0.30 \pm 0.11$  & $4.35 \pm 0.11$ \\
$[1.5, 4.0]$ & $0.42 \pm 0.16$  & $4.28 \pm 0.16$ \\
\hline
\end{tabular}
\label{table:fc_error}
\end{table}

\begin{table}[h!]
\centering
\caption{Estimated bias and resolution for PC muons with and without a selection requirement of at least one meter of reconstructed track. Resolution and bias are determined from fits of Gaussian distributions to $E_{\mu}^{\text{MCS}}$ fractional error in bins of true $E_{\mu}$. The fit is centered on the most probable value and extended by the RMS of the simulated distribution in each direction.}
\begin{tabular}{|c|cc|cc|}
\hline
$E_{\mu}$ Bin (GeV) 
& \multicolumn{2}{c|}{PC events with 1\,m muon track} 
& \multicolumn{2}{c|}{All PC events} \\
\cline{2-5}
& Bias (\%) & Resolution (\%) 
& Bias (\%) & Resolution (\%) \\
\hline

$[0.3, 0.4]$ & $-0.19 \pm 0.43$ & $7.33 \pm 0.43$ 
              & $-1.37 \pm 0.42$ & $11.57 \pm 0.45$ \\

$[0.4, 0.5]$ & $-0.80 \pm 0.33$ & $8.03 \pm 0.34$ 
              & $-1.49 \pm 0.54$ & $12.68 \pm 0.57$ \\

$[0.5, 0.6]$ & $-0.73 \pm 0.34$ & $9.80 \pm 0.38$ 
              & $-1.15 \pm 0.52$ & $13.97 \pm 0.58$ \\

$[0.6, 0.7]$ & $-0.15 \pm 0.47$ & $10.31 \pm 0.51$ 
              & $-0.51 \pm 0.63$ & $13.73 \pm 0.69$ \\

$[0.7, 0.8]$ & $-0.21 \pm 0.46$ & $10.89 \pm 0.51$ 
              & $-0.64 \pm 0.60$ & $14.24 \pm 0.65$ \\

$[0.8, 0.9]$ & $-0.12 \pm 0.54$ & $11.62 \pm 0.58$ 
              & $-0.98 \pm 0.65$ & $14.36 \pm 0.70$ \\

$[0.9, 1.0]$ & $-0.45 \pm 0.49$ & $12.16 \pm 0.53$ 
              & $-1.27 \pm 0.59$ & $14.49 \pm 0.64$ \\

$[1.0, 1.1]$ & $-0.86 \pm 0.49$ & $12.70 \pm 0.53$ 
              & $-1.70 \pm 0.61$ & $14.96 \pm 0.67$ \\

$[1.1, 1.3]$ & $-1.17 \pm 0.51$ & $13.87 \pm 0.57$ 
              & $-1.80 \pm 0.62$ & $16.19 \pm 0.68$ \\

$[1.3, 1.5]$ & $-1.16 \pm 0.59$ & $14.60 \pm 0.64$ 
              & $-1.92 \pm 0.75$ & $16.59 \pm 0.82$ \\

$[1.5, 2.0]$ & $-2.11 \pm 0.69$ & $16.99 \pm 0.78$ 
              & $-2.91 \pm 0.74$ & $19.05 \pm 0.84$ \\

$[2.0, 4.0]$ & $-4.81 \pm 0.79$ & $29.00 \pm 1.12$ 
              & $-5.86 \pm 0.80$ & $32.25 \pm 1.13$ \\

\hline
\end{tabular}
\label{table:pc_error}
\end{table}

$E_{\mu}^{\mathrm{MCS}}$ can be used to estimate the energy of exiting muons in partially contained (PC) events. Figure~\ref{fig:pc_error} shows the fractional error in bins of true muon energy for PC events with at least one meter of reconstructed muon track. A Gaussian distribution is fit to each bin to estimate the bias and resolution, which are reported in Table~\ref{table:pc_error}. This table also shows the fitted bias and resolution of $E_{\mu}^{\text{MCS}}$ on the full selection of PC muons without the 1\,m requirement for comparison. As with figure~\ref{fig:fc_error}, the total area of the Gaussian fit is allowed to vary so that it can capture the Gaussian-like core of the distribution, leaving a description of the tail to the RMS statistic of the histogram. The fractional error distributions show non-Gaussian tails, which is largely caused by variations in energy resolution as a function of reconstructed muon track length. Intuitively, longer tracks yield better resolutions, and the distribution in each $E_{\mu}$ bin can be considered an integral of fractional error distributions of varying width over all reconstructed muon lengths. The overall bias is less than about 2\% below 2\,GeV and increases to 5\% at 4\,GeV. The resolution varies from 7\% to 17\% below 2\,GeV and degrades to 29\% at 4\,GeV. This performance represents a significant improvement over the previous work~\cite{mcs_2017}, which observed longer non-Gaussian tails and, as a result, a bias as large as 50\% above 2\,GeV. These improvements are believed to be the result of the model improvements discussed in section~\ref{sec:mcs_intro}.

This performance is competitive with other muon energy reconstruction methods at MicroBooNE, both for contained muons at high energy and for all exiting muons. For example, Ref.~\cite{rnn_emu} uses a residual neural network combined with range- and calorimetry-based reconstructions to achieve muon energy resolutions of 3.5\% and 25\% on FC and PC event selections, respectively. This suggests that this MCS-based estimator may be useful to improve the overall muon energy resolution through a combined estimator that combines information from multiple approaches. It is also useful to have a complementary energy estimator to help validate the reconstruction quality of high-energy muon tracks by looking for consistency across estimated energies.
\section{Model validation}\label{sec:validation}

\begin{figure}[hbtp!]
     \centering
     \includegraphics[clip,trim={1cm 1.0cm 2.0cm 1.0cm},width=0.82\textwidth]{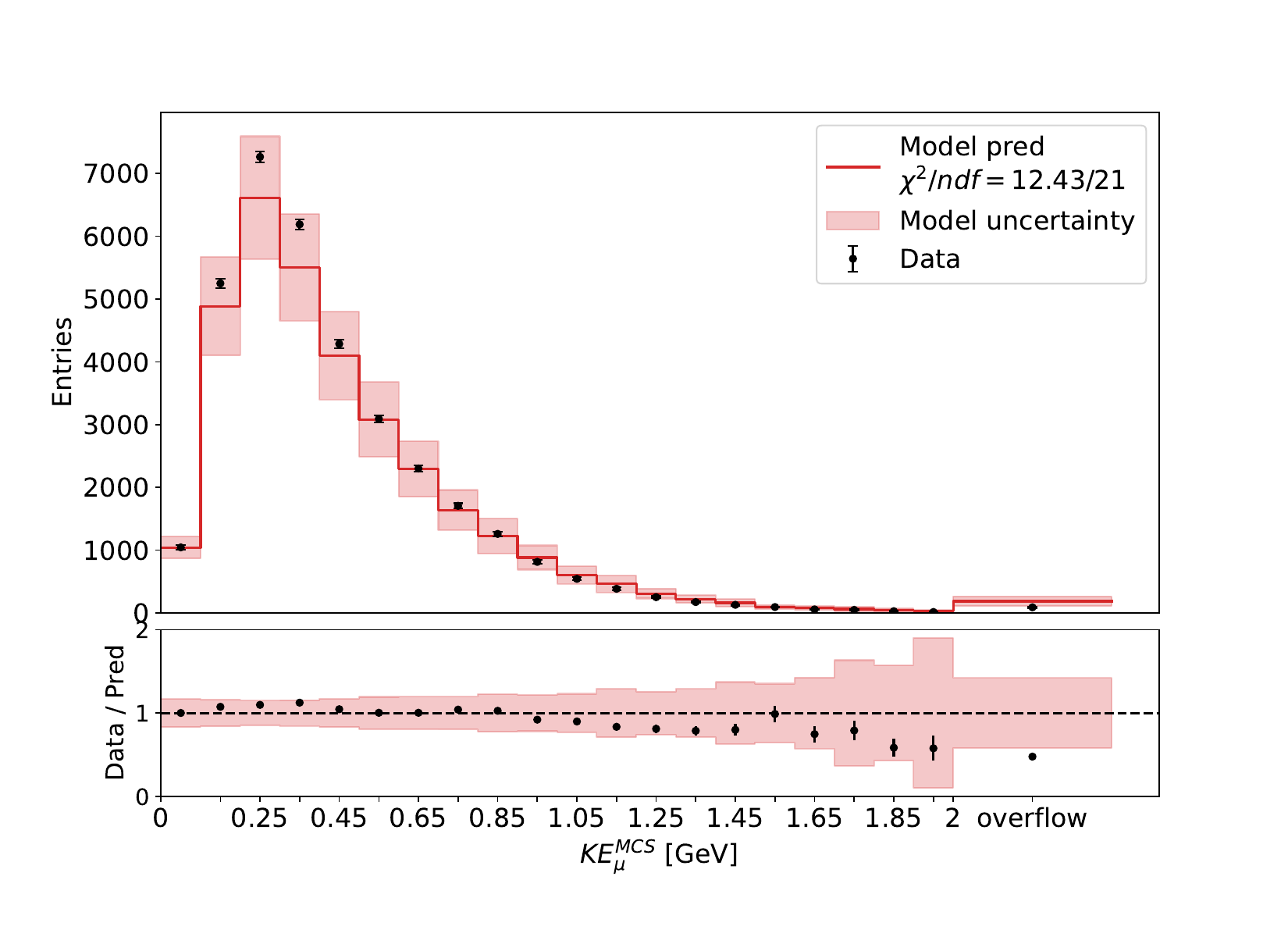}
     \put(-138,172){\small MicroBooNE $6.4 \times 10^{20}$ POT}
     \caption{Comparison of the distributions of $KE_{\mu}^{\text{MCS}}$ between data and simulation for FC muon events. Uncertainties account for statistics as well as systematics associated with the modeling of the neutrino flux, neutrino interaction modeling, and detector response.
     }
    \label{fig:fc_val}
\end{figure}


\begin{figure}[hbtp!]
     \centering
     \includegraphics[clip,trim={1cm 0.8cm 2.0cm 0.0cm},width=0.82\textwidth]{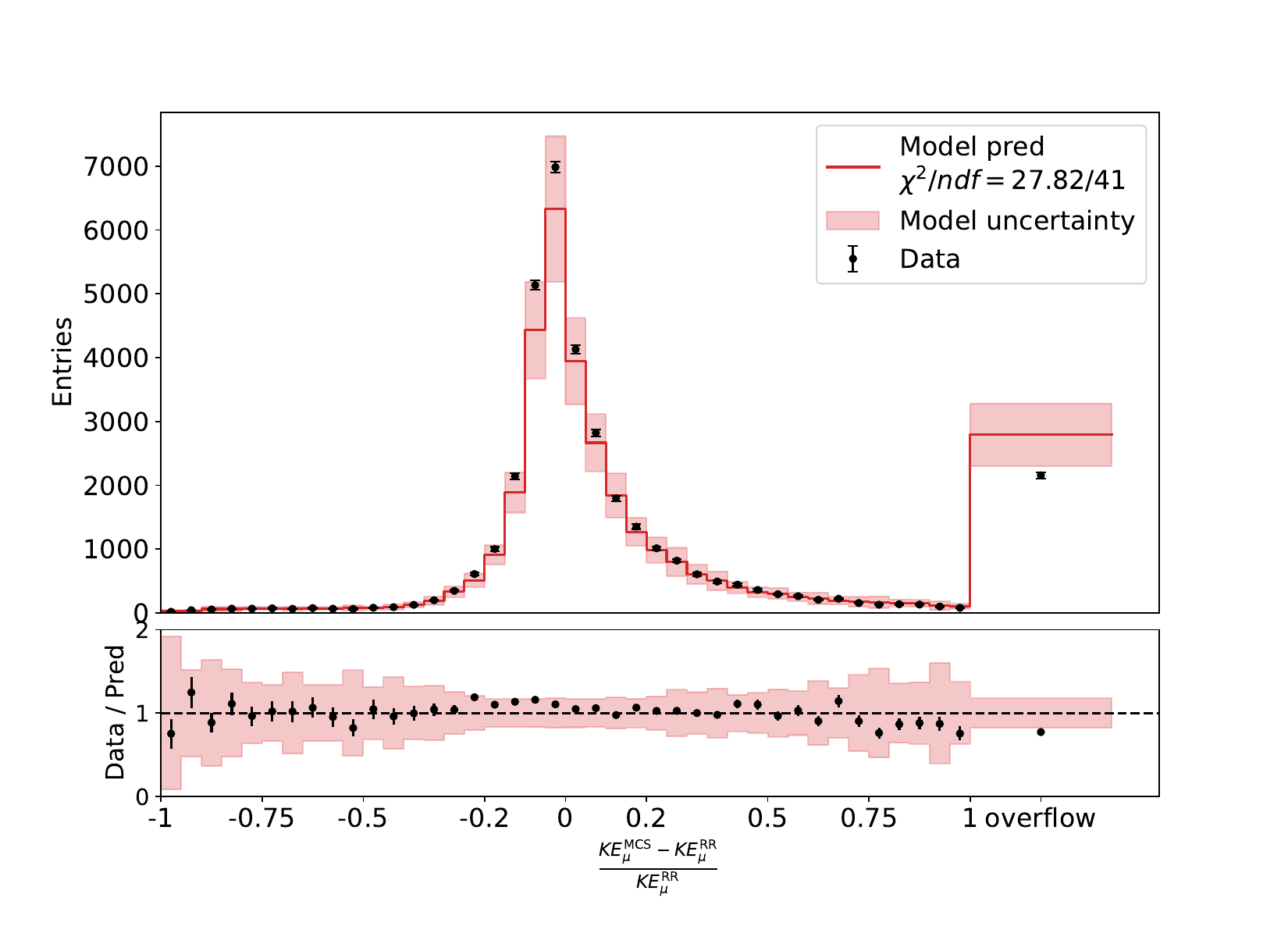}
     \put(-135,172){\small MicroBooNE $6.4 \times 10^{20}$ POT}
     \caption{Comparison of the distributions of the fractional difference between $KE_{\mu}^{\text{MCS}}$ and $KE_{\mu}^{\text{RR}}$ between data and simulation for FC muon events. Uncertainties account for statistics as well as systematics associated with the modeling of the neutrino flux, neutrino interaction modeling, and detector response.}
    \label{fig:fc_val_rr_error}
\end{figure}

So far we have only considered the performance of $E_{\mu}^{\text{MCS}}$ on simulation. In this section, we present validation of the algorithm and its tuning through comparison with real data. The data-model comparison of the estimated muon kinetic energy $KE_{\mu}^{\text{MCS}}$ for FC events is shown in figure~\ref{fig:fc_val}. The level of tension is evaluated through a $\chi^{2}$ test statistic computed from the data vector $D$, the model vector $M$, and their combined covariance matrix $C$:
\begin{equation}
    \chi^{2} = (D-M)^{T} C^{-1} (D-M).
\end{equation}
The covariance matrix includes statistical uncertainties as well as systematic uncertainties associated with the modeling of the neutrino flux~\cite{uboone_flux}, neutrino interaction and secondary interaction modeling~\cite{genie-tune-paper,geant_4}, and detector response~\cite{sce2,sce1,recombination,wiremod,light_systematics}. The observed $\chi^{2}/\text{ndf}$ of $12.4/21$ is below one and corresponds to a $p$-value of 0.93, indicating that there is sufficient uncertainty coverage in the model to explain the data-model differences present in $KE_{\mu}^{\text{MCS}}$. The comparison shows a data excess at low reconstructed muon energy and a data deficit at high reconstructed muon energy, which is consistent with previous data-model comparisons in MicroBooNE using range-based muon energy estimators that do not leverage MCS~\cite{numuCC_selection}. This suggests that these features may originate from some source of mismodeling in common between these distinct $E_{\mu}$ estimators, such as a mismodeling in the neutrino flux prediction or neutrino interaction model.

Comparisons of $KE_{\mu}^{\text{MCS}}$ between data and simulation for FC events are aided by the fact that the muon's kinetic energy can be estimated from its residual range. This estimate is labeled as $KE_{\mu}^{\text{RR}}$ and can serve as a proxy for true $KE_{\mu}$ to within 5\% precision, as determined from comparisons between $KE_{\mu}^{\text{RR}}$ and $KE_{\mu}$ in simulation. Figure~\ref{fig:fc_val_rr_error} shows the distribution of the fractional difference between $KE_{\mu}^{\text{MCS}}$ and $KE_{\mu}^{\text{RR}}$, expressed as a fraction of $KE_{\mu}^{\text{RR}}$, in both data and simulation. There is very good agreement between the two distributions, with a $\chi^{2}/\text{ndf}$ below one. This provides a good indication that the treatment of the $E_{\mu}^{\text{MCS}}$ estimator between data and simulation is well within uncertainties.


\begin{figure}[hbtp!]
     \centering
     \includegraphics[clip,trim={1.0cm 0.9cm 2.0cm 2.3cm},width=0.60\textwidth]{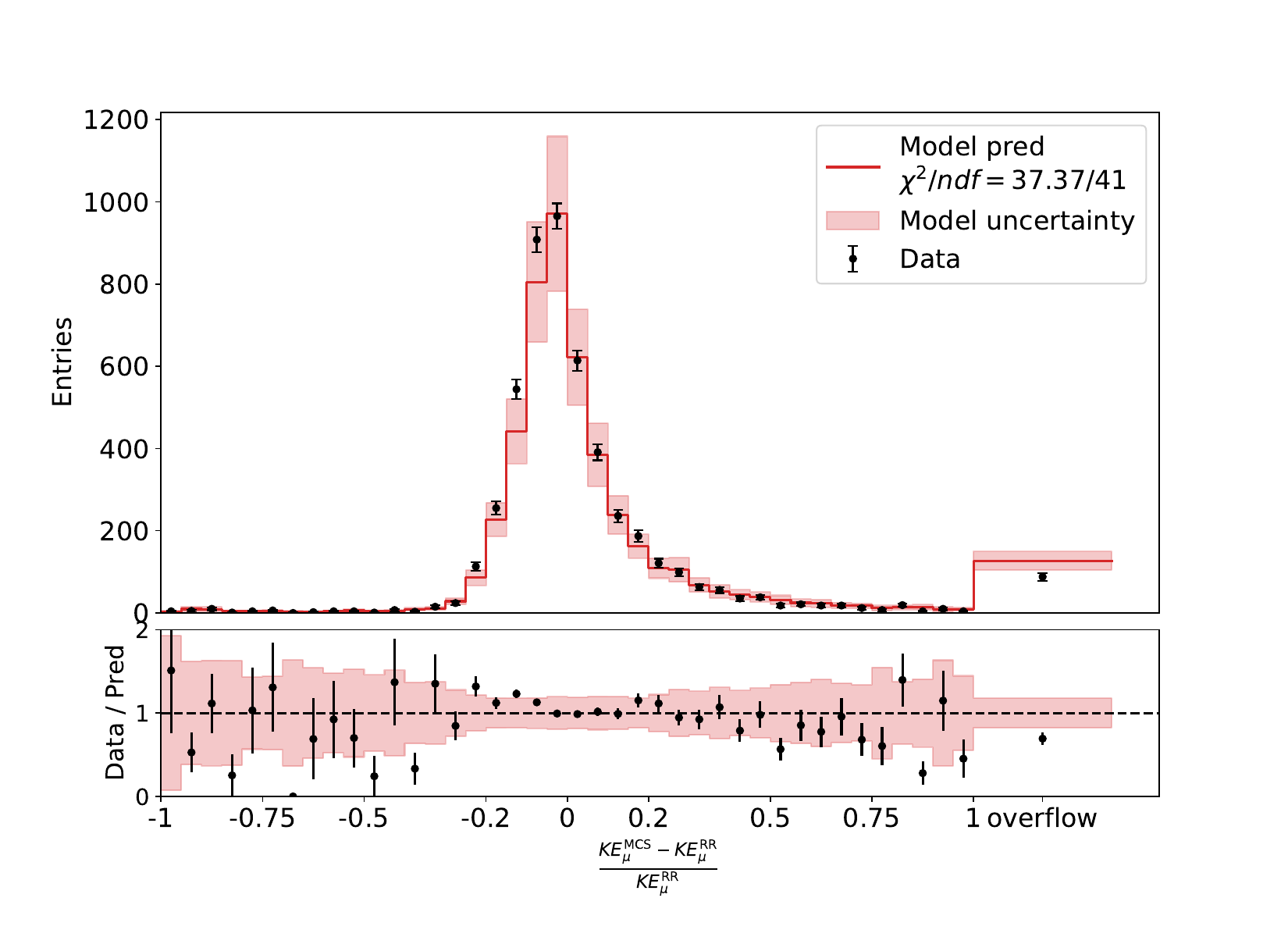}
     \includegraphics[clip,trim={1.0cm 0.9cm 2.0cm 2.3cm},width=0.60\textwidth]{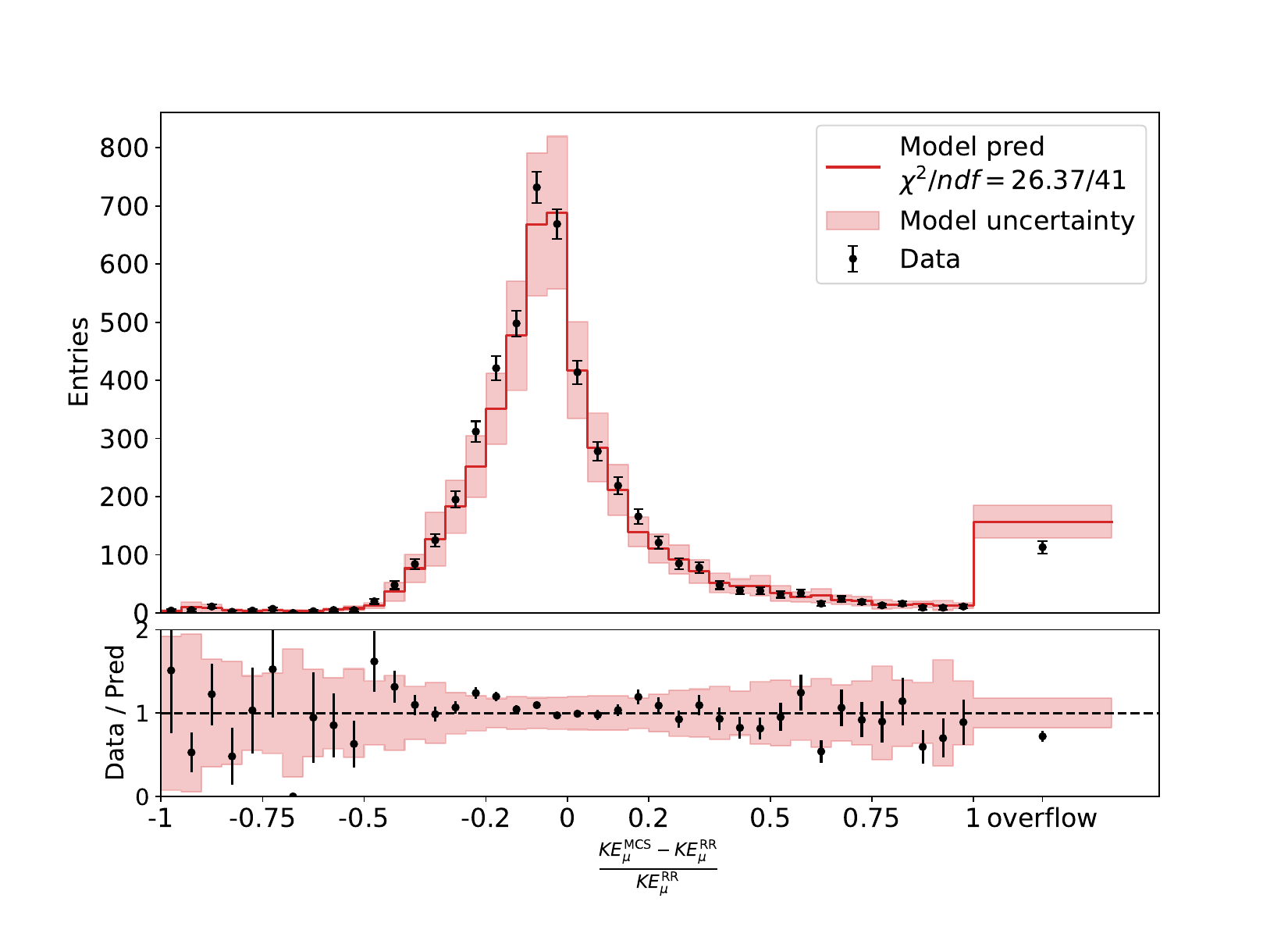}
     \includegraphics[clip,trim={1.0cm 0.9cm 2.0cm 2.3cm},width=0.60\textwidth]{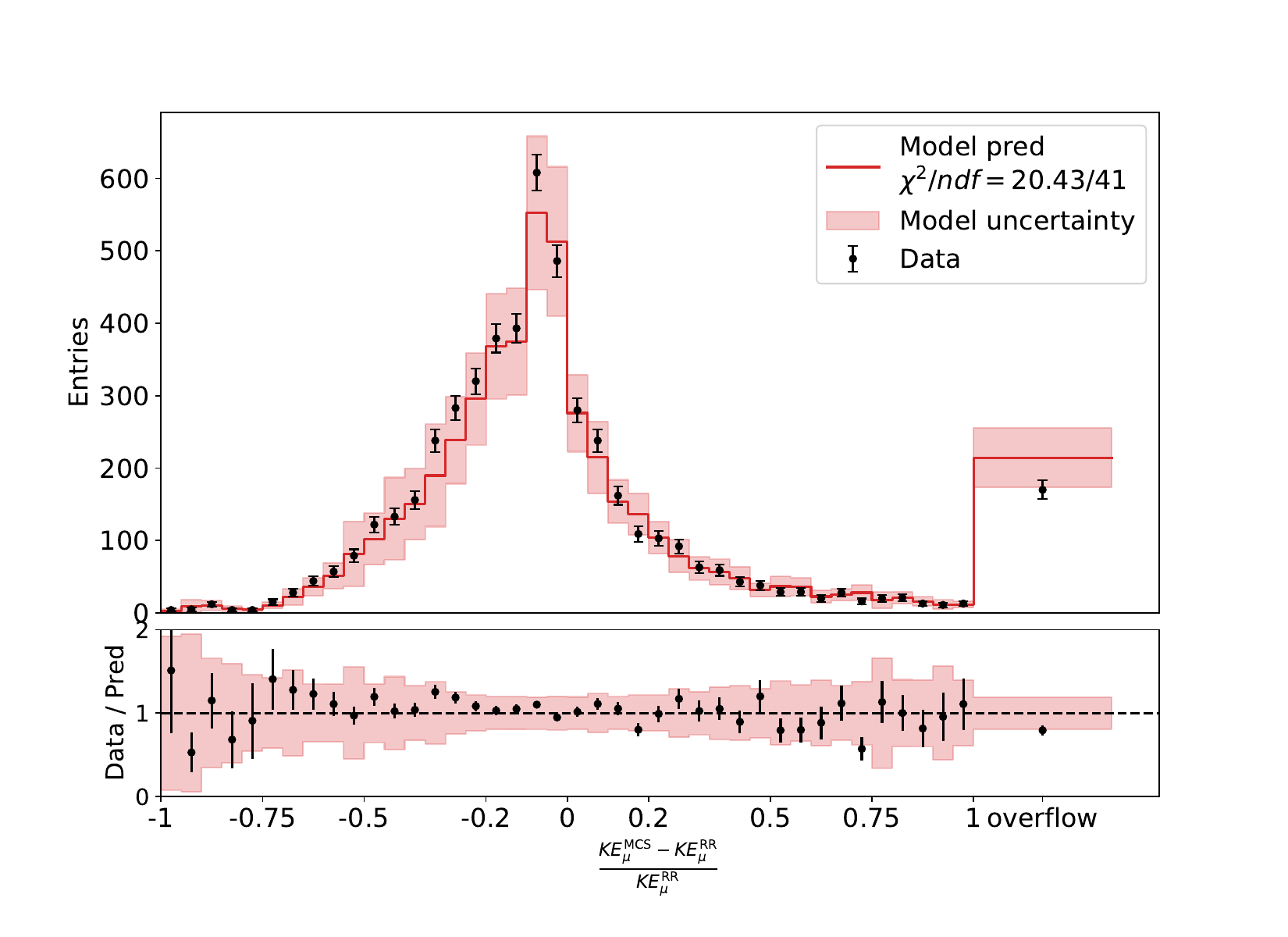}
     \put(-110,120){\scalebox{1.0}{\scriptsize MicroBooNE $6.4 \times 10^{20}$ POT}}
     \caption{Comparison of the distributions of the fractional difference between $KE_{\mu}^{\text{MCS}}$ and $KE_{\mu}^{\text{RR}}$ between data and simulation for FC muon events. To mimic performance on exiting muons, the calculation of $KE_{\mu}^{\text{MCS}}$ is performed without information from the final two, four, and six track segments in the top, middle, and bottom plots, respectively. Because of the high computation time required to re-simulate events under each detector response variation, the detector response uncertainties calculated without any segments removed are used instead. Other uncertainties, including statistical, neutrino flux modeling, and neutrino interaction modeling, are re-computed for each distribution.}
    \label{fig:mcs_segs_omitted}
\end{figure}

It is also possible to use FC events to partially recreate the circumstances of PC events. The end of a muon track is omitted from the information used to determine $KE_{\mu}^{\text{MCS}}$, mimicking the situation of an exiting muon where the track end is not observed. Meanwhile, $KE_{\mu}^{\text{RR}}$ is still calculated using the full track length, allowing it to continue to serve as a reasonable proxy of true $E_{\mu}$. This allows for the fractional difference in energy estimates to be examined under pseudo-partially-contained circumstances. Figure~\ref{fig:mcs_segs_omitted} shows these data-model comparisons when each of two, four, and six track segments are omitted in the computation of $KE_{\mu}^{\text{MCS}}$ from events containing at least eight muon segments so that an angle measurement is always calculable. Because of limits in production resources, detector response uncertainties are not recalculated in each case, and instead uncertainties from the full track length are used. This is expected to underestimate the detector response uncertainties, as the $KE_{\mu}^{\text{MCS}}$ estimator has more variance for exiting muons than for contained muons. Each comparison shows good data-model agreement with $\chi^{2}/\text{ndf}$ all below one.

\begin{figure}[hbtp!]
     \centering
     \includegraphics[clip,trim={1.0cm 1.0cm 2.0cm 1.0cm},width=0.82\textwidth]{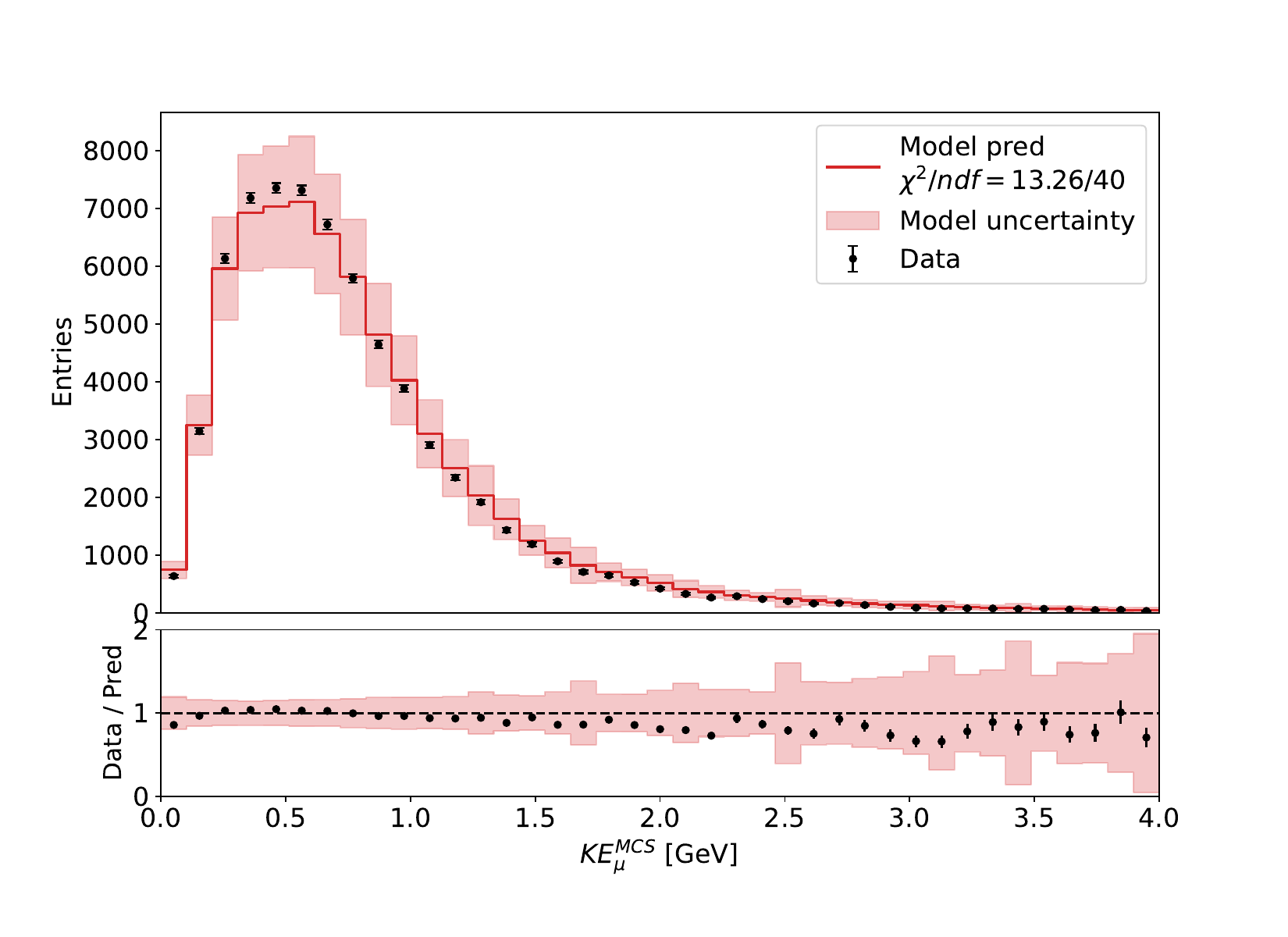}
     \put(-135,172){\small MicroBooNE $6.4 \times 10^{20}$ POT}
     \caption{Comparison of the distributions of $KE_{\mu}^{\text{MCS}}$ in data and simulation for PC muon events. Uncertainties account for statistics as well as systematics associated with the modeling of the neutrino flux, neutrino interaction modeling, and detector response.}
    \label{fig:pc_val}
\end{figure}

Finally, we can also directly compare $KE_{\mu}^{\text{MCS}}$ distributions between data and simulation for actual partially contained events, shown in figure~\ref{fig:pc_val}. The observed $\chi^{2}/\text{ndf}$ of $13.3/41$ is well below one and corresponds to a $p$-value above 0.99, demonstrating very strong agreement within model uncertainties. Combined with the above studies on FC and pseudo-PC events, this shows that the $E_{\mu}^{\text{MCS}}$ estimator is reliably described in simulation, validating its use on data.
\section{Conclusion}\label{sec:conclusion}

We present a new method for using the measurement of multiple Coulomb scatterings to estimate a muon's energy. This method introduces a more detailed PDF to describe the distribution of scattering angles expected at a given energy, a new approach to more accurately determine those angles' resolutions in LArTPC detectors, and a new parametrization of those angles that separates detector drift and wire-plane directions and allows differences in reconstruction quality to be well described.

At fixed $E_{\mu}$ and reconstructed track length, the $E_{\mu}^{\text{MCS}}$ estimator is predominantly Gaussian distributed with minimal skew, improving on the performance of previous work in MicroBooNE~\cite{mcs_2017}. The $E_{\mu}^{\text{MCS}}$ estimator also contains minimal bias, less than 1\% for contained muons, 2\% for exiting muons with energy below 2\,GeV, and 5\% for exiting muons with energy above 2\,GeV. The $E_{\mu}^{MCS}$ estimator shows an improved resolution that is as low as 4.3\% for contained muons, and between 7\% and 17\% for exiting muons that are longer than a meter and have energy less than 2\,GeV. The excellent performance in simulation is supported by a series of successful model validation tests. These investigate data-model differences over FC and PC distributions of $E_{\mu}^{\text{MCS}}$, as well as for the fractional difference between the $E_{\mu}^{\text{MCS}}$ and $E_{\mu}^{\text{RR}}$ estimators for FC and pseudo-PC event selections. In each case, model uncertainties fully cover the data-model differences, demonstrating that simulation provides a reliable estimate of the performance of $E_{\mu}^{\text{MCS}}$ in data.

The use of multiple Coulomb scattering measurement in muon energy estimation uniquely enables reliable reconstruction of exiting muons, extending the phase space available for analyses to higher energies and wider angles. This work demonstrates the success of new techniques that construct a more refined model of MCS in LArTPC detectors. Similar approaches may be useful in the implementation of MCS-based muon energy reconstruction in other LArTPC detectors, including those of the short baseline neutrino and deep underground neutrino experiment programs~\cite{sbn, dune_enu}.

\acknowledgments

This document was prepared by the MicroBooNE collaboration using the resources of the Fermi National Accelerator Laboratory (Fermilab), a U.S. Department of Energy, Office of Science, Office of High Energy Physics HEP User Facility. Fermilab is managed by Fermi Forward Discovery Group, LLC, acting under Contract No. 89243024CSC000002. MicroBooNE is supported by the
following: 
the U.S. Department of Energy, Office of Science, Offices of High Energy Physics and Nuclear Physics; 
the U.S. National Science Foundation; 
the Swiss National Science Foundation; 
the Science and Technology Facilities Council (STFC), part of United Kingdom Research and Innovation (UKRI);
the Royal Society (United Kingdom);
the UKRI Future Leaders Fellowship;
the NSF AI Institute for Artificial Intelligence and Fundamental Interactions;
and the European Union’s Horizon 2020 research and innovation programme under the Marie Sk\l{}odowska-Curie grant agreement No. 101003460 (PROBES). Additional support for the laser calibration system and cosmic ray tagger was provided by the Albert Einstein Center for Fundamental Physics, Bern, Switzerland. We also acknowledge the contributions of technical and scientific staff to the design, construction, and operation of the MicroBooNE detector as well as the contributions of past collaborators to the development of MicroBooNE analyses, without whom this work would not have been possible. For the purpose of open access, the authors have applied a Creative Commons Attribution (CC BY) public copyright license to any Author Accepted Manuscript version arising from this submission.

\begin{appendices}
    \section{Iterative tune details}\label{sec:appendix}

An iterative tuning approach is used to give more control to the tune to avoid parameter degeneracy and undesired function behavior. The primary Gaussian width $\sigma_{1}$ is strongly tied to the Highland formula so it is tuned first. The area fraction $A$ is tuned second as it is significantly constrained by the value of $\sigma_{1}$ and the maximum amplitude of the observed angle distribution. Finally, $\sigma_{2}$ is tuned, and captures the long tails in the distribution. Before each round of tuning, double-Gaussian PDFs are fit to the angle distributions across energy bins, with any previously tuned parameter determined by that tune, and all remaining parameters left free to vary. The fits of these freely varying parameters form the inputs for the next parameter tune. Tunes of $\sigma_{1}(E_{\mu})$ and $\sigma_{2}(E_{\mu})$ from equation~\ref{eqn:double_gaussian} are performed by fitting parameters in scaling functions $\kappa_{1}(E_{\mu})$ and $\kappa_{2}(E_{\mu})$, respectively. These scaling functions take the form of a quartic polynomial suppressed by a decaying exponential term, which allows them to capture reconstructed angle distributions while also converging to a detector-resolution-dominated prediction at high $E_{\mu}$.
\begin{align}
    u & = \frac{E_{\mu}}{E_{0}}, \\
    \kappa(E_{\mu}) & = 1 + (\alpha_{0} + \alpha_{1} u + \alpha_{2} u^{2} + \alpha_{3} u^{3} + \alpha_{4} u^{4} )e^{-u},
\end{align}
where the $\alpha_{i}$ and $E_{0}$ are free parameters in the scaling function. The area fraction $A(E_{\mu})$ is parametrized with a sigmoid function, which allows for a gentle transition from a larger primary Gaussian contribution at low energies to a slightly smaller primary Gaussian contribution at high energies.
\begin{equation}
    A(E_{\mu}) = \beta_{0} + (\beta_{1}-\beta_{0}) \left ( 1 - \frac{1}{1+e^{-\beta_{2}(E_{\mu}-\beta_{3})}} \right ),
\end{equation}
where the $\beta_{i}$ are free parameters in the sigmoid function. The step-by-step tuning procedure is outlined below.
\begin{enumerate}
    \item Initial round of double-Gaussian fits: In each bin $i$ with average energy $E_{\mu \text{ }i}$, a double-Gaussian PDF is fit to the reconstructed scattering angle distribution in simulation, determining the fit parameters $\sigma_{1 \text{ }i}^{(0)}$, $\sigma_{2 \text{ }i}^{(0)}$, and $A_{i}^{(0)}$.
    
    \item Tune of $\sigma_{1}(E_{\mu})$: Primary Gaussian widths $\sigma_{1 \text{ }i}^{(0)}$ are compared at each bin $i$ to the Highland formula prediction in equation~\ref{eqn:highland_full}. The scaling function $\kappa_{1}(E_{\mu})$ is tuned to achieve the best agreement.
    
    \item Second round of double-Gaussian fits: In each bin $i$ with average energy $E_{\mu \text{ }i}$, a double-Gaussian PDF is fit to the reconstructed scattering angle distribution in simulation, determining the fit parameters $\sigma_{2 \text{ }i}^{(1)}$, and $A_{i}^{(1)}$. The primary Gaussian width is not fit, but is instead fully determined by the tune of $\kappa_{1}(E_{\mu})$.

    \item Tune of $A(E_{\mu})$: Area fractions $A_{i}^{(1)}$ are compared at each bin $i$ to the Highland formula prediction in equation~\ref{eqn:highland_full}. The parametrized area fraction $A(E_{\mu})$ is tuned using a sigmoid function to achieve the best agreement.

    \item Third round of double-Gaussian fits: In each bin $i$ with average energy $E_{\mu \text{ }i}$, a double-Gaussian PDF is fit to the reconstructed scattering angle distribution in simulation, determining the fit parameter $\sigma_{2 \text{ }i}^{(2)}$. The primary Gaussian width and area fraction are not fit, but are instead fully determined by the tunes of $\kappa_{1}(E_{\mu})$ and $A(E_{\mu})$, respectively.

    \item Tune of $\sigma_{2}(E_{\mu})$: Secondary Gaussian widths $\sigma_{2 \text{ }i}^{(2)}$ are compared at each bin $i$ to the Highland formula prediction in equation~\ref{eqn:highland_full} enlarged by a constant scaling. The scaling function $\kappa_{2}(E_{\mu})$ is tuned to achieve the best agreement.
\end{enumerate}

\section{MCS algorithm implementation}

So far the MCS algorithm has been described conceptually, without much attention to the details of implementation. This section attempts to provide some guidance for readers trying to reproduce this work for other analyses. The main body of the algorithm code is provided on Github~\cite{mcs_github} along with reconstructed 3D space-points of a muon track over which the MCS algorithm can be run. These space-points were generated from a dedicated particle trajectory reconstruction algorithm, reducing the impact of artifacts such as delta rays, crossing tracks, and noisy reconstructions. In principle this set of points can be replaced with any suitable point cloud describing a muon trajectory in a LArTPC, however algorithm performance will benefit from an effort to clean-up input data by removing artifacts such as those previously mentioned. Within the MCS algorithm there is a further data pre-processing step that attempts to find a shortest path through the point cloud and remove points not associated with it, which serves to further remove undesirable reconstruction artifacts.

Once the input data is fully prepared, a multi-step process is used to ultimately determine the scattering angles between segments. First, principal component analysis (PCA) is used to determine the particle's overall direction and to sort points along this direction. Then, points are separated into bins, each 14cm long, along the principal component direction. PCA is used to fit a particle direction to the points in each bin, and orthogonal local coordinates $x$ and $y$ are determined following equation~\ref{eqn:local_coords}, to maximally align the local $x$ direction with the drift direction. From these, angles $\theta_{xz}$ and $\theta_{yz}$ are measured. The likelihood function, given in equation~\ref{eqn:likelihood_fn}, takes the measured angles as inputs, as well as the distance to each angle location, so that the estimated starting energy can be propagated to each angle location to form the PDF at that location. This propagation is performed by converting an estimated energy to a residual range, subtracting the distance to the segment from this range, and then converting back to a new energy estimate. This approach is consistent with using the Bethe-Bloch formula~\cite{bethe-bloch} to estimate the energy loss. The final step in the algorithm is to maximize the total likelihood by varying the starting muon energy estimate, so that the optimal energy estimate can be obtained.
\end{appendices}

\bibliographystyle{JHEP}
\bibliography{multiple_coulomb_scattering}

\end{document}